\pdfoutput=1
\documentclass[namedreferences]{solarphysics}
\usepackage[optionalrh,solaromanenum]{spr-sola-addons} 
\usepackage{color}    
\usepackage{url}

\usepackage[margin=1.25in]{geometry}
\usepackage[pdfborder={0 0 0 },urlcolor=blue,breaklinks]{hyperref}
\ifx \urlurl    \undefined \def \urlurl#1{\href{http://#1}{\textsf{#1}}}\fi
\ifx \doiurl    \undefined \def \doiurl#1{\href{http://dx.doi.org/#1}{\textsf{\textsf{DOI}}}}\fi
\ifx \adsurl    \undefined \def \adsurl#1{\href{http://adsabs.harvard.edu/abs/#1}{\textsf{\textsf{ADS}}}}\fi
\ifx \arxivurl  \undefined \def \arxivurl#1{\href{http://arxiv.org/abs/#1}{\textsf{\textsf{arXiv}}}}\fi

\usepackage{graphicx}
\usepackage{caption}
\usepackage{listings}
\usepackage{lineno}
\usepackage{hyperref}
\hypersetup{colorlinks=true, citecolor=black, filecolor=black, linkcolor=black}
\PassOptionsToPackage{linktocpage}{hyperref}

\begin{document}
\begin{article}
\begin{opening}
\title{The {\it Helioseismic and Magnetic Imager} (HMI) Vector Magnetic Field Pipeline: SHARPs -- Space-weather HMI Active Region Patches}
\author
        {M.G.~\surname{Bobra}$^{1}$\sep
        X.~\surname{Sun}$^{1}$\sep
        J.T.~\surname{Hoeksema}$^{1}$\sep
        M.~\surname{Turmon}$^{2}$\sep
        Y.~\surname{Liu}$^{1}$\sep
        K.~\surname{Hayashi}$^{1}$\sep
        G.~\surname{Barnes}$^{3}$\sep 
        K.D.~\surname{Leka}$^{3}$\\}

\institute{$^{1}$ W.W. Hansen Experimental Physics Laboratory, Stanford University, Stanford, CA, USA  \\
	email: \href{mailto:jthoeksema@sun.stanford.edu}{jthoeksema@sun.stanford.edu} \\
	$^{2}$ Jet Propulsion Laboratory, Pasadena, CA, USA \\
	$^{3}$ Northwest Research Associates, Inc., Boulder, CO, USA \\}
	
\runningauthor{M.G. Bobra {\textit{et al.}}}
\runningtitle{{\sf SHARP}s - Space-weather HMI Active Region Patches}
                
\begin{abstract}
A new data product from the \textit{Helioseismic and Magnetic Imager}
(HMI) onboard the \textit{Solar Dynamics Observatory} (SDO) called Space-weather
HMI Active Region Patches ({\sf SHARP}s) is now available. SDO/HMI is the first
space-based instrument to map the full-disk photospheric vector magnetic
field with high cadence and continuity. The {\sf SHARP} data series provide maps
in patches that encompass automatically tracked magnetic concentrations for
their entire lifetime; map quantities include the photospheric vector magnetic
field and its uncertainty, along with Doppler velocity, continuum intensity,
and line-of-sight magnetic field.  Furthermore, keywords in the {\sf SHARP} data
series provide several parameters that concisely characterize the magnetic-field 
distribution and its deviation from a potential-field configuration.
These indices may be useful for active-region event forecasting and for
identifying regions of interest.  The indices are calculated per patch and
are available on a twelve-minute cadence. Quick-look data are available
within approximately three hours of observation; definitive science products
are produced approximately five weeks later. {\sf SHARP} data are available at
\href{http://jsoc.stanford.edu}{{\sf jsoc.stanford.edu}} and maps are available in either of two different
coordinate systems.  This article describes the {\sf SHARP} data products and presents
examples of {\sf SHARP} data and parameters.  
\end{abstract}


\end{opening}

 \section{Introduction}
 \label{s:Introduction} 

This article describes a data product from the \textit{Solar Dynamics Observatory's
Helioseismic and Magnetic Imager} (SDO/HMI) called Space-weather HMI Active
Region Patches ({\sf SHARP}s). {\sf SHARP}s follow each significant patch of solar
magnetic field from before the time it appears until 
after it disappears. 
The {\sf SHARP} data series presently include 16 indices
computed from the vector magnetic field in active-region patches.
These parameters, many of which have been associated with enhanced flare productivity,  
are automatically calculated for each solar active region using HMI vector
magnetic-field data with a 12-minute cadence. The indices and other 
keywords can be used to select regions and time intervals for further study.  
The active-region patches are 
automatically identified and tracked for their entire lifetime \cite{Turmon2013}.
In addition to the indices, the four
{\sf SHARP} data series include the photospheric vector magnetic-field 
data for the patches, as well as co-registered maps of Doppler velocity,
continuum intensity, line-of-sight magnetic field, and other quantities.

Measurements of the photospheric magnetic field provide insight into
understanding and possibly predicting eruptive phenomena in the solar
atmosphere, such as flares and coronal mass ejections. For example,
it is generally accepted that large, complex, and rapidly evolving
photospheric active regions are the most likely to produce eruptive events
\cite{zirin,priest}. As such, it is an active area of research to seek a
correlation (or its rejection) between eruptive events and quantitative
parameterizations of the photospheric magnetic field. Many studies have found
a relationship between solar-flare productivity and various indices: magnetic helicity
(\textit{e.g.} \opencite{tian}; \opencite{torokkliem}; \opencite{labonte}), free
energy proxies (\textit{e.g.} \opencite{moore44}), magnetic shear angle (\textit{e.g.} 
\opencite{hagyard1984}; 
\opencite{lekabarnes2003}, \citeyear{lekabarnes2}, \citeyear{lekabarnes2007}), 
magnetic topology (\textit{e.g.} \opencite{Cui2006}; \opencite{lekabarnes2006}, \opencite{grust}), or
the properties of active-region polarity inversion lines (\textit{e.g.} \opencite{mason}; \opencite{falconer56};
\opencite{schrijverR}). 
However, when \inlinecite{lekabarnes2003} conducted a discriminant analysis
of over a hundred parameters calculated from vector magnetic-field measurements of seven active regions,
they could identify ``no single, or even
small number of, physical properties of an active region that is sufficient
and necessary to produce a flare." 
Larger statistical samples show correlations between some vector-field non-potentiality parameters and overall flare productivity
\cite{lekabarnes2007,Yang2012}, as well as correlations between the parameters themselves.
Still, characteristics have yet to be identified that uniquely distinguish imminent flaring in an active region.

The {\sf SHARP} data series will
provide a complete record of all visible solar active regions
since 1 May 2010.  {\sf SHARP} data are stored in a database and readily accessible 
at the Joint Science Operations Center (JSOC). 
JSOC data products from SDO, as well as source code for the modules, can
be found at \href{http://jsoc.stanford.edu}{{\sf jsoc.stanford.edu}}. Continuously updated plots of near-real-time
parameters are available online (see Table~\ref{tab:urls} for URLs).  We describe
how the {\sf SHARP} series are created and show results for two representative
active regions.  We also present examples of four active-region parameters
for 12 X-, M-, and C-class flaring active regions.

\section{Methodology: SHARP Data and Active Region Parameters}
\label{s:method}
 
Data taken onboard SDO/HMI are downlinked to the ground, automatically processed
through the HMI data pipeline, and made available at \href{http://jsoc.stanford.edu}{{\sf jsoc.stanford.edu}} organized in data series \cite{Schou-calib2012,Scherrer2012}.  Conceptually, a JSOC data
series consists of a sequence of {\em records}, each of which includes i)
a table of keywords and ii) associated data arrays, called {\em segments}. 
A record exists for each
time step or unique set of prime keyword(s). Keywords and data array segments are
merged by the JSOC into FITS files in response to a user's request to download
(or export) the data series.  {\sf SHARP} data for export can be selected by time,
given in the keyword {\sc t\_rec},
and the region number, {\sc harpnum}; additionally, requests for data from the JSOC can also take 
advantage of simple SQL database queries on keywords to select data of interest.
A complete overview of JSOC Data Series is available on the JSOC wiki
(see Table~\ref{tab:urls}).

Certain HMI data series are processed on two time
scales: near real time (NRT) and definitive. NRT data are processed quickly,
ordinarily within three hours of the observation time, but with preliminary
calibrations.  Section \ref{sec:NRT} describes the differences between definitive and NRT HARPs.
Although most NRT data series are not archived and go offline
after approximately three months, the NRT {\sf SHARP} data since 14 September 2012 are archived. 
NRT data are primarily intended for quick-look monitoring or as a forecasting tool. 
This section briefly describes the elements of
the HMI data pipeline necessary to create the definitive {\sf SHARP} data. A more detailed
explanation of the HMI vector magnetic-field pipeline processing is given by
\inlinecite{hoeksema2013} and references therein.

\begin{itemize}
\item In each 135-second interval HMI samples six points across the Fe\,{\sc i}
6173.3\,\AA\ spectral line and measures six polarization states: $I \pm Q$,
$I\pm U$, and $I\pm V$, generating 36 4096\,$\times$\,4096 full-disk filtergrams.

\item To reduce noise and minimize the effects of solar oscillations, a
tapered temporal average is performed every 720 seconds using 360 filtergrams
collected over a 1350-second interval to produce 36 corrected, filtered, co-registered
images \cite{Couvidat-wavelength2012}.

\item A polarization calibration is applied and the four Stokes polarization states, 
[$I\,Q\,U\,V$], are determined at each wavelength, giving a total of 24
images at each time step \cite{Schou-polarization2012} that are available
in the data series {\sf hmi.S\_720s}.


{\begin{table}
\caption{Relevant Uniform Resource Locators}
\begin{flushleft}
\begin{tabular}{ p{7.4cm}  p{6.0cm}}
Uniform Resource Locator & Description \\ \hline
\urlurl{jsoc.stanford.edu/data/hmi/sharp/dataviewer}
  & Continuously updated plots of near-real-time {\sf SHARP} parameters. \\
\urlurl{jsoc.stanford.edu/doc/data/hmi/sharp/sharp.htm}
  & Description of the {\sf SHARP} data product. \\
\urlurl{jsoc.stanford.edu/jsocwiki/DataSeries}
  & A complete overview of the Joint Science and Operations Center (JSOC) Data Series. \\ 
\urlurl{jsoc.stanford.edu/jsocwiki/PipelineCode}
  & Guide to HMI pipeline code and processing notes. \\
\urlurl{jsoc.stanford.edu/jsocwiki/Lev1qualBits}
  & Description of bits in {\sc quality} keyword. \\ 
\parbox{7.4cm}{\vspace{4pt}
 \urlurl{jsoc.stanford.edu/cvs/JSOC/proj/sharp/apps/sharp.c} \\
 \href{http://jsoc.stanford.edu/cvs/JSOC/proj/sharp/apps/sw_functions.c}{{\sf
../cvs/JSOC/proj/sharp/apps/sw\_functions.c}}}
  & The {\sf SHARP} data are created via this publicly available C module (sharp.c) that 
includes a library of active region parameter calculations (sw\_functions.c). \\
\urlurl{jsoc.stanford.edu/jsocwiki/sharp\_coord}
  & A technical note on {\sf SHARP} coordinate systems, mapping, and vector transformations \cite{Sun2013}. \\
\urlurl{jsoc.stanford.edu/jsocwiki/HARPDataSeries}
  & Description of the HARP data deries \cite{Turmon2013} \\
\urlurl{hmi.stanford.edu/magnetic}
  & Portal to HMI magnetic field data, image catalogs, coverage maps, and documentation \\
\urlurl{www.lmsal.com/sdouserguide.html}
  & Comprehensive guide to SDO Data Analysis \\
\end{tabular}
\end{flushleft}
\caption*{Listed below are URLs relevant for finding the {\sf SHARP} data, codes,
documentation, and data visualizations. These URLs will be maintained for
at least the duration of the SDO mission.}
\label{tab:urls}
\end{table}} 


{\begin{figure}
\centering
\caption{Detection of HARPs}
\includegraphics[width=\textwidth]{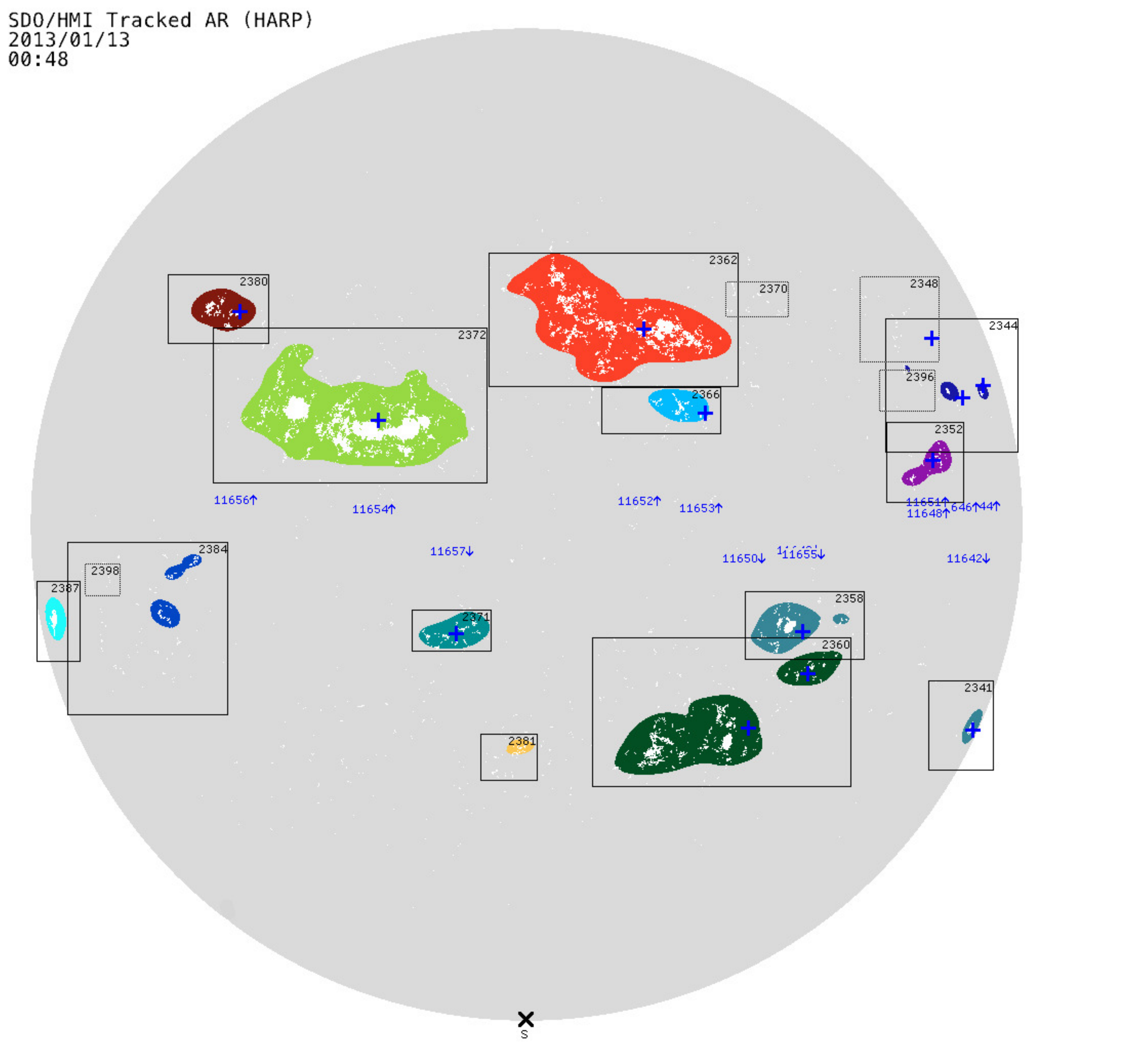}
\caption*{The results of the active-region automatic
detection algorithm applied to the data on 13 January 2013 at 00:48 TAI. NOAA
active-region numbers are labeled in blue near the Equator, next to arrows
indicating the hemisphere; the HARP number is indicated inside the rectangular
bounding box at the upper right.
Note that HARP 2360 (lower right, in green) includes two NOAA active regions,
11650 and 11655. The colored patches show coherent magnetic structures that comprise the HARP. White
pixels have a line-of-sight field strength above a line-of-sight magnetic-field threshold \cite{Turmon2013}.
Blue ``$+$'' symbols indicate coordinates that correspond to the reported center of a NOAA active region. The
temporal life of a definitive HARP starts when it rotates onto the visible
disk or two days before the magnetic feature is first identified in the
photosphere. As such, empty boxes, \textit{e.g.} HARP 2398 (on the left), represent patches of
photosphere that will contain a coherent magnetic structure at a future time.}
\label{fig:harp}
\end{figure}}


\item Active-region patches are automatically detected and tracked in the photospheric
line-of-sight magnetograms \cite{Turmon2013}. The detection algorithm
identifies both a rectangular bounding box on the CCD image that encompasses the
entire region and, within this box, creates a bitmap that both encodes membership in
the coherent magnetic structure and indicates strong-field
pixels. Specifically, the bitmap array assigns a value to each pixel inside
the bounding box, depending on whether it i) resides inside or outside
the active region, and ii) corresponds to weak or strong line-of-sight
magnetic field. This coding scheme permits non-contiguous active-region patches.

\item The tracking module numbers each HMI Active Region Patch (HARP) and
generates a time series of bitmaps large enough to contain the maximum 
known heliographic extent of the region. 
Each numbered HARP (keyword {\sc harpnum}) corresponds to 
one active region or AR complex (see Figure \ref{fig:harp}). The HARP database
generally captures more patches of solar magnetic activity than the NOAA active
region database because coherent regions that are small in extent or have no
associated photometric sunspot are detected and tracked by our code; such faint HARPs
often have no NOAA correspondence. A HARP may include zero, one, or multiple NOAA
active regions (for example, see HARP 2360 in Figure \ref{fig:harp}); about
one-third of HARPs correspond to a single NOAA region. 
The bitmap array described above is in the {\sc bitmap} segment of
the data series {\sf hmi.Mharp\_720s}. 

\item The full-disk Stokes data are inverted using the Very Fast Inversion
of the Stokes Vector ({\sf VFISV}) code, which assumes a Milne--Eddington
model of the solar atmosphere, to yield vector magnetic field data
\cite{vfisv2010,centeno2013}. Inverted data are available in the data series
{\sf hmi.ME\_720s\_fd10}.
Full disk inversions are being computed for all HMI data since 1 May 2010. An improvement 
made to the inversion code in May 2013 \cite{hoeksema2013} to use time-dependent information about the HMI filter profiles
introduces measurable systematic differences in inversion results.
Data in the interval from 1 August 2012\,--\,24 May 2013 were
processed before the improvement.
Some care must be taken when comparing data computed with different versions of the analysis code (see the entry under
\href{http://jsoc.stanford.edu/jsocwiki/PipelineCode}{{\sf PipelineCode}} referenced in Table \ref{tab:urls}.

\item The azimuthal component of the vector magnetic field is disambiguated
using the Minimum Energy Code (ME0) to resolve the 180$^\circ$ ambiguity
\cite{metcalf,leka2009}. 
Through 14 January 2014 {\sf SHARP} regions have been disambiguated individually
using {\tt fd10} data inside a rectangle that extends beyond the HARP
bounding box by the number of pixels given in the {\sc ambnpad} keyword.
Disambiguation results for each {\sc harpnum} at each time step are stored in the 
{\sc disambig} segment of the {\sf hmi.Bharp\_720s} data series. All
pixels inside the rectangular bounding box are annealed in the patchwise
{\sf SHARP} disambiguation; however, pixels below a noise threshold are also smoothed
\cite{Barnes2013,hoeksema2013}. 
Since 19 December 2013 we have disambiguated the entire disk and use that
data from the consistently derived {\sc disambig}
segment of the {\sf hmi.B\_720s} data series for definitive {\sf SHARP}s
observed from 15 January 2014 onward.

\item Finally, to complete the {\sf SHARP} data series the analysis pipeline collects maps 
of HMI observables and computes a set of active region summary parameters
using a publicly available module (see Table~\ref{tab:urls} and Section \ref{sec:Parameters}). 

\end{itemize}

\section{SHARP Coordinates: CCD Cutouts and Cylindrical Equal Area Maps}\label{sec:CEA}

{\begin{table}
\caption{{\sf SHARP} Data Series}
\begin{flushleft}
\begin{tabular}{l p{10.64cm} }
Data Series Name & Description \\\hline
{\sf hmi.sharp\_720s} & Definitive data with 31 map segments in CCD coordinates wherein the 
vector {\bf {\textit B}} is comprised of azimuth, inclination, and field strength. \\
{\sf hmi.sharp\_cea\_720s} & Definitive data with 11 segments wherein all quantities 
have been remapped to a heliographic Cylindrical Equal-Area coordinate system 
centered on the patch
and the 
vector {\bf {\textit B}} has been transformed into the components $B_r$, $B_{\theta}$, and $B_{\phi}$. \\
{\sf hmi.sharp\_720s\_nrt} & Near-real-time data; otherwise same as {\sf hmi.sharp\_720s}. \\
{\sf hmi.sharp\_cea\_720s\_nrt} & Near-real-time data; otherwise same as {\sf hmi.sharp\_cea\_720s}. \\
\end{tabular}
\end{flushleft}
\caption*{Listed below are four series that contain {\sf SHARP} data. {\sf SHARP} active
region parameters are stored as keywords for these series. For a 
list of parameters, see Table~\ref{tab:SpaceweatherFormulae}.}
\label{tab:sharpseries}
\end{table}} 

{\begin{figure}
\caption{{\sf SHARP} data for NOAA Active Region\,11166}
\renewcommand{\tabcolsep}{0.0018\textwidth}
\begin{tabular}{ccc}
\includegraphics[angle=90,width=0.496\textwidth]{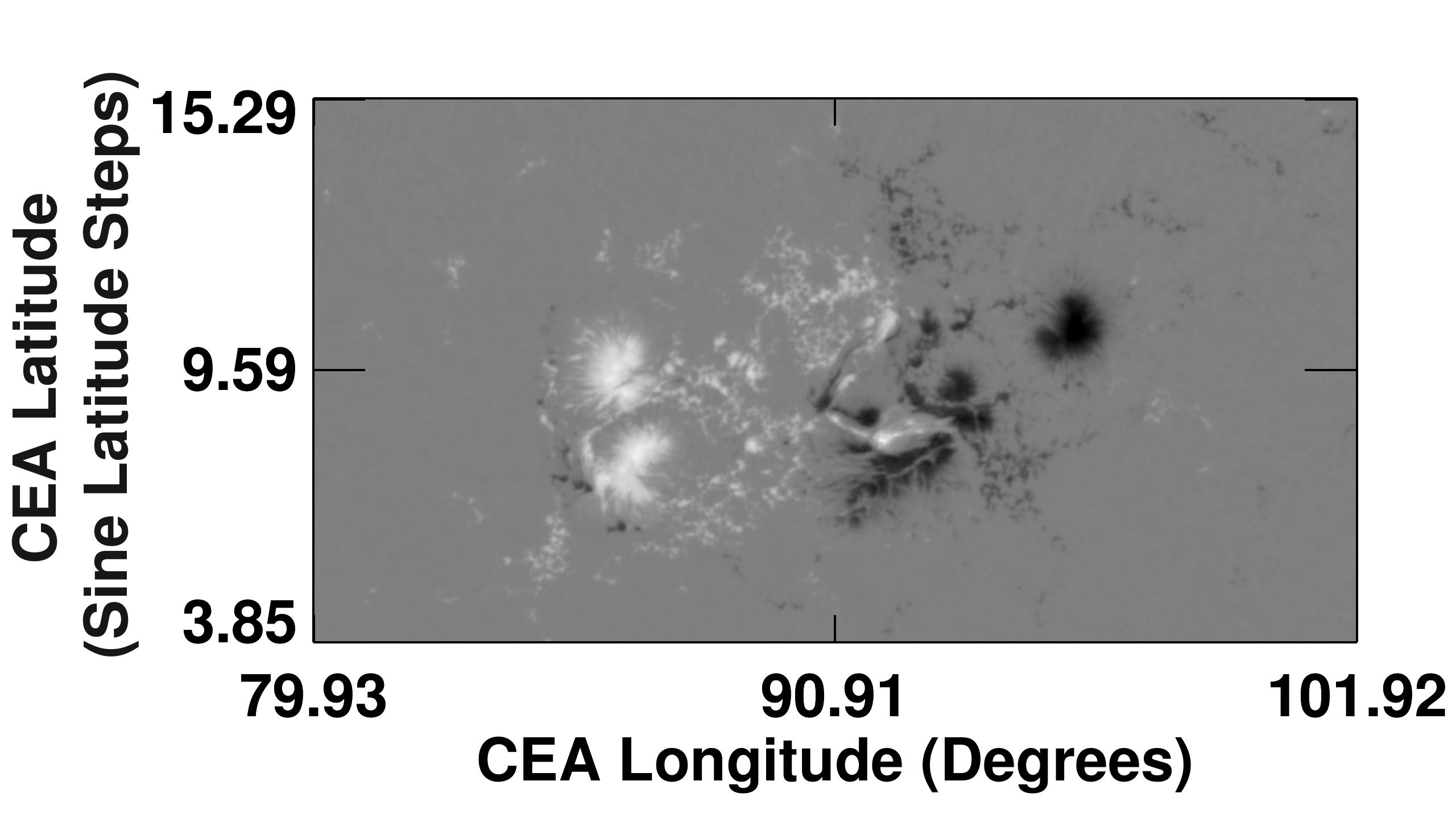} &
\includegraphics[angle=90,width=0.496\textwidth]{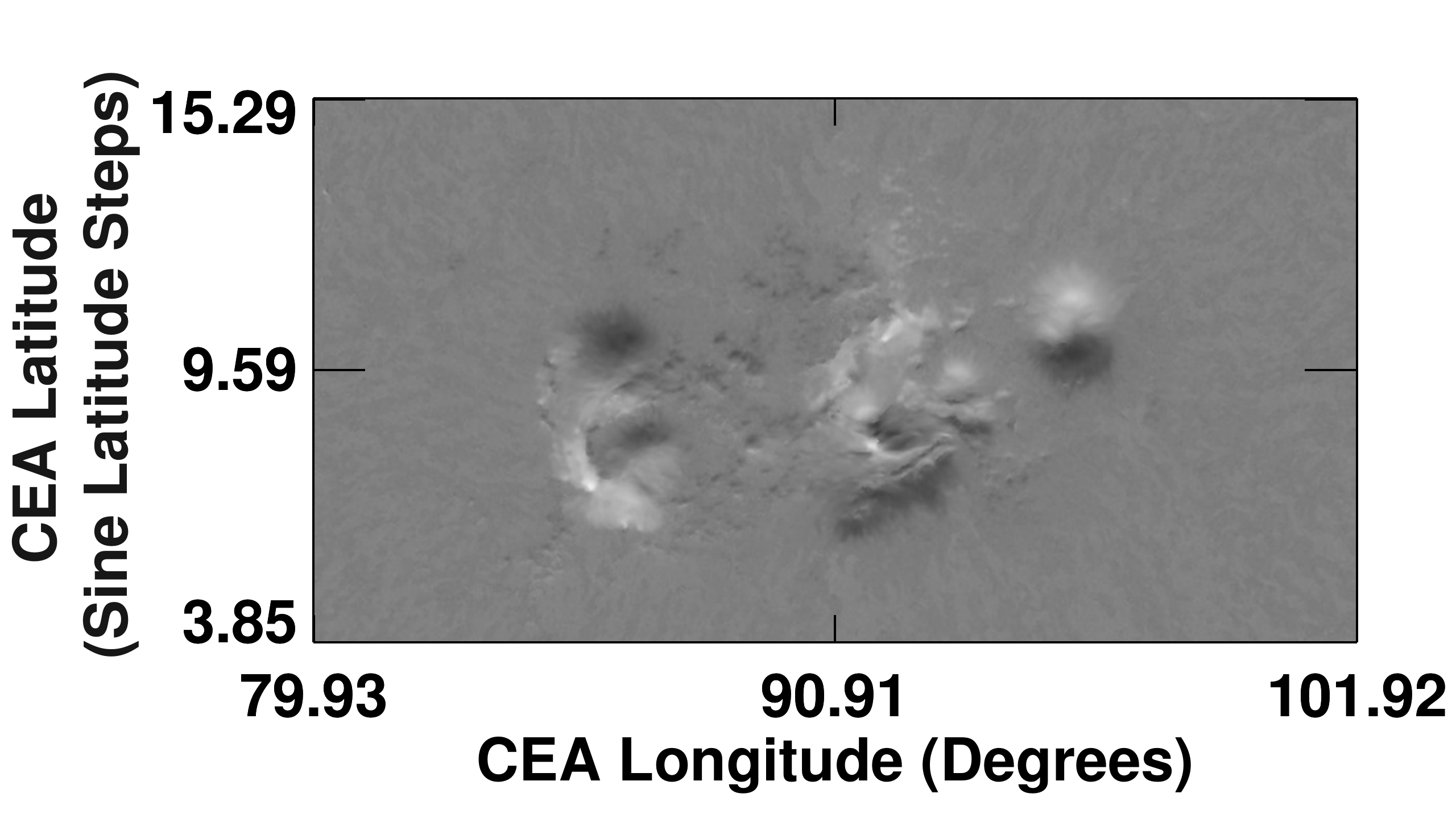} \\
\includegraphics[angle=90,width=0.496\textwidth]{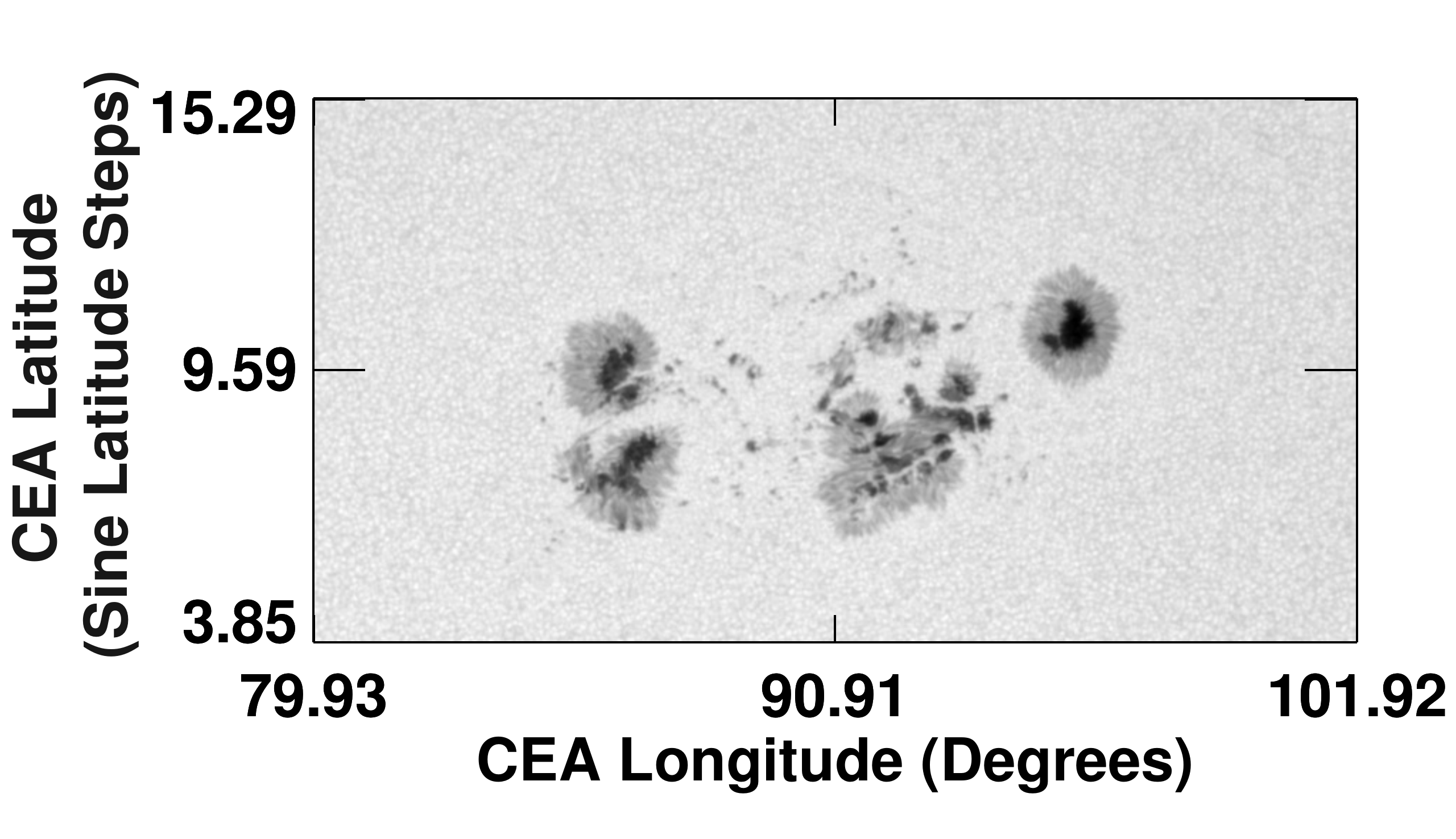} &
\includegraphics[angle=90,width=0.496\textwidth]{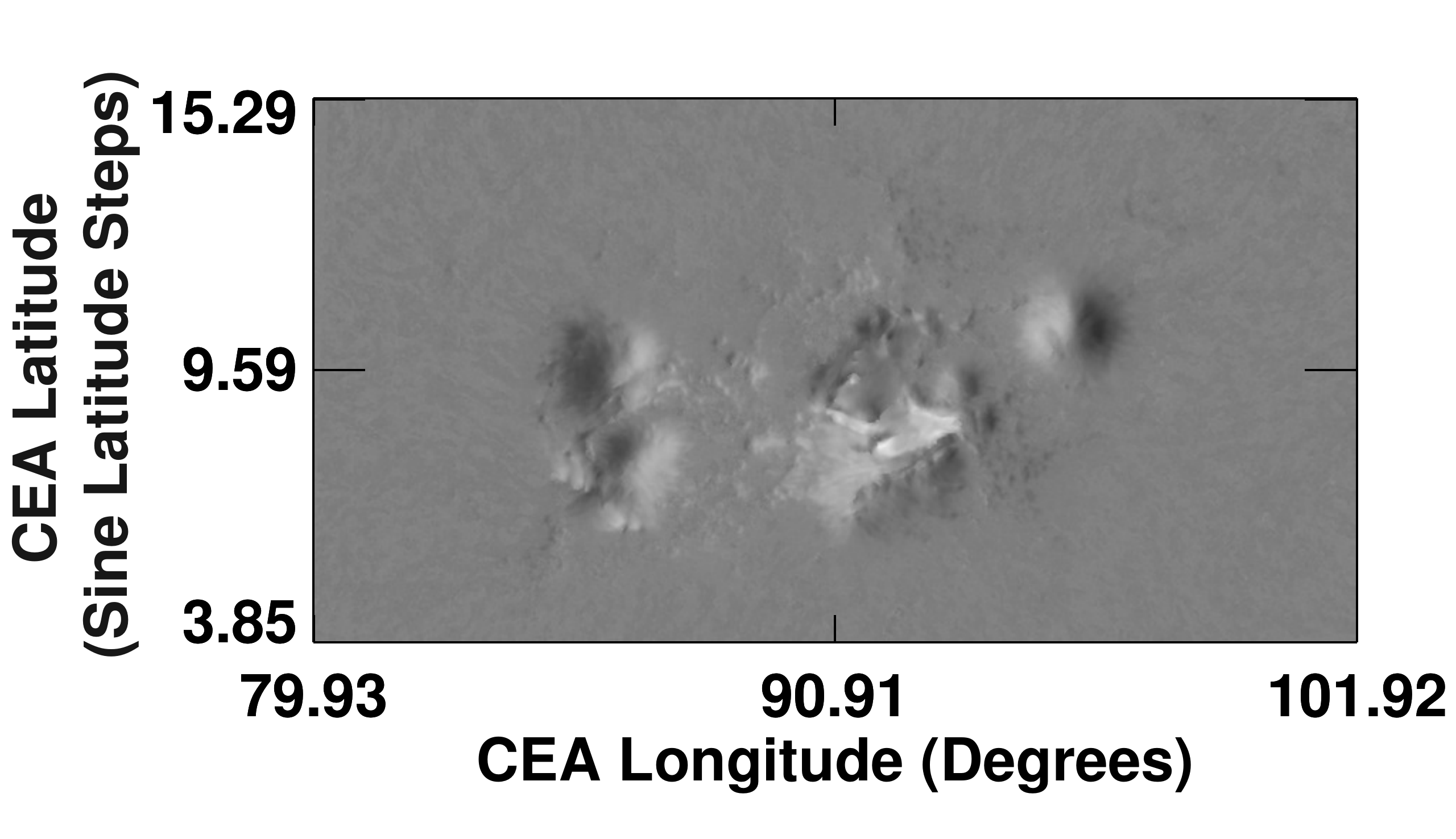} \\
\end{tabular}
\caption*{The first three panels, clockwise from upper left, show the inverted
and disambiguated data wherein the vector {\bf {\textit B}} has been remapped to a
Cylindrical Equal-Area projection and decomposed into $B_r$, $B_{\theta}$,
and $B_{\phi}$, respectively, for HARP 401 (NOAA AR\,11166)
on 9 March 2011 at 23:24:00 TAI. The color table is scaled between $\pm$2500
Gauss for all three magnetic-field arrays. The lower-left
panel shows the computed continuum intensity for the same region at the same time.
The patch is centered at longitude $90.91^\circ$, latitude $9.59^\circ$ 
in Carrington Rotation 2107. CEA longitude and latitude are described in the text.
}
\label{fig:example}
\end{figure}}

HMI data series use standard World Coordinate System (WCS) for solar images \cite{thompson}.
{\sf SHARP} data series are available in either of two coordinate systems: one
is effectively cut out directly from corrected full-disk images, which are in
helio-projective Cartesian CCD image coordinates,
and the other is remapped from CCD coordinates to a heliographic Cylindrical Equal-Area
(CEA) projection centered on the patch.  Table \ref{tab:sharpseries} lists the four
available {\sf SHARP} data series. 

For standard CCD-cutout {\sf SHARP}s, the pipeline module collects 31 maps, including many
of the primary HMI observable data segments (line-of-sight magnetogram, Dopplergram, 
continuum intensity, and vector magnetogram), other inversion and 
disambiguation quantities, uncertainty arrays, and the HARP bitmap. Using the HARP 
bounding box as a stencil, the module extracts the corresponding arrays of 
observable data. The first six tables in Appendix \ref{s:Appendix} give a description of 
each of the cut-out {\sf SHARP} series segment maps.

Additional processing is applied to the CEA versions of the {\sf SHARP}s to
convert selected segments from CCD pixels in plane-of-the-sky coordinates to 
a heliographic coordinate system in the photosphere. 
Table \ref{tab:ceaseg} in Appendix A lists the 11 segment maps
that are available in CEA coordinates.


{\begin{figure}
\centering
\caption{Pixels Contributing to Active Region Parameter Calculation}
\renewcommand{\tabcolsep}{0.0018\textwidth}
\begin{tabular}{ccc}
\includegraphics[angle=90,width=0.33\textwidth]{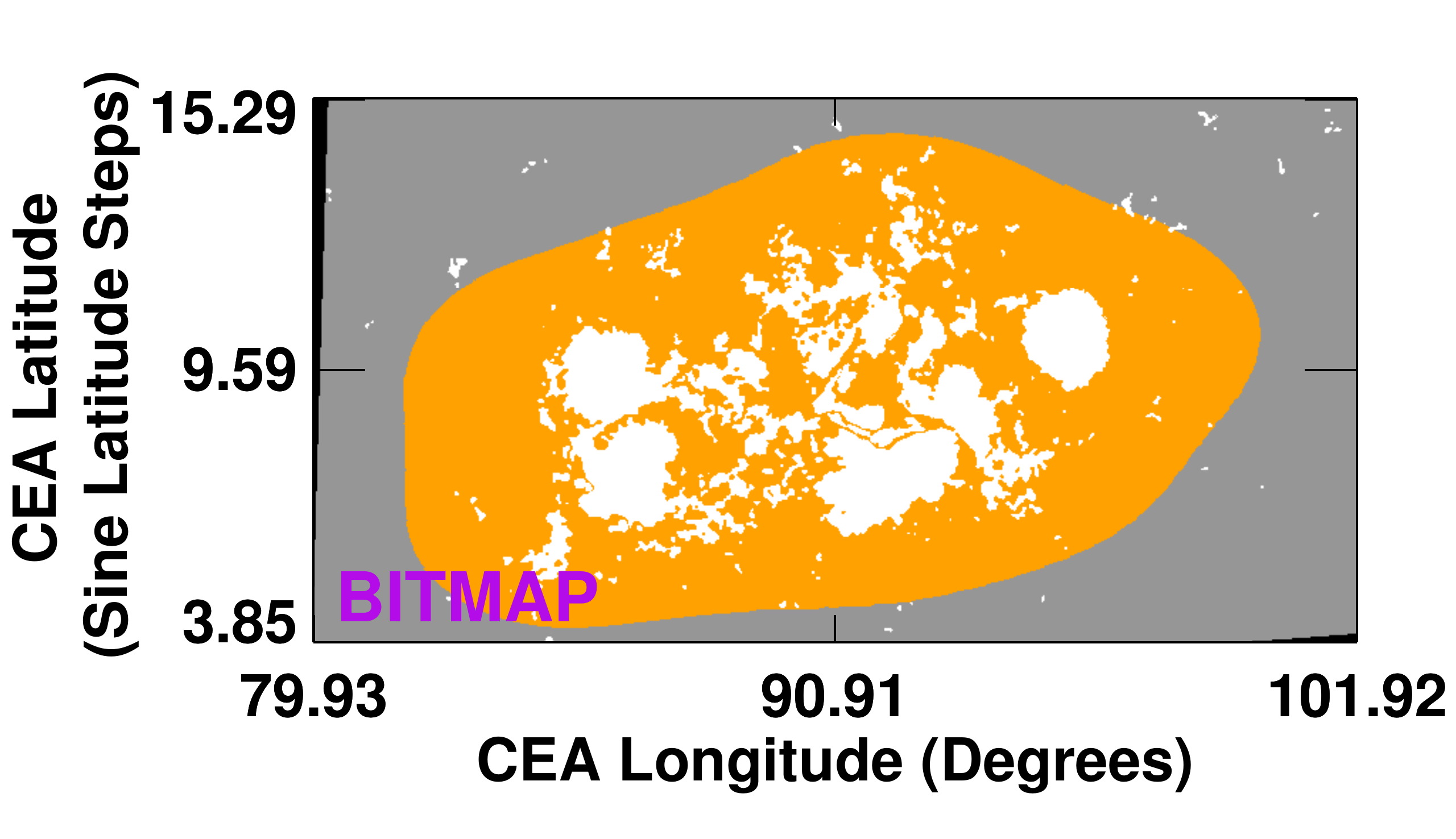} &
\includegraphics[angle=90,width=0.33\textwidth]{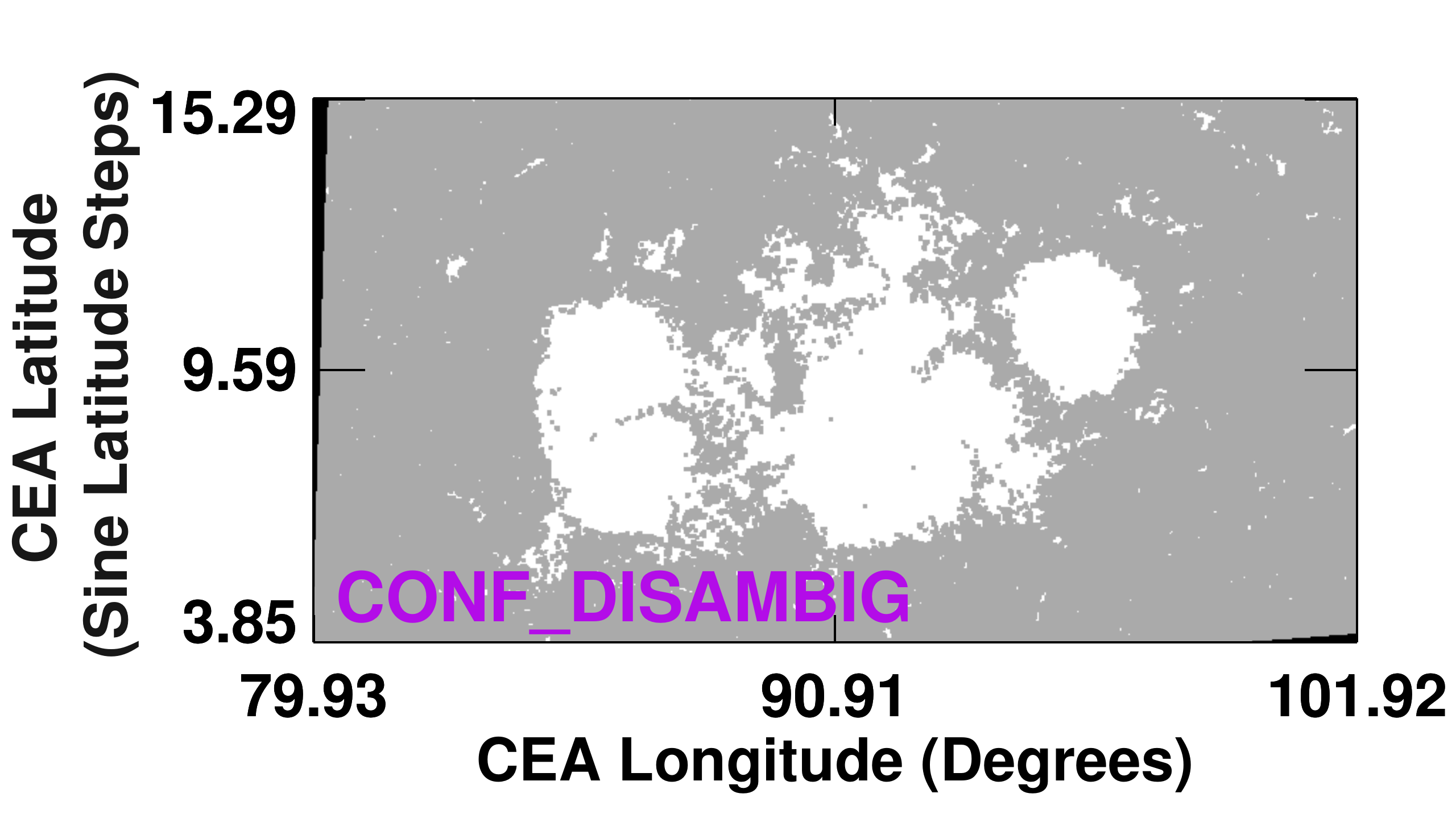} &
\includegraphics[angle=90,width=0.33\textwidth]{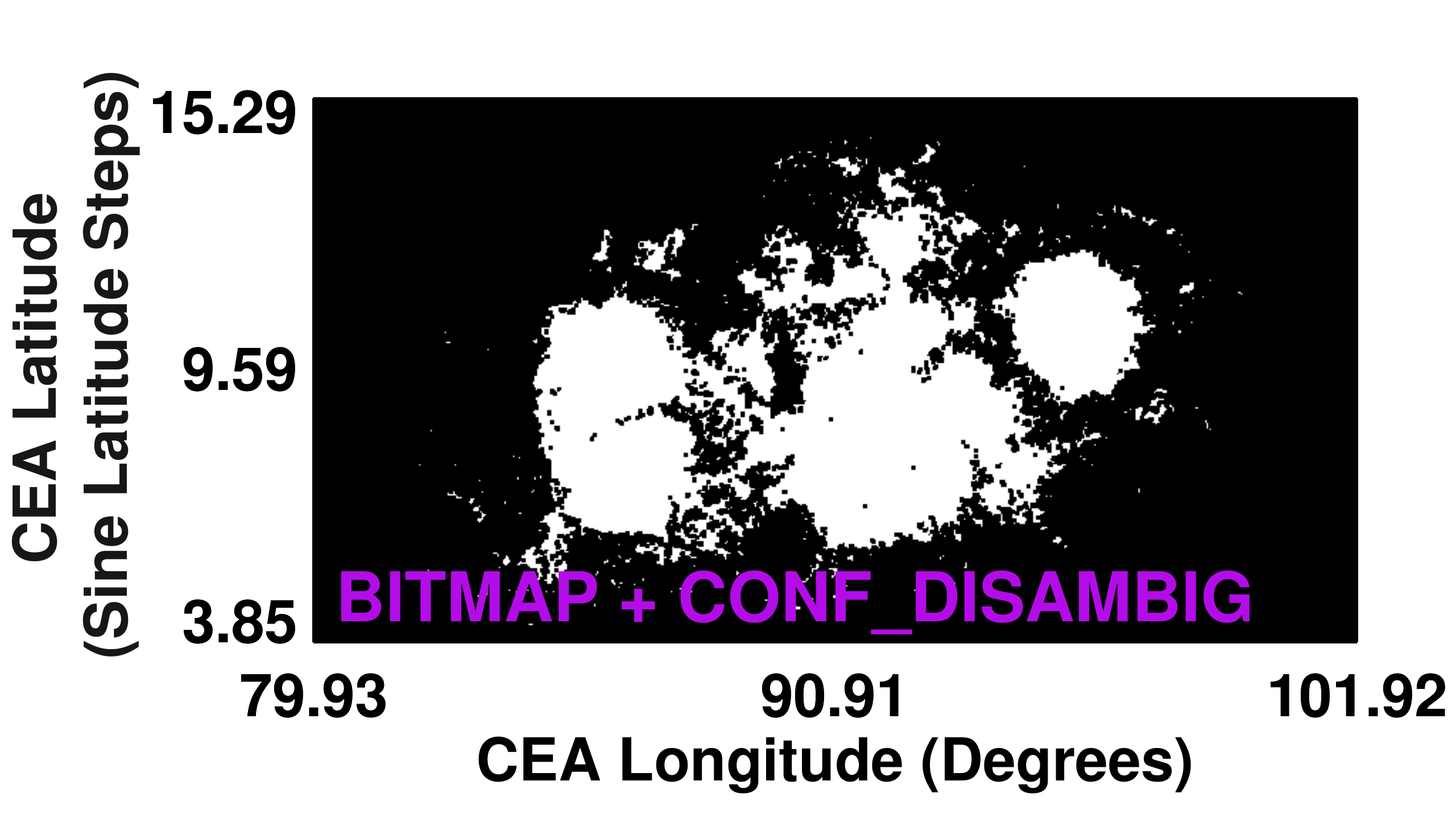} \\
\end{tabular}
\caption*{Only pixels that are both within the
HARP (shaded orange in map segment {\sc bitmap}, left) and 
above the high-confidence disambiguation threshold 
(shown in white in the middle panel where segment {\sc conf\_disambig} = 90) 
contribute to the active region parameters (represented in the rightmost panel). 
This example from {\sf hmi.sharp\_720s\_cea} shows HARP 401 (NOAA AR\,11166) on 9 March 2011
at 23:24:00 TAI, where the quantities have been remapped to a Cylindrical
Equal-Area coordinate system. Black areas at the edge of the {\sc bitmap} and {\sc conf\_disambig}
images fall outside the maximal CCD HARP bounding box; therefore, the azimuthal ambiguity
resolution has not been applied to these areas. As in Figure \ref{fig:example},
the axes are labeled in CEA coordinates, as described in the text.}
\label{fig:bitmap1}
\end{figure}}


The expression relating the final CEA map coordinate [$x, y$] to the heliographic
longitude and latitude [$\phi$, $\lambda$] follows Equations (79) and (80) of
\inlinecite{calabretta}, compliant with the World Coordinate System (WCS)
standard (\textit{e.g.} \opencite{thompson}). The remapping uses the patch center as
reference point, thus effectively de-rotating the patch center to $\phi=0$,
$\lambda=0$ 
before CEA projection
in order to minimize distortion (see Section 2.5 of \opencite{calabretta}).
As a consequence, the correspondence between what are labeled CEA 
degrees and the familiar Carrington latitude and longitude is complex.
The Carrington coordinates of the patch center are indicated 
in the keywords {\sc crval1} and {\sc crval2}.
The {\sf SHARP} CEA pixels have a linear dimension in the $x$-direction of 0.03 heliographic 
degrees in the rotated coordinate system and an area on the photosphere of 
$1.33 \times 10^5$\,km$^2$.  The size in the $y$-direction is defined by the CEA requirement
that the area of each pixel be the same, so the pixels are equally spaced in 
the sine of the angular distance from the great circle that defines the $x$-axis 
and the step size is fixed such that the pixel dimension is equal to 0.03 degrees 
at patch center.
In Figures \ref{fig:example} and \ref{fig:bitmap1} the axes are labeled 
in CEA degrees with the center point having the Carrington longitude and latitude values.
In our remapping process the CEA grid is oversampled by interpolating the nearby CCD values and
then smoothed with a Gaussian filter to the final sampling.  
Details are provided by \inlinecite{Sun2013}.  

The remapping of the uncertainty images, as well as the {\sc bitmap} and {\sc
conf\_disambig} maps, is done a little differently. For these the center
of each pixel in the remapped CEA coordinate system is first located in the
original CCD image; then the nearest neighboring pixel in the original image
is identified and the value for that nearest original CCD pixel is reported.

For the CEA version, the native three-component vector magnetic-field output from the inversion --
expressed as field strength [$B$], inclination [$\gamma$], and azimuth [$\psi$] in the image plane --
is transformed to the components $B_r$, $B_{\theta}$, and $B_{\phi}$ in standard
heliographic spherical coordinates 
[$\hat{\textit{\textbf e}}_{r}$, $\hat{\textit{\textbf e}}_{\theta}$, $\hat{\textit{\textbf e}}_{\phi}$] following Equation~(1) of \inlinecite{gh}. 
Figure \ref{fig:example} shows the three components of the vector magnetic field and 
the computed continuum intensity for HARP 401 on 9 March 2011 at 23:24 TAI in CEA coordinates.
We note that because 
($\hat{\textit{\textbf e}}_{r}$, $\hat{\textit{\textbf e}}_{\theta}$, $\hat{\textit{\textbf e}}_{\phi}$) 
is a spherical coordinate system with the rotation axis at the pole
($\hat{\textit{\textbf e}}_{x}$, $\hat{\textit{\textbf e}}_{y}$, $\hat{\textit{\textbf e}}_{z}$) 
is a planar cylindrical equal-area coordinate system centered on the patch, 
the unit vectors ($\hat{\textit{\textbf e}}_{\theta}$, $\hat{\textit{\textbf e}}_{\phi}$) 
do not precisely align with ($\hat{\textit{\textbf e}}_{x}$, $\hat{\textit{\textbf e}}_{y}$) 
except at the center of the patch.
In general, only along the $y$-axis passing through patch center do
$\hat{\textit{\textbf e}}_{\phi}$ and $\hat{\textit{\textbf e}}_{y}$ align.
See Figure 2 of \inlinecite{calabretta} for an illustrative example. 
For more information on {\sf SHARP} coordinate systems, mapping, and vector transformations, see \inlinecite{Sun2013}.

{\begin{table}
\caption{Active-Region Parameter Formulae}
\begin{flushleft}
\renewcommand{\arraystretch}{1.5}
\renewcommand{\tabcolsep}{0.1cm}
\begin{tabular}{l p{3.5cm} l p{4.9cm} l l }
Keyword & Description & 
Unit\tabnote{The HMI vector-magnetogram data are in units of Mx\,cm$^{-2}$, whereas the active-region parameters use
units of Gauss. Currently, the filling factor is set to unity, so the two units have the same meaning.}
 & Formula\tabnote{Constant terms are not shown.} & Statistic & \parbox{1.4cm}{Error\\Keyword}\\\hline
{\sc usflux} & Total unsigned flux & {\tiny Mx} & $\Phi = \sum|B_{z}|dA$ & Integral & {\sc errvf} \\
{\sc meangam} & Mean angle of field from radial & {\tiny Degree} & 
 $\overline{\gamma} = \frac{1}{N} \sum \arctan\left(\frac{B_h}{B_z}\right)$ 
 & Mean & {\sc errgam}\\
{\sc meangbt} & Horizontal gradient of total field & {\tiny G\,Mm$^{-1}$} &
 $\overline{\left|{\nabla B_{\rm tot}}\right|} = \frac{1}{N} \sum \sqrt{\left(\frac{\partial B}{\partial x}\right)^2 + \left(\frac{\partial B}{\partial y}\right)^2}$
 & Mean & {\sc errbt}\\
{\sc meangbz} & Horizontal gradient of vertical field & {\tiny G\,Mm$^{-1}$}& 
 $\overline{\left|{\nabla B_z}\right|} = \frac{1}{N} \sum \sqrt{\left(\frac{\partial B_z}{\partial x}\right)^2 + \left(\frac{\partial B_z}{\partial y}\right)^2}$ 
 & Mean & {\sc errbz} \\
{\sc meangbh} & Horizontal gradient of horizontal field & {\tiny G\,Mm$^{-1}$} & 
 $\overline{\left|{\nabla B_h}\right|} = \frac{1}{N} \sum \sqrt{\left(\frac{\partial B_h}{\partial x}\right)^2 + \left(\frac{\partial B_h}{\partial y}\right)^2}$ 
 & Mean & {\sc errbh}\\
{\sc meanjzd} & Vertical current density & {\tiny mA\,m$^{-2}$} & 
 $\overline{J_z} \propto \frac{1}{N} \sum \left(\frac{\partial B_y}{\partial x} - \frac{\partial B_x}{\partial y}\right) $
 & Mean & {\sc errjz}\\
{\sc totusjz} & Total unsigned vertical current & {\tiny A} & 
 ${J_{z_{total}}} =  \sum |J_{z}|dA$ 
 & Integral & {\sc errusi}\\
{\sc meanalp} & Characteristic twist parameter, $\alpha$ & {\tiny Mm$^{-1}$} & 
 $ \alpha_{total} \propto \frac{\sum J{_z} \cdot B_z}{\sum B{^2_z}} $ 
 & Mean & {\sc erralp} \\
{\sc meanjzh} & Current helicity ($B_{z}$ contribution) & {\tiny G$^{2}$\,m$^{-1}$} & 
 $\overline{H_c} \propto \frac{1}{N} \sum B_z \cdot J_z $ 
 & Mean & {\sc errmih}\\
{\sc totusjh} & Total unsigned current helicity & {\tiny G$^{2}$\,m$^{-1}$} & 
 ${H_{c_{total}}} \propto \sum |B_z \cdot J_z|$ 
 & Sum & {\sc errtui} \\
{\sc absnjzh} & Absolute value of the net current helicity & {\tiny G$^{2}$\,m$^{-1}$} & 
 ${H_{c_{abs}}} \propto \left| \sum B_z \cdot J_z \right|$ 
 & Sum & {\sc errtai}\\
{\sc savncpp} & Sum of the modulus of the net current per polarity & {\tiny A} & 
 $J_{z_{sum}} \propto \Big\vert \displaystyle\sum\limits^{B{_z^+}} J{_z}dA \Big\vert + \Big\vert \displaystyle\sum\limits^{B{_z^-}} J{_z}dA \Big\vert $ 
 & Integral & {\sc errjht}\\
{\sc meanpot} & Proxy for mean photospheric excess magnetic energy density & {\tiny erg\,cm$^{-3}$} & 
 $ \overline{\rho} \propto \frac{1}{N} \sum \left( \vec{\textit{\textbf B}}^{\rm Obs} - \vec{\textit{\textbf B}}^{\rm Pot} \right)^2 $ 
 & Mean & {\sc errmpot}\\
{\sc totpot} & Proxy for total photospheric magnetic free energy density & {\tiny erg\,cm$^{-1}$} & 
 $ \rho_{tot} \propto  \sum \left( \vec{\textit{\textbf B}}^{\rm Obs} - \vec{\textit{\textbf B}}^{\rm Pot} \right)^2 dA $ 
 & Integral & {\sc errtpot}\\
{\sc meanshr} & Shear angle & {\tiny Degree} & 
 $ \overline{\Gamma} = \frac{1}{N} \sum \arccos \left( \frac{\vec{\textit{\textbf B}}^{\rm Obs} \cdot \vec{\textit{\textbf B}}^{\rm Pot}}{|B^{\rm Obs}|\,|B^{\rm Pot}|} \right)$  
 & Mean & {\sc errmsha}\\
{\sc shrgt45} & Fractional of Area with Shear $> 45^\circ$ &  & Area with Shear $>45^\circ$ / HARP Area & Fraction & \\
\end{tabular}
\end{flushleft}
\caption*{Active-region parameters are stored as keywords in each {\sf SHARP} series. 
This table lists each active-region parameter keyword with a brief description and
formula. The keyword for the error associated with each parameter is given in the 
last column. Each parameter represents either a mean, sum, or integral of the
distribution in the high-confidence part of the HARP; this is indicated in the 
Statistic column. The active-region parameters were generally adapted from 
\inlinecite{lekabarnes2} except as noted in the text. WCS-standard keywords such as
{\sc cdelt1}, {\sc rsun\_obs}, and {\sc rsun\_ref}, as well as fundamental constants, 
were used to convert to the units specified in the eponymous column. Calculations are
performed on the {\sc cmask} high-confidence pixels in the CEA {\sf SHARP}. Derivations of the
errors can be found at the {\sf SHARP} web page (see Table \ref{tab:urls}). 
Further description of the parameters can be found in Section \ref{sec:AR410}.
}
\label{tab:SpaceweatherFormulae}
\end{table}}

\section{SHARP Summary Parameters}\label{sec:Parameters}

The {\sf SHARP} module calculates summary parameters
every twelve minutes on the inverted and disambiguated data using the
vector field and other quantities in the CEA projection. 
The {\sf SHARP} series presently contain sixteen summary parameters,
as detailed in Table \ref{tab:SpaceweatherFormulae}. 
This initial
list parametrizes some of the features of solar active
regions that have been associated with enhanced flare productivity
(\textit{e.g.} \opencite{lekabarnes2003}, \citeyear{lekabarnes2007}, and references therein) 
and includes different kinds of indices such as the total magnetic flux, the spatial gradients of the
field, the characteristics of the vertical current density, current helicity, 
and a proxy for the integrated free magnetic energy. 
Until now, indices based on vector-field values have not been 
available with the coverage, cadence, and continuity afforded by HMI. With 
previously available data, none of the parameters were found to be necessary 
or sufficient to forecast a flaring event \cite{lekabarnes2007}.
As of this writing, the {\sf SHARP} indices focus on 
low-order statistical moments of observables and readily derived quantities. 
As the {\sf SHARP} database develops further, new quantities
will be added, including ones that characterize the magnetic inversion
lines, the relevant fractal indices, and models of the coronal field (see
Section \ref{s:analysis} for further discussion).

The pixels that contribute to any given index calculation are selected by examining
two data segment maps: {\sc bitmap} and {\sc conf\_disambig}. The {\sc bitmap} segment, 
an example of which is shown in the left panel of 
Figure \ref{fig:bitmap1}, identifies pixels within the HARP
({\sc bitmap} $\geq 33$). 
Pixels with strong line-of-sight magnetic field strength are shown in white, whether inside or outside the
orange HARP area.
The {\sc conf\_disambig} segment 
has a high value for clusters of pixels
above the spatially 
and temporally
dependent disambiguation noise threshold 
($\approx150$\,G, {\sc conf\_disambig} = 90; see Table \ref{tab:disseg} and 
\opencite{hoeksema2013}).
Only data that are both within the 
HARP and above the high-confidence threshold 
contribute to the {\sf SHARP} parameter calculation; 
the number of contributing CEA pixels is given in the keyword {\sc cmask}.
The rightmost panel of Figure \ref{fig:bitmap1} shows the pixels that contribute to 
the active region parameters for HARP 401 (NOAA AR\,11166) 
on 9 March 2011 at 23:24:00 TAI. 
The indices in all four {\sf SHARP} series are computed from the CEA data.


{\begin{figure}
\centering
\caption{{\sf SHARP} Parameters 401}
\renewcommand{\tabcolsep}{0.01\textwidth}
\begin{tabular}{rr}
\includegraphics*[angle=90,height=0.154\textheight,width=0.48\textwidth]{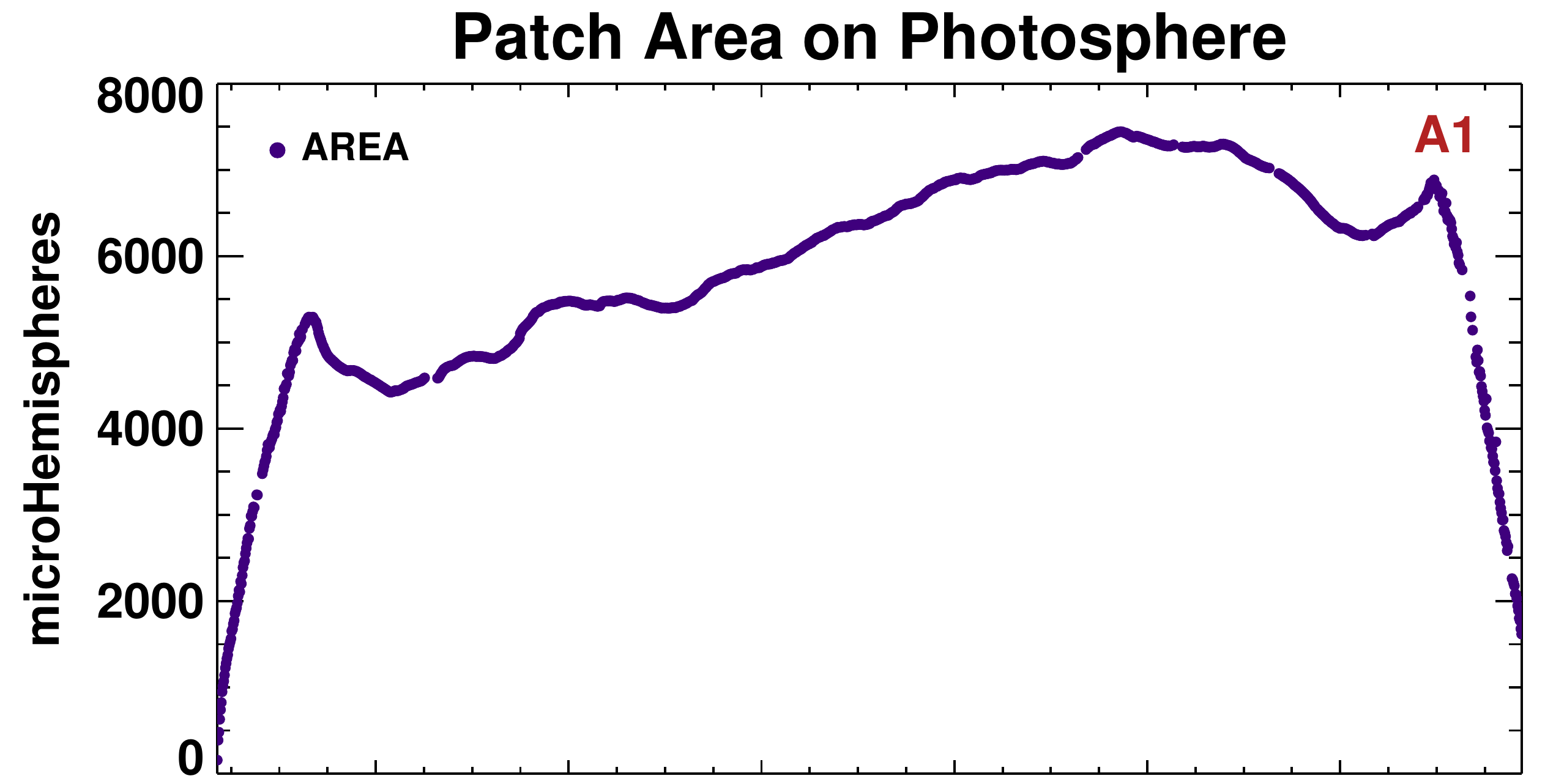} &
\includegraphics*[angle=90,height=0.154\textheight,width=0.48\textwidth]{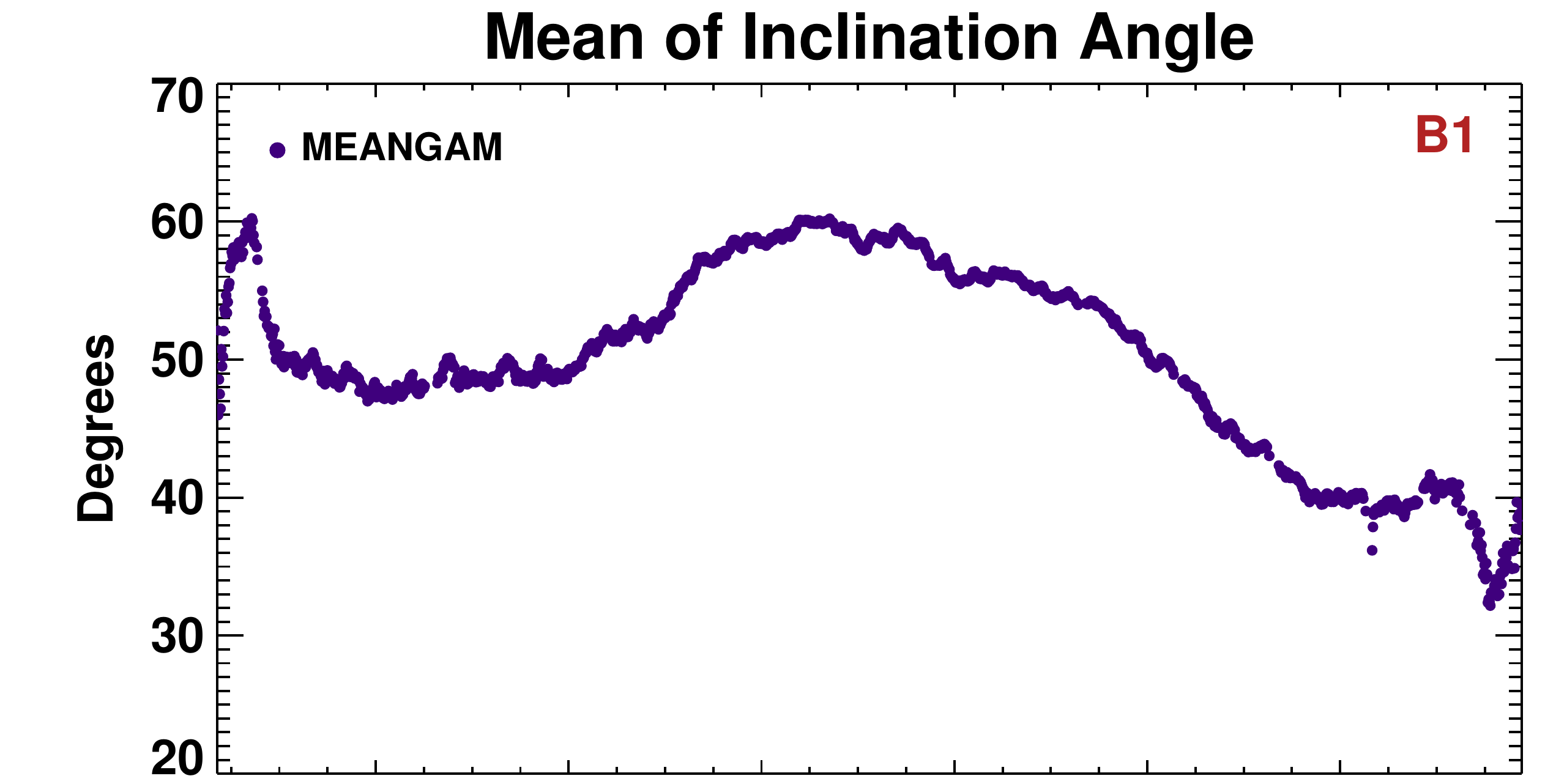} \\
\includegraphics*[angle=90,height=0.154\textheight,width=0.48\textwidth]{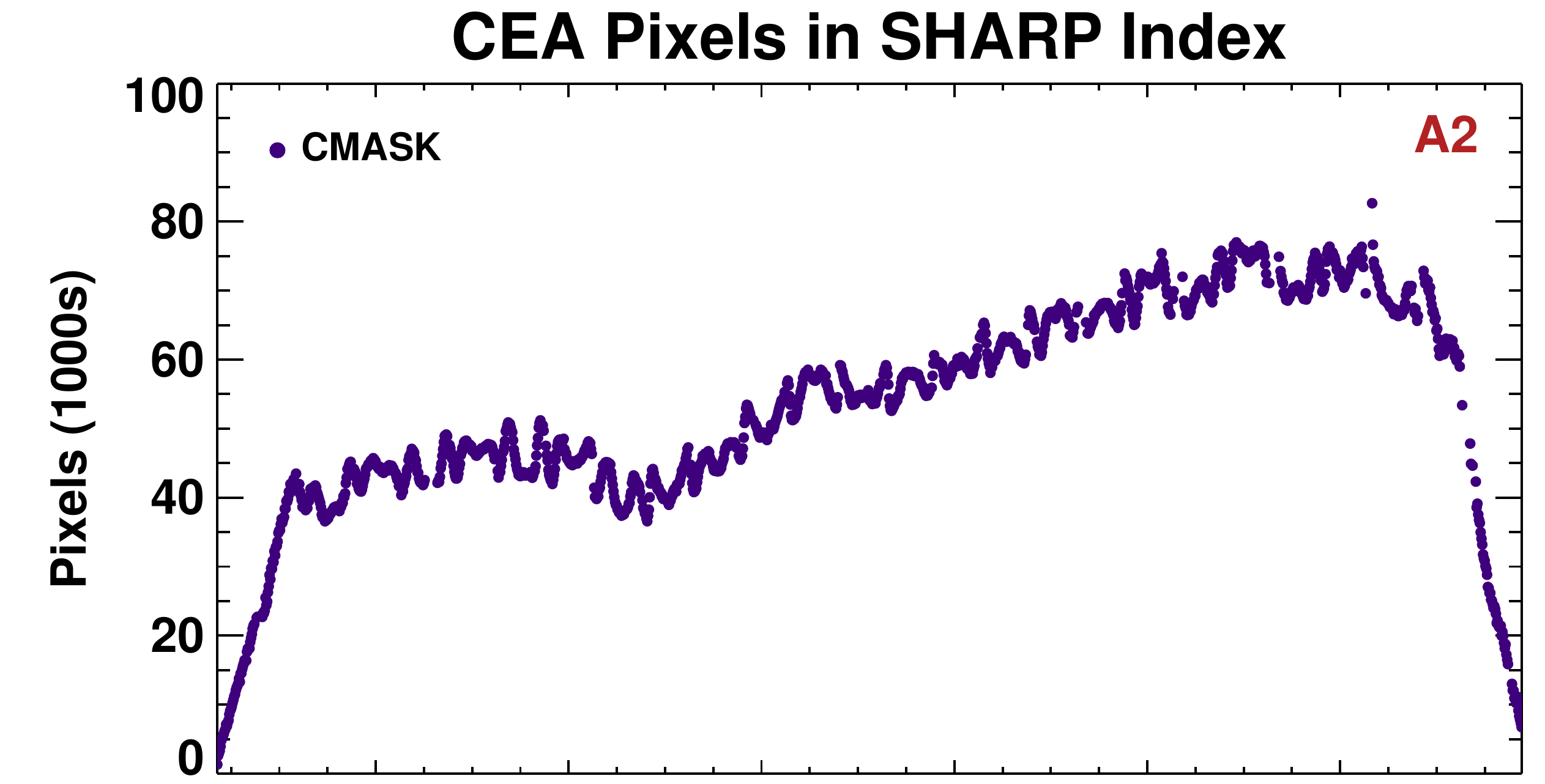}  &
\includegraphics*[angle=90,height=0.154\textheight,width=0.48\textwidth]{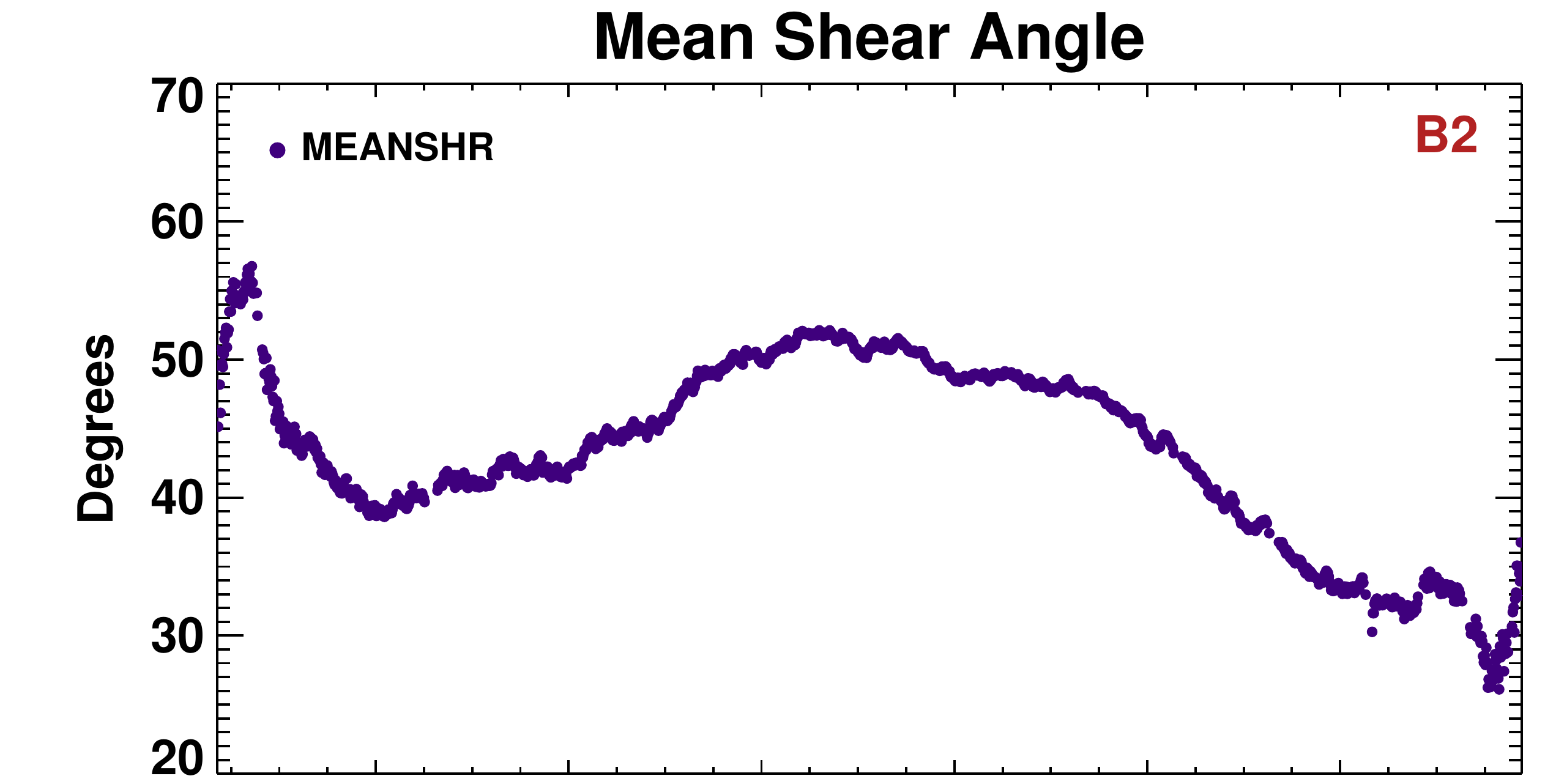} \\
\includegraphics*[angle=90,height=0.154\textheight,width=0.48\textwidth]{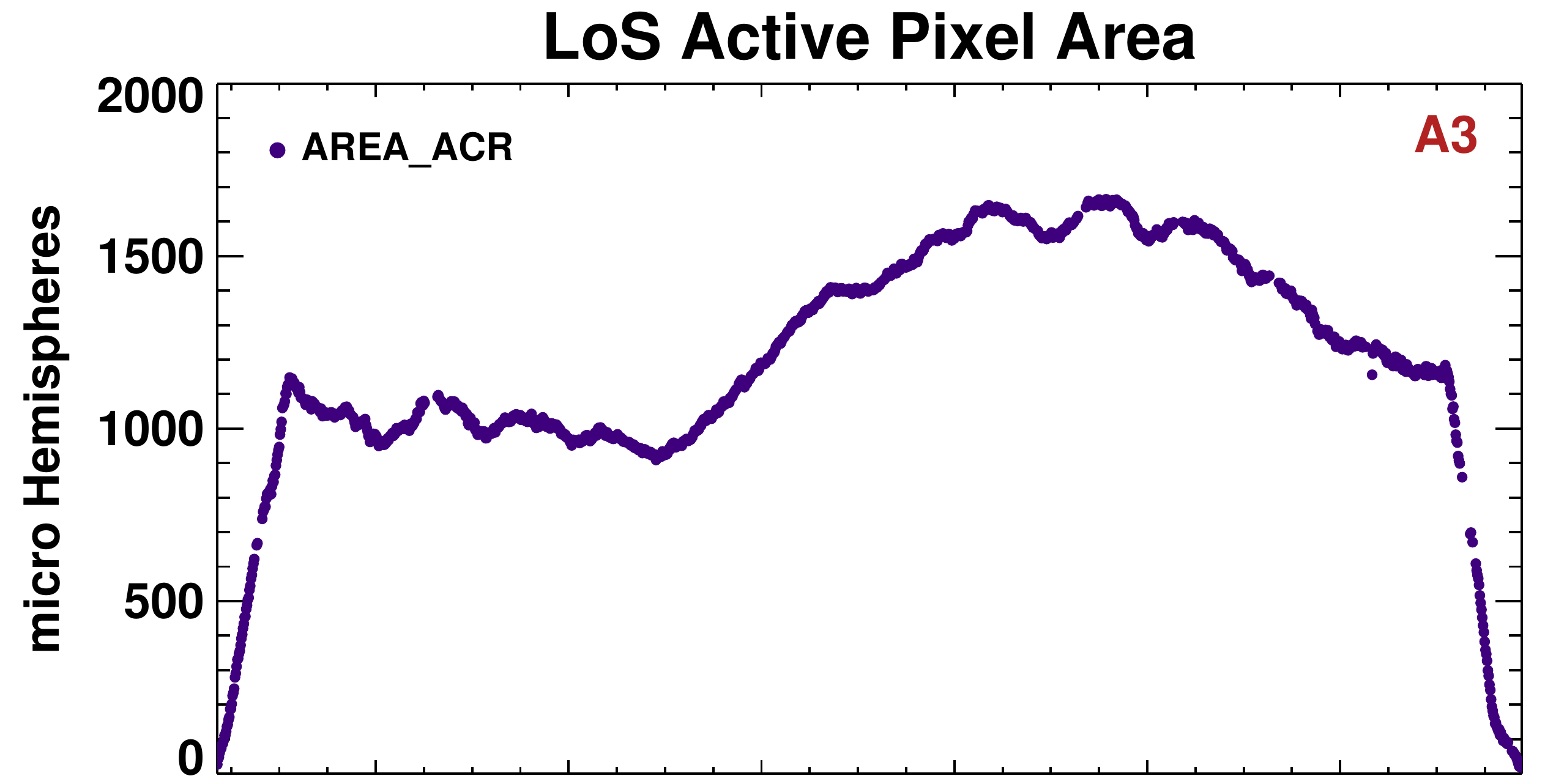} &
\includegraphics*[angle=90,height=0.154\textheight,width=0.48\textwidth]{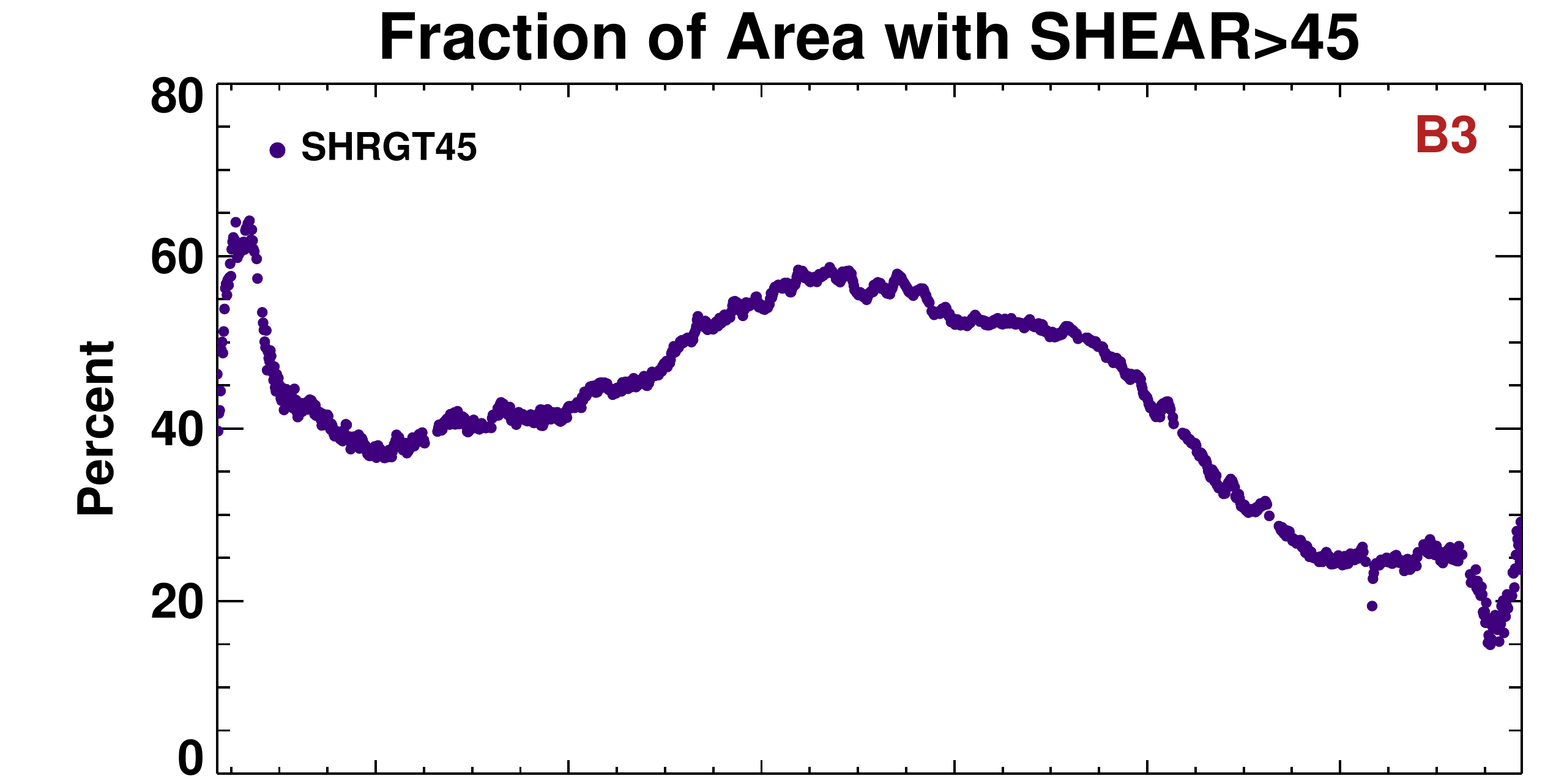} \\
\includegraphics[angle=90,height=0.154\textheight,width=0.48\textwidth]{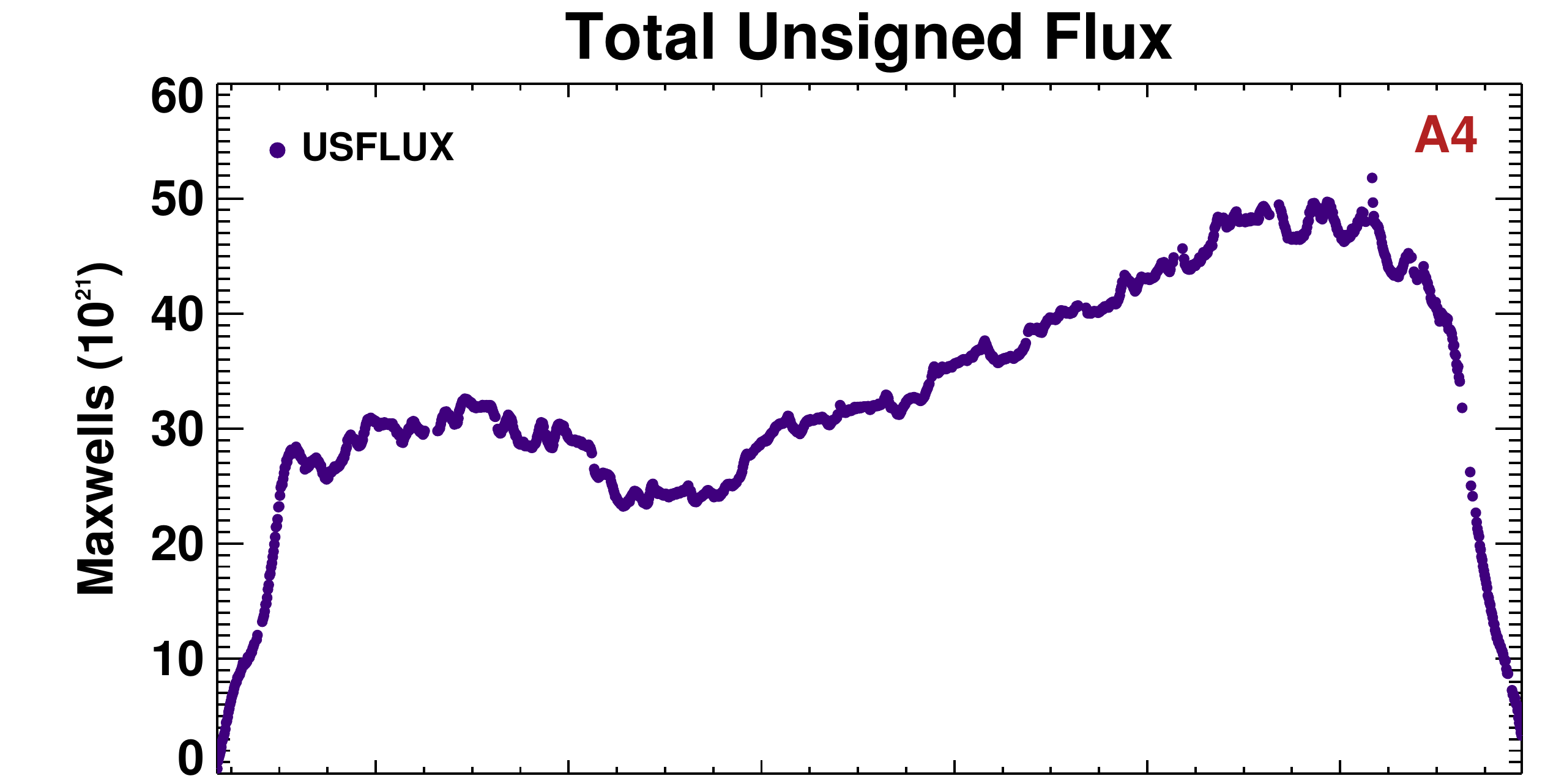} &
\includegraphics*[angle=90,height=0.154\textheight,width=0.48\textwidth]{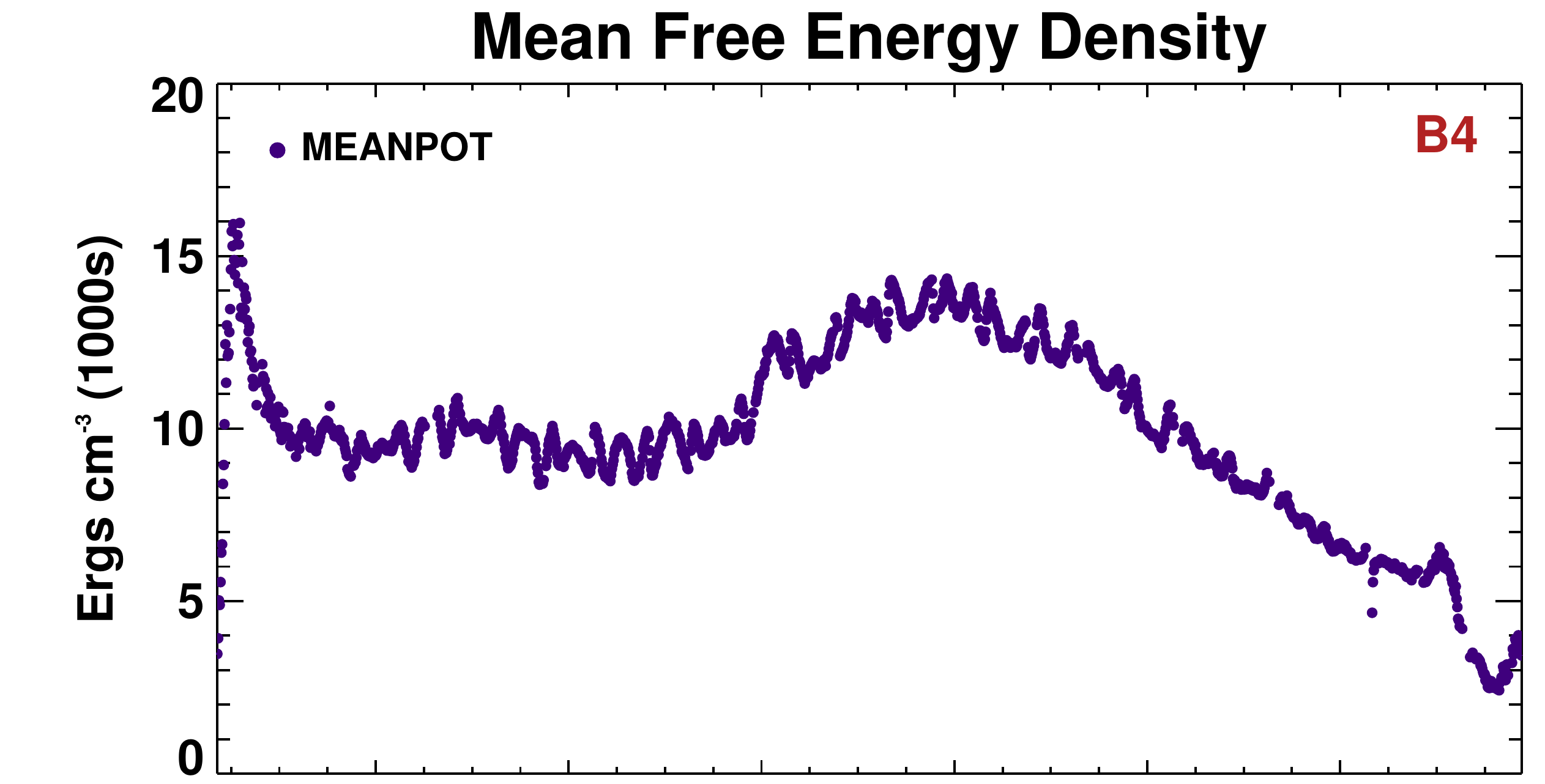} \\
&
\includegraphics[angle=90,height=0.154\textheight,width=0.48\textwidth]{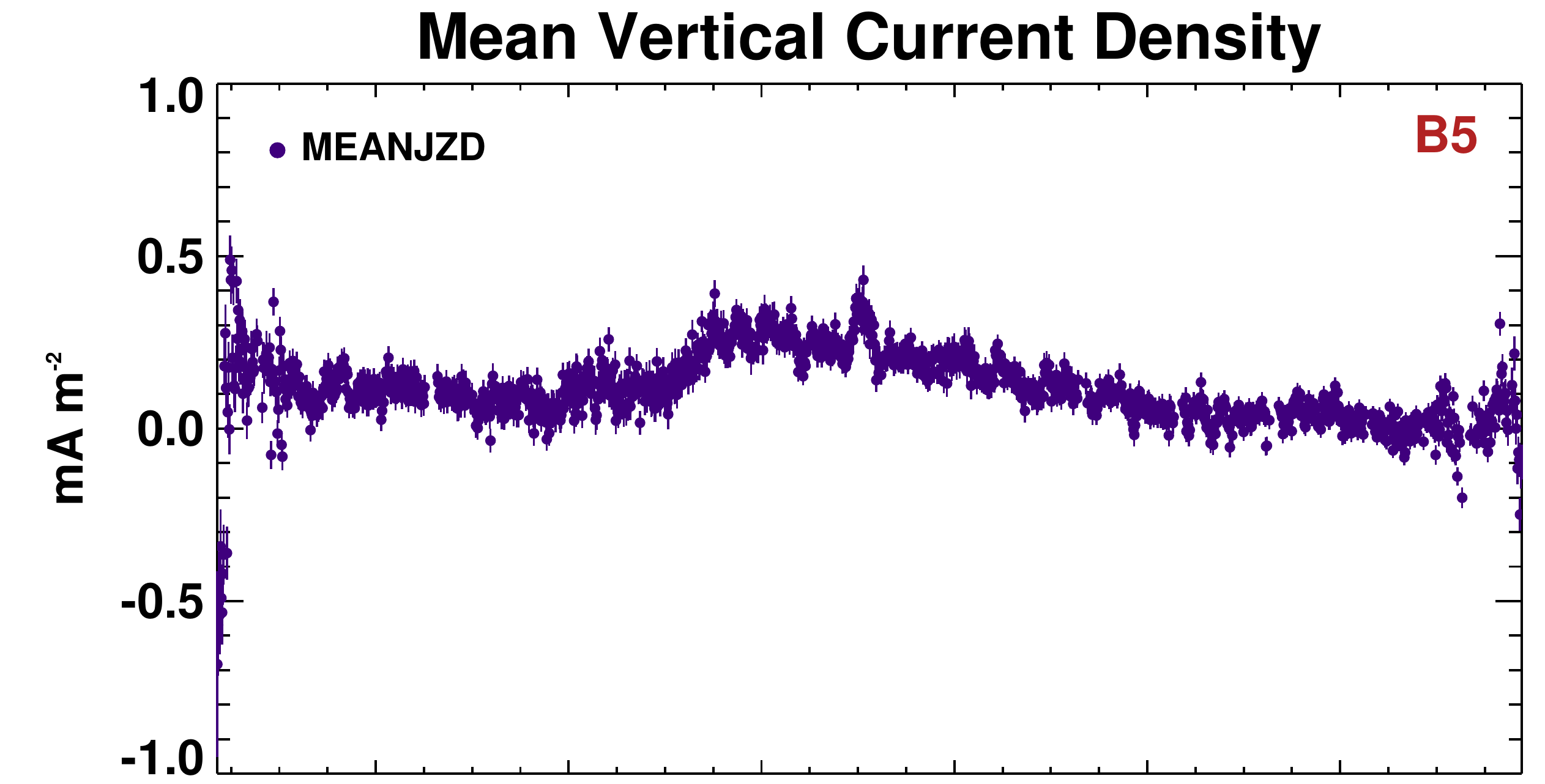} \\
\vspace{12pt}
\end{tabular}
\caption*{{\sf SHARP} Active-Region Parameters for HARP 401, 2\,--\,15 March 2011.
Column A on the left shows four quantities: Panel A1, {\sc area}; A2, {\sc cmask};
A3, {\sc area\_acr}; and A4, {\sc usflux}; Column B on the right shows five quantities: 
Panel B1, {\sc meangam}; B2, {\sc meanshr}; B3, {\sc shrgt45}; B4, {\sc meanpot}; and 
B5, {\sc meanjzd}.}
\label{fig:params401}
\end{figure}}

{\begin{figure}
\centering
\caption{{\sf SHARP} Parameters 401}
\renewcommand{\tabcolsep}{0.01\textwidth}
\begin{tabular}{rr}
\includegraphics*[angle=90,height=0.154\textheight,width=0.48\textwidth]{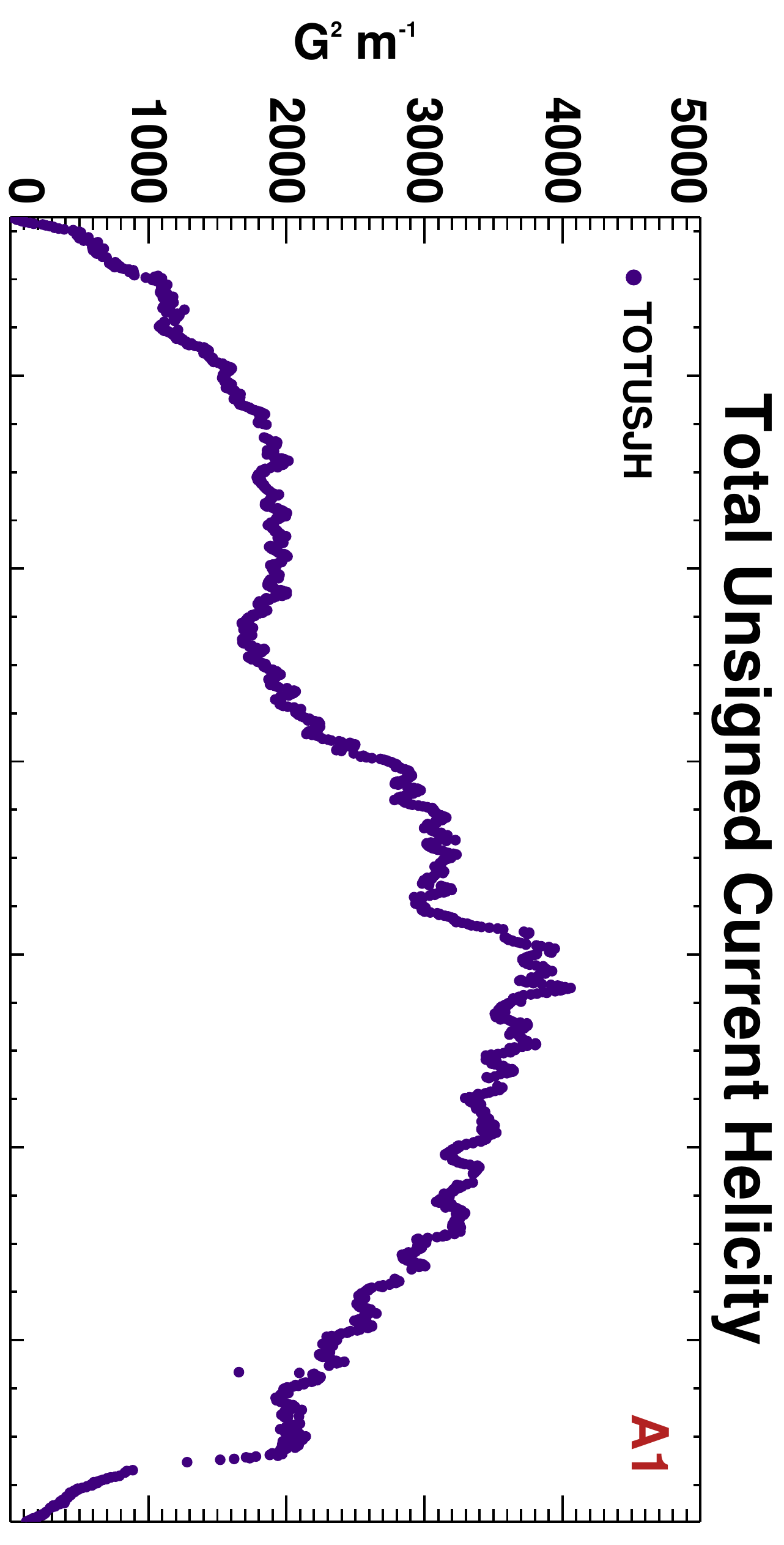} &
\includegraphics*[angle=90,height=0.154\textheight,width=0.48\textwidth]{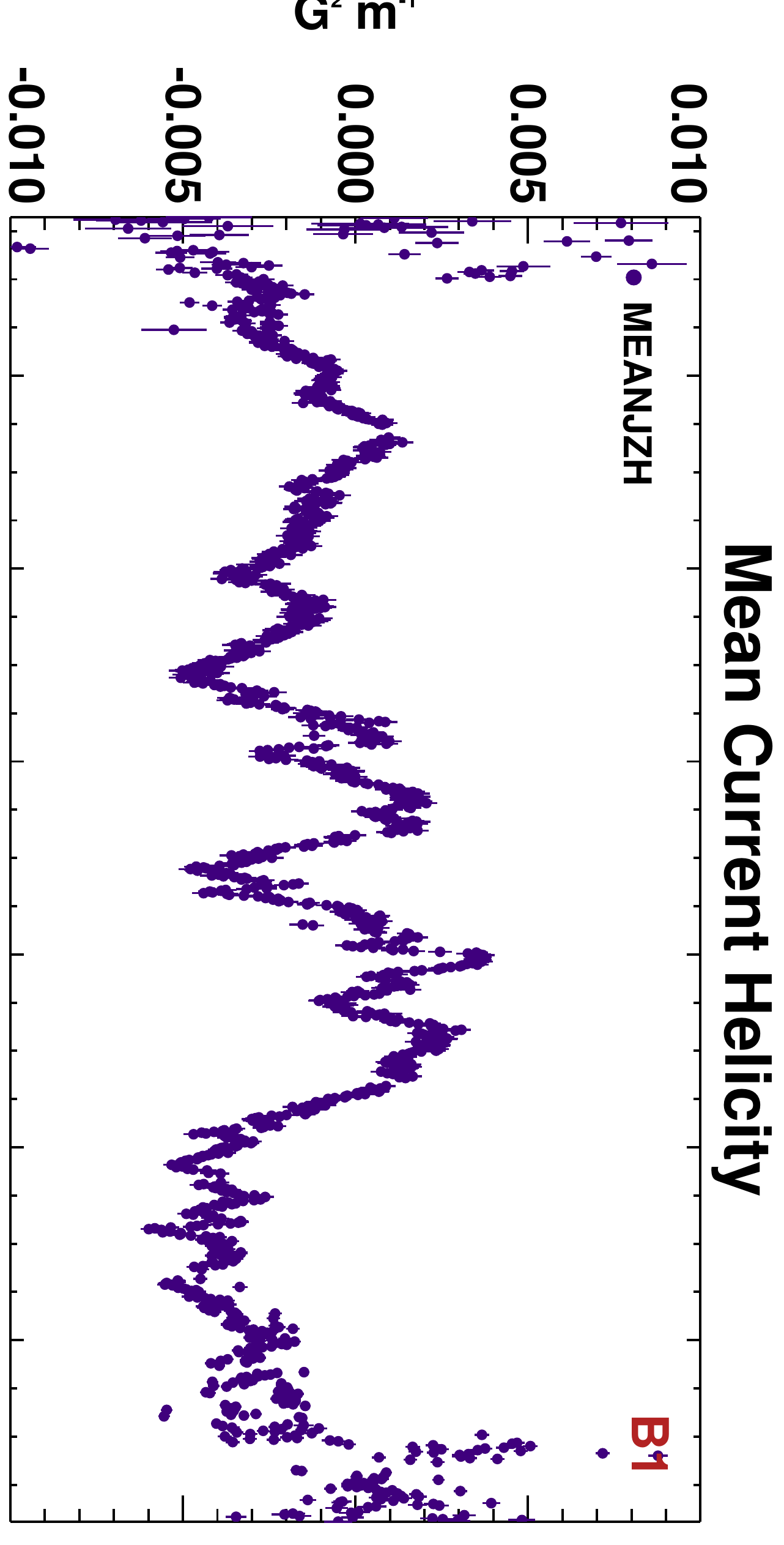} \\
\includegraphics*[angle=90,height=0.154\textheight,width=0.48\textwidth]{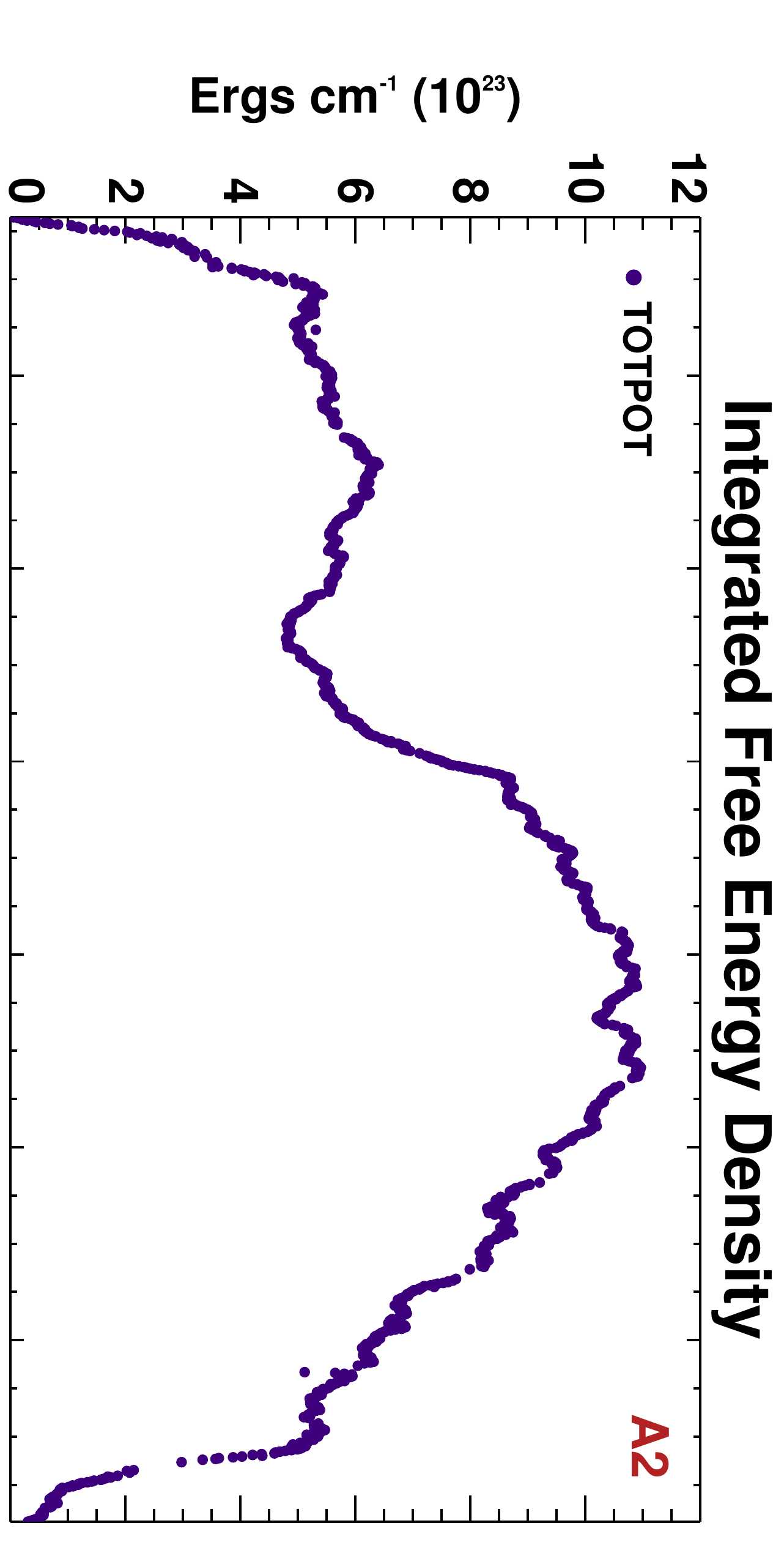}  &
\includegraphics*[angle=90,height=0.154\textheight,width=0.48\textwidth]{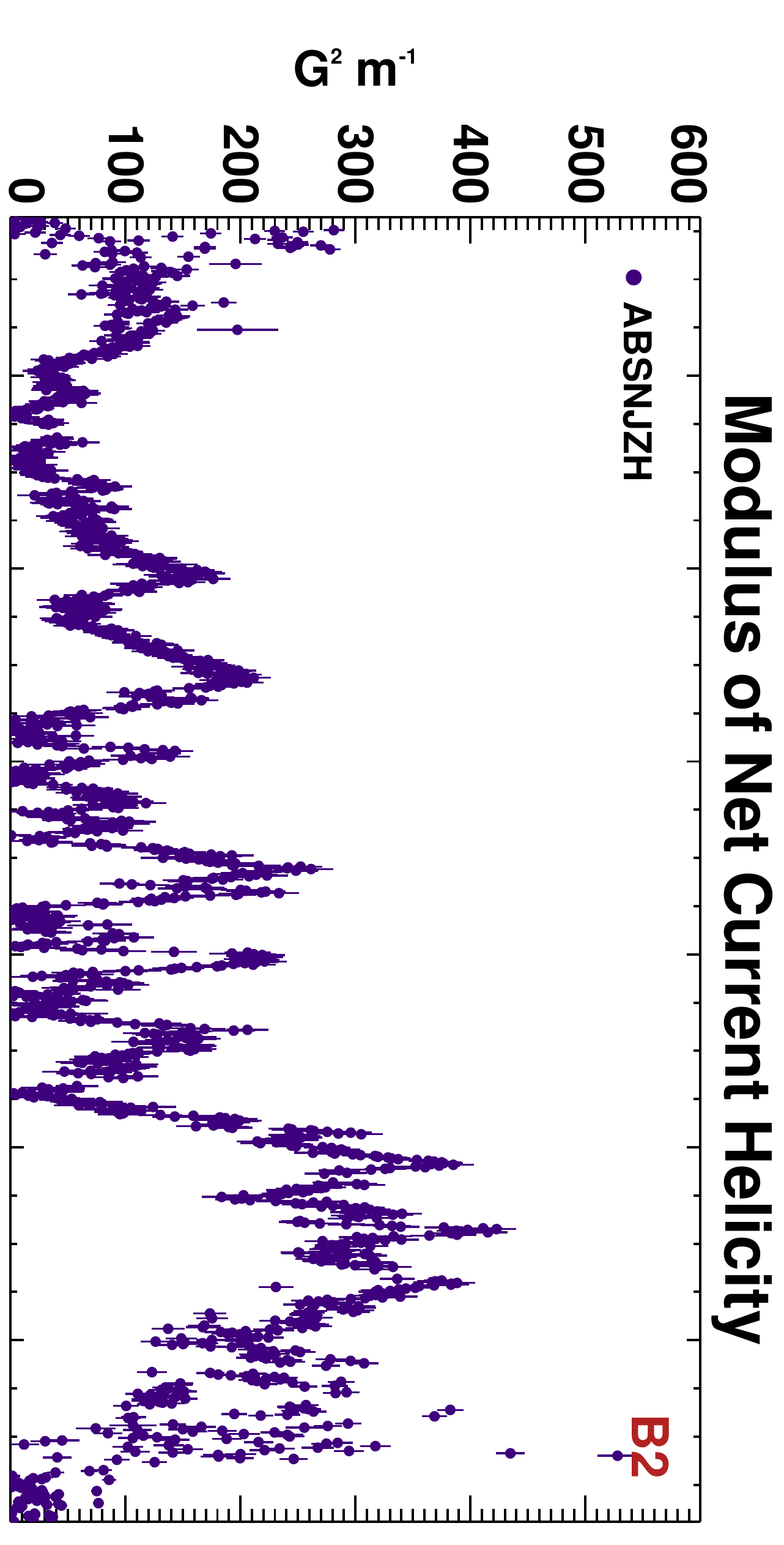} \\
\includegraphics*[angle=90,height=0.154\textheight,width=0.48\textwidth]{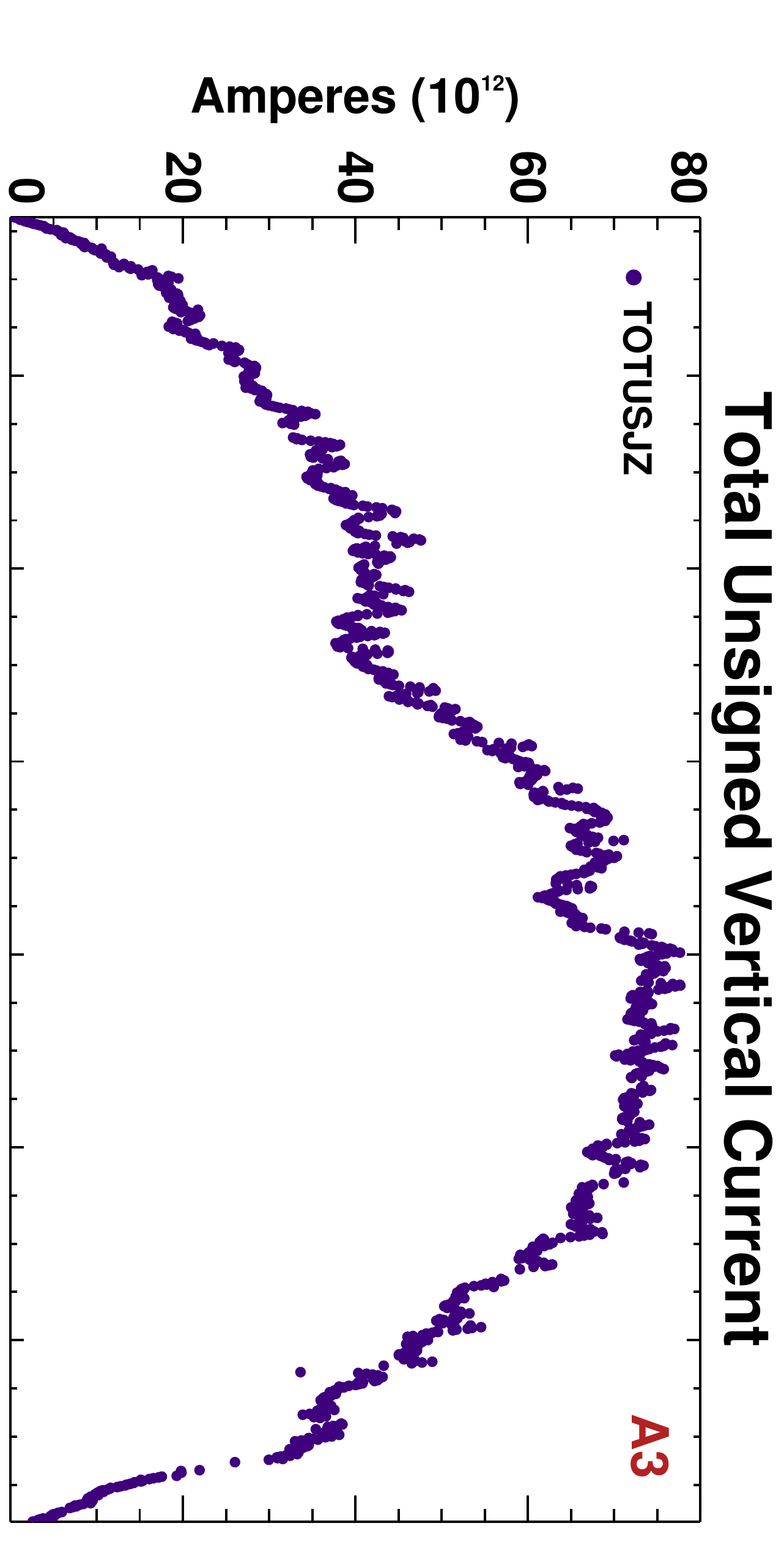} &
\includegraphics*[angle=90,height=0.154\textheight,width=0.48\textwidth]{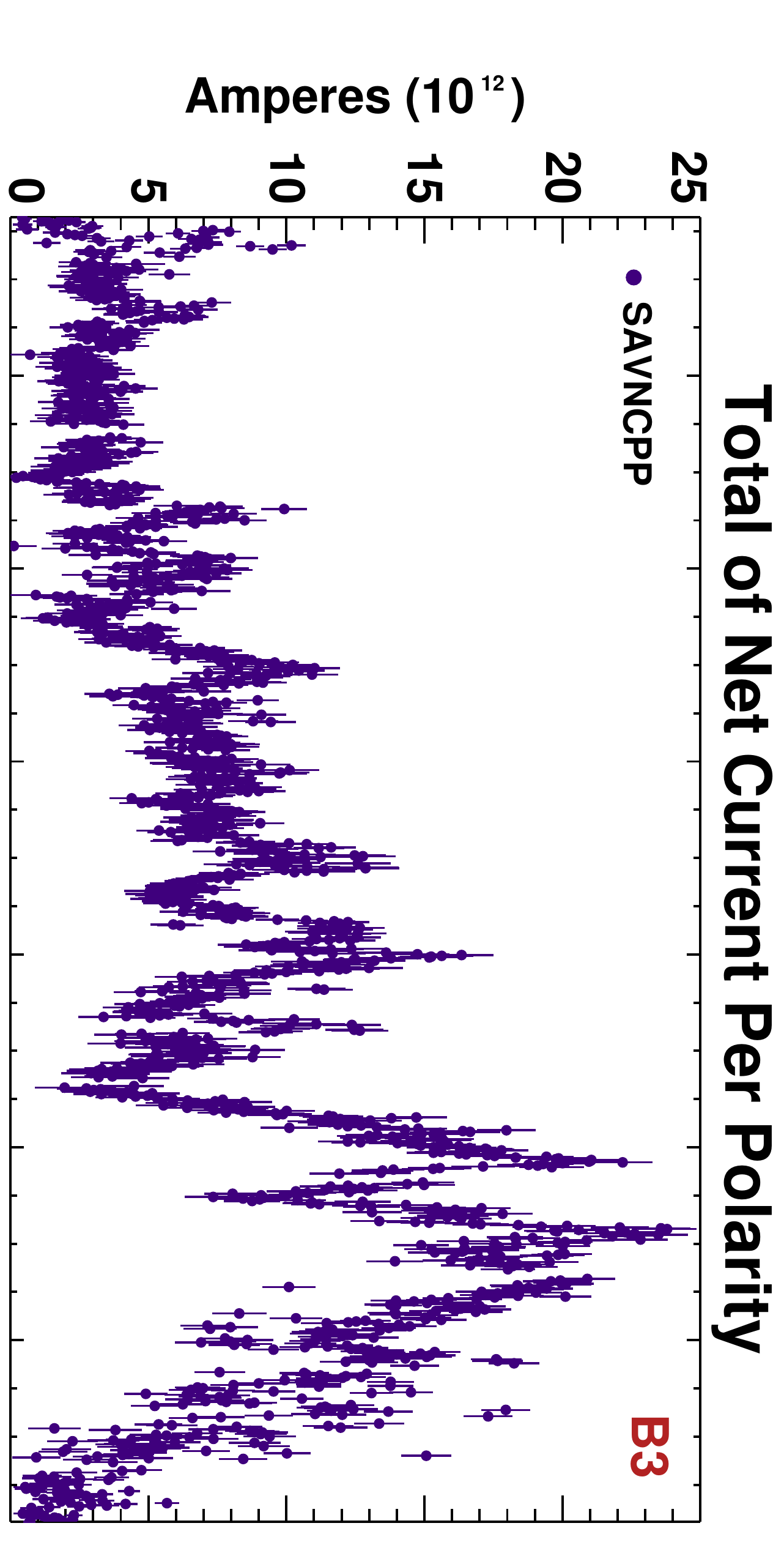} \\
\includegraphics*[angle=90,height=0.154\textheight,width=0.48\textwidth]{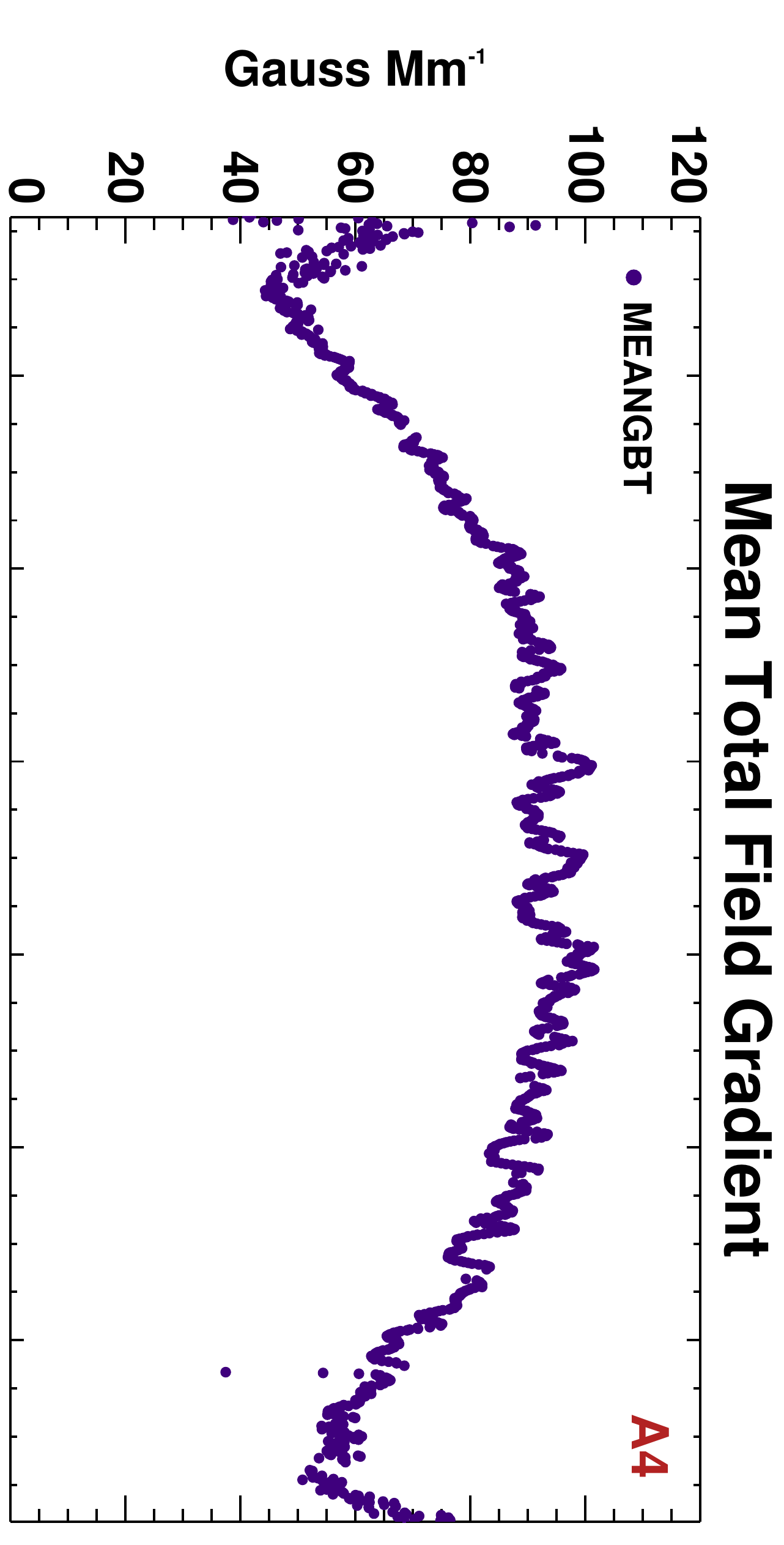} &
\includegraphics*[angle=90,height=0.154\textheight,width=0.48\textwidth]{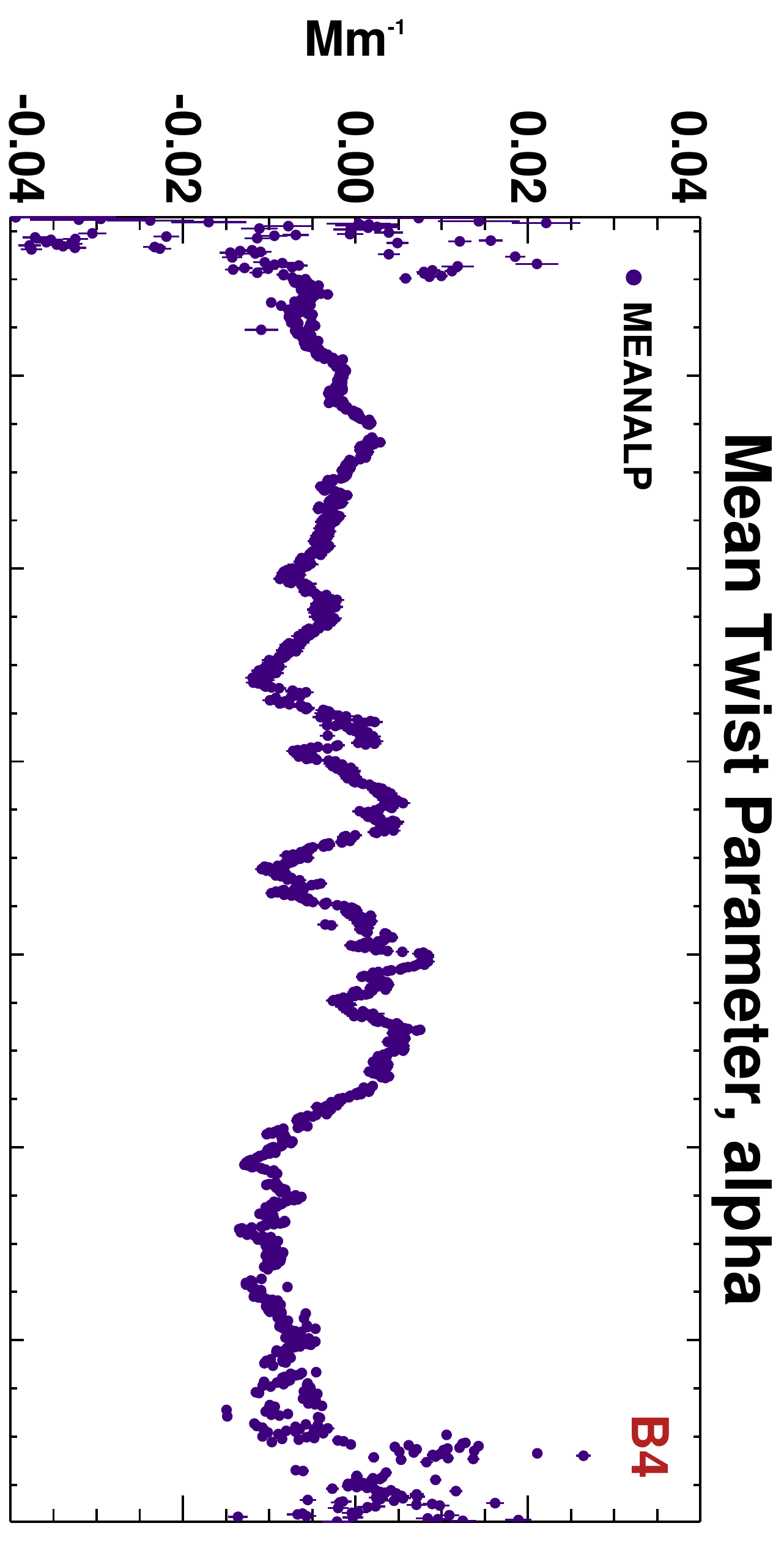} \\
\includegraphics[angle=90,height=0.154\textheight,width=0.48\textwidth]{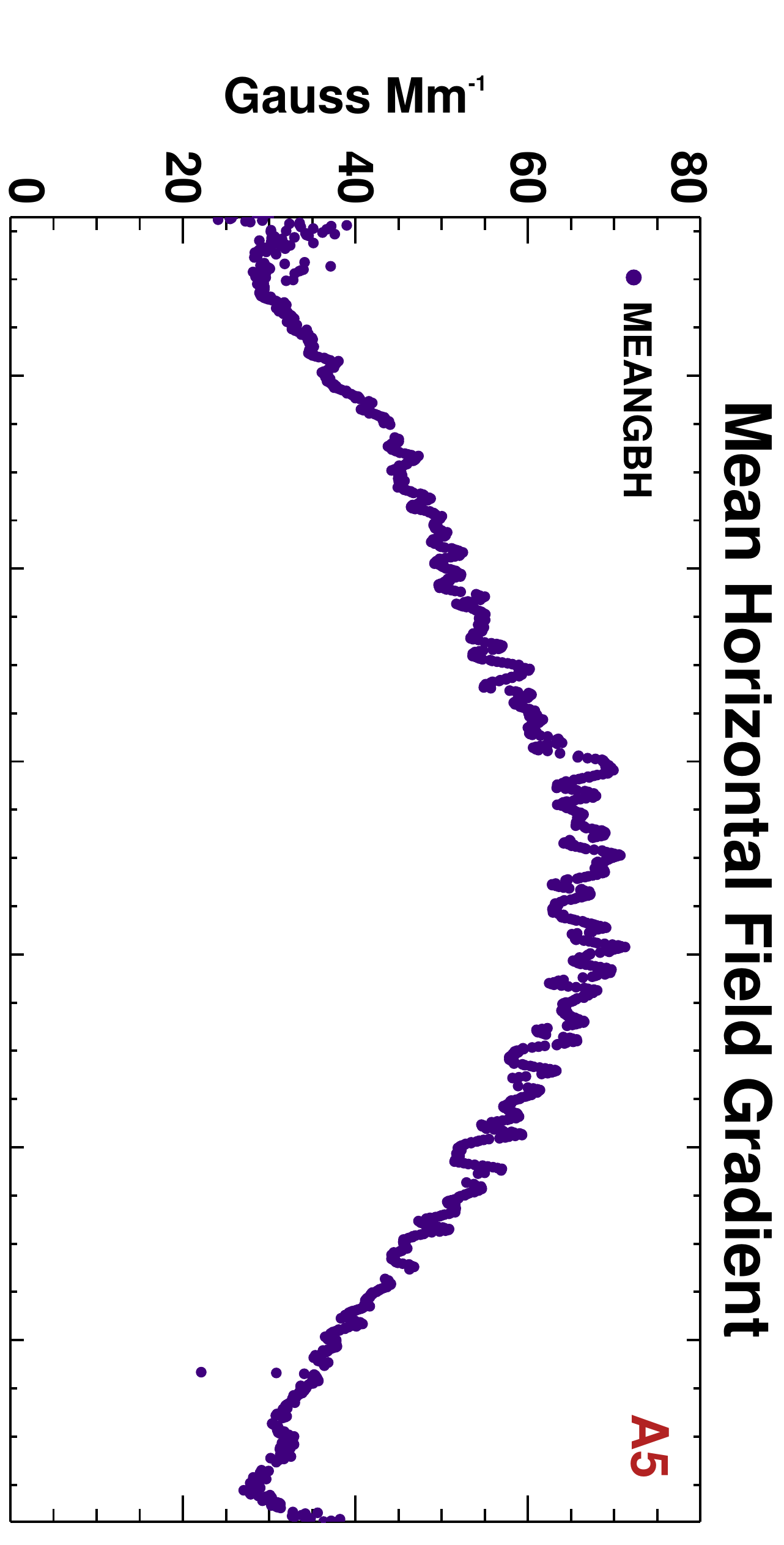} &
\includegraphics[angle=90,height=0.154\textheight,width=0.48\textwidth]{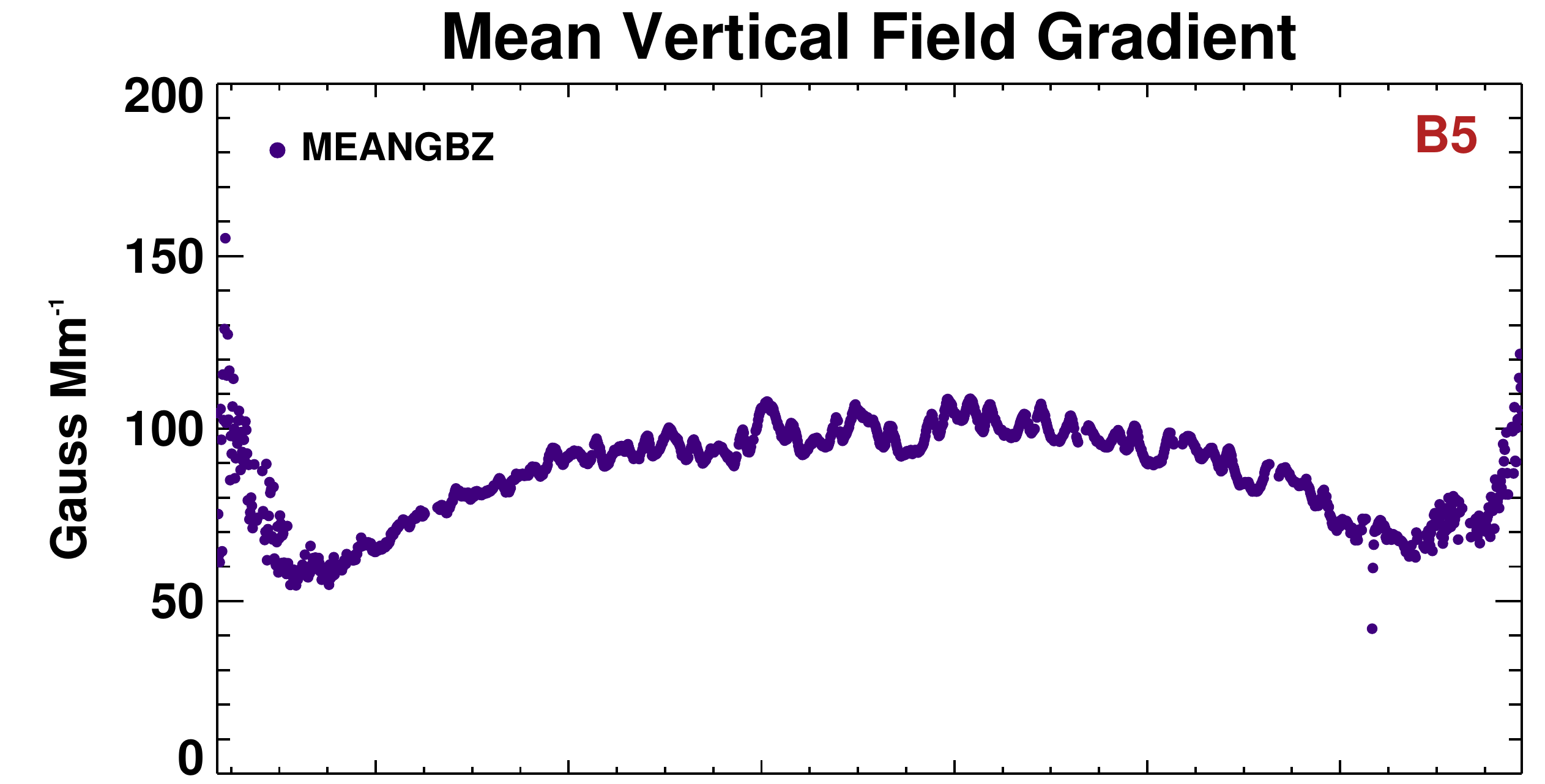} \\
\vspace{12pt}
\end{tabular}
\caption*{Additional {\sf SHARP} Active-Region Parameters for HARP 401, 2\,--\,15 March 2011.
Column A on the left shows five quantities: Panel A1, {\sc totusjh}; A2, {\sc totpot};
A3, {\sc totusjz}; A4, {\sc meangbt}; and A5, {\sc meangbh}. 
Column B on the right shows five quantities: 
Panel B1, {\sc meanjzh}; B2, {\sc absnjzh}; B3, {\sc savncpp}; B4, {\sc meanalp}; and 
B5, {\sc meangbz}.}
\label{fig:params401b}
\end{figure}}

\section{SHARP Parameters for an Illustrative Region: HARP 401}\label{sec:AR410}

The {\sf SHARP} indices are common active-region parameters described in the
literature, as discussed in the previous section, and the formulae are given
in Table \ref{tab:SpaceweatherFormulae}. Figures \ref{fig:params401} and
\ref{fig:params401b} show the {\sf SHARP} indices for HARP 401 from the time it
first rotated onto the disk on 2 March 2011 through its final disappearance
on 15 March.  Computed quantities from Table \ref{tab:SpaceweatherFormulae}
are plotted with error bars, except those that are areas or pixel counts. 
In most cases the error bars are smaller than the size of the dots because 
formal errors are small and systematic errors are not reflected.  
We have excluded data points with poor status bits set in the 
{\sc quality} keyword, which provides information about data reliability 
(see Table \ref{tab:otherkeywords} and \href{http://jsoc.stanford.edu/jsocwiki/Lev1qualBits}{{\sf Lev1qualBits}} referenced
in Table \ref{tab:urls} for more information about {\sc quality}).

The photospheric {\sc area} (Figure \ref{fig:params401} Panel A1, top left)
is determined by the HARP module
using the HMI line-of-sight magnetic field measurements. The
{\sc area} includes everything inside the orange patch in the left panel of
Figure \ref{fig:bitmap1}. This established active region rotates onto the disk on 2 March
and grows steadily as it crosses the disk. The patch reaches a maximum area
of $\approx7500$ microhemispheres on 11 March before it starts to decrease as
it rotates off the disk.
The panel below (Figure \ref{fig:params401} Panel A2) shows the total number of high-confidence pixels
that contribute to the {\sf SHARP} index calculation, {\sc cmask}, \textit{i.e.}  the pixels in
white in the right panel of Figure \ref{fig:bitmap1}.  Once the region is on
the disk, the number of {\sc cmask} pixels increases from about 40\,000 to 
nearly 80\,000. The number of contributing pixels changes with the size of
the region and also depends on the noise threshold that varies with location
on the disk and velocity of SDO relative to the Sun (see Section 7.1 of \opencite{hoeksema2013}). 
A histogram of the total-field noise level (not shown) increases and broadens near 
$60^\circ$ from central meridian, consequently increasing the number of pixels above the
noise threshold relative to disk center.

For comparison, Figure \ref{fig:params401} Panel A3 shows the area of the strong active pixels determined
from the line-of-sight field during the initial identification of the
HARP region.  This area, {\sc area\_acr}, associated with the white pixels
inside the orange patch on the left of Figure \ref{fig:bitmap1}, is smaller
than the area associated with the high-confidence pixels in the center panel
of that figure. The area of strong field shows a steady 40\,\% increase
during the new flux emergence on 7 -- 8 March. 
The total unsigned flux [{\sc usflux}] computed from the 
radial component of the vector
magnetic field appears in Figure \ref{fig:params401} Panel A4, at the bottom of 
the left column. The total flux, initially about $3 \times 10^{22}$\,Mx, 
decreases by 20\,\% on 6 March, recovers by a similar amount late
on 7 March, and then gradually builds to about $5 \times 10^{22}$\,Mx on 13
March. Variations in {\sc usflux} in this time interval do not exactly track
changes in the area of the region, the number of pixels in the computation,
or the strong-pixel area, indicating that the strength of the field in
the region is also changing. Correlated daily variations in {\sc usflux}
and {\sc cmask} are associated with SDO's geosynchronous orbital velocity.
The episode of flux emergence during 7 and 8 March is reflected in a number
of the quantities. The largest flare produced by HARP 401, an X\,1.5 flare,
peaked at 23:23 TAI on 9 March, about the time that the active-pixel area first
reaches a maximum. Numerous C-class and M-class flares occurred during the
lifetime of the region.

The systematic change in the transverse-field noise level is reflected in the
trend of the mean value of the inclination angle ({\sc meangam}) shown in Panel B1
at the top right of Figure \ref{fig:params401}. The plot shows both the
evolution of the region and a position-dependent trend that results from the
different strengths and noise levels in the circular and linear polarization
signals. (See \inlinecite{Borrero2011} for a relevant discussion of the effects of 
noise on the interpretation of vector field measurements.)
At disk center, the vertical magnetic-field component [B$_z$] is
closest to the lower-noise line-of-sight direction that depends on the stronger
Stokes-$V$; the horizontal component [B$_h$] reflects the sensitivity to noise in
Stokes-$Q$ and -$U$. In weak-field pixels this tends to bias the inclination angle
away from 0$^\circ$. The relative contributions of noise to the vertical
and horizontal field components change with center-to-limb angle [$\mu$].
As a consequence the ratio B$_z$/B$_h$ in the weak-field pixels increases,
decreasing the horizontal bias in the reported inclination. {\sc meangam}
reaches a maximum of $\approx60^\circ$ from radial near disk center and shows two
broad minima at $45^\circ$ and $40^\circ$ when the region is near the east
and west limbs, respectively, where the noise contributions to
the vertical and horizontal field components are roughly the same. 

The mean shear angle [{\sc meanshr}] in Figure \ref{fig:params401} Panel B2 
shows a similar variation across the disk, with a maximum a little over
$50^\circ$ near central meridian passage and broad minima below $40^\circ$
and $35^\circ$ in the East and West, respectively. The shear angle is calculated by 
determining the angle between the observed field [B$^{\rm Obs}$] and a potential 
field [B$^{\rm Pot}$]. 
To compute the parameters that require a potential-field model, we 
use the discretized Green's function based on Equation
(2.14) of \inlinecite{sakurai1982}, which is the potential due to a submerged
monopole at a depth of $\Delta/\sqrt{2\pi}$. In that case, $\Delta$
is the size of a pixel, which preserves the total flux of $B_z$. However,
using that depth yields a $B_z$ map that is blurry compared to the original
observational data, which, in turn, yields blurry calculated $B_x$ and
$B_y$ maps. Therefore, we choose a smaller $\Delta$ that corresponds to 0.001
pixels. Since this yields a sharper $B_z$ map, with a resolution similar to
the original observational data, the calculated $B_x$ and $B_y$ maps are of
a higher resolution as well. We preserve the original observational data for
the $z$-component of the potential magnetic field. 
Figure \ref{fig:params401} Panel B3, the fraction of {\sc cmask} pixels with 
hear greater than $45^\circ$ [{\sc sheargt45}] shows a pattern very similar 
to the mean shear and mean inclination angle.  Trends in the large-scale 
averages are affected by what is happening in the weak and intermediate field 
strength pixels near the noise level and the systematic change in reported 
field direction from center to limb.  There is a few percent decrease in the 
fraction of strong-shear pixels over the course of 9 March, prior to the 
X-class flare, which may or may not be significant.

Figure \ref{fig:params401} Panel B4 presents the mean value of the free energy
density averaged over the patch, {\sc meanpot}. {\sc meanpot} shares
evolutionary characteristics of the shear and inclination angle.
Figure \ref{fig:params401} Panel B5 (bottom right) shows the evolution of the mean vertical current
density [{\sc meanjzd}].  The point-to-point scatter and the uncertainties
in this quantity are relatively larger than for most of the other {\sf SHARP}
parameters.  The mean vertical-current density more than doubles from about
0.1 to 0.25 mA\,m$^{-2}$ on 7 March when new flux began to rapidly emerge.
The vertical current is computed using derivatives of the horizontal
magnetic-field components.  To compute any of the parameters that require
a computational derivative, we use a second-order finite-difference method
with a nine-point stencil centered on each of the {\sc cmask} pixels.

We now consider Figure \ref{fig:params401b}, which shows additional {\sf SHARP}
parameters for the same HARP 401. Figure \ref{fig:params401b} Panels A1 and A2 on the upper left show
the total unsigned current helicity [{\sc totusjh}] and a proxy for the
integrated total free-energy density [{\sc totpot}]. Both quantities show a
sustained increase on 7 March when new flux was emerging. The total current
helicity showed a sharp increase from 3100 to 3900 G$^2$\,m$^{-1}$ on 9 March
leading up to the X-class flare.  The integrated free-energy density is the
difference between the observed and potential magnetic-field energy integrated
over the region. {\sc totpot} nearly doubles from $5 \times 10^{23}$ to $9 \times
10^{23}$ erg\,cm$^{-1}$ on 7 March; however, no obvious signal associated with
the flare or its immediate aftermath is reflected in the free energy density
plot. In fact {\sc totpot} continues to increase gradually until 11 March.

The total unsigned vertical current ({\sc totusjz} in Figure \ref{fig:params401b} Panel A3) changes
dramatically during the life of HARP 401. Like the current helicity and
integrated free energy density, it reaches a plateau on 5 March and then
increases rapidly on 7 and 8 March from $4 \times 10^{13}$ to $7 \times 10^{13}$ A. 
A dip and rapid rise occur on 9 March before the X-class flare, after which the 
current stabilizes for several days.

Figure \ref{fig:params401b} Panels A4, A5 (bottom left), and B5 (bottom right) show
the temporal dependence of the horizontal gradients of the field. Each index is the mean
value of the gradient computed at the {\sc cmask} pixels in
the patch. 
Figure \ref{fig:params401b} Panel A4 shows the mean horizontal gradient of the total field
magnitude [{\sc meangbt}]. There is a fairly clear daily periodicity associated
with the spacecraft velocity and the number of pixels in {\sc cmask}. The
daily variation is superposed on a broad peak near central meridian at
about 100 G\,Mm$^{-1}$. 
The same shape is evident in Figure \ref{fig:params401b} Panel A5, which shows
the horizontal gradient of the horizontal component of the field [{\sc meangbh}].
The peak is a little sharper, ranging from $\approx20$\,--\,65~G\,Mm$^{-1}$ during
the disk passage of the region. Figure \ref{fig:params401b} Panel B5 (on the lower right), 
shows that the horizontal gradient of
the vertical component of the field [{\sc meangbz}] is less sharply peaked
near central meridian and has a more pronounced daily variation. Consideration
of other regions (see the discussion of HARP 2920 and Figure \ref{fig:params2920})
suggests that the broad shape tends to follow that of {\sc
cmask} and {\sc area}; so, perhaps the mean gradient of the vertical field is more
heavily influenced by the contributions of the variable number of weak-field
pixels than are the means of the total or horizontal field gradient.

Figure \ref{fig:params401b} Panel B1 (upper right) shows 
the mean of the contribution to the current helicity from the vertical 
components of the magnetic field and the current density [{\sc meanjzh}]. 
We cannot calculate the other terms that contribute to the
total helicity because HMI cannot determine the field gradient in the vertical
direction. The mean current helicity is generally negative for this region
through much of its lifetime and shows relatively strong variability while the
region is evolving rapidly from 6\,--\,11~March. Starting 12~March the helicity
was relatively large in magnitude, at -0.004 G$^2$\,m$^{-1}$, but stable. Indices
plotted in the next three panels, B2, B3, and B4, are related to physical quantities associated
with helicity, and thus all share a similar temporal profile. The sum of the
absolute values of the net current helicity [{\sc absnjzh}] 
is shown in Figure \ref{fig:params401b} Panel B2; 
the sum of absolute values of the net current determined separately in
the positive and negative B$_z$ regions [{\sc savncpp}] appears in 
Figure \ref{fig:params401b} Panel B3; and the
mean of the magnetic field twist, $\alpha$, of the region [{\sc meanalp}]
is in Figure \ref{fig:params401b} Panel B4. 
All exhibit some degree of daily variation. Periodic variations
are particularly strong on 6, 7, 9, and 11~March. All experience a steep
increase in magnitude on 11\,--\,12~March, after which the indices remain
fairly stable. The sum of the net currents in the two polarity regions
[{\sc savncpp}] peaks above $2 \times 10^{13}$\,A on 13 March.

The average twist parameter [{\sc meanalp}] posed a challenge. 
The simple definition of twist, $\alpha$=$\frac{J_z}{B_z}$, is noisy for individual pixels when the
field is low and near the noise level (\textit{cf.} \opencite{leka1999}). Simply averaging the computed $\alpha$
in the high-confidence {\sf SHARP} region pixels results in a meaningless scatter of
points from one time step to the next, suggesting that a higher threshold may 
be more appropriate. Instead we calculate a parameter intended to reflect the 
mean twist of the field in the entire active region. 
A variety of methods have been proposed \cite{Pevtsov1995,Leka1996,leka1999,falconer2002} based on 
fits to differences from a linear force-free field, moments of the distribution of $\alpha$, 
and taking ratios of spatial averages determined in parts of the active region. None of 
the methods is clearly superior. For the {\sf SHARP} index {\sc meanalp} we adopt the
B$_z^2$-weighted $\alpha$ method proposed by \inlinecite{hagino2004} in which 
one simply computes the sum of the product of $J_z B_z$ at the {\sc cmask}
pixels and divides by the sum of $B_z^2$.

\section{Selected Parameters for a Second Region, HARP 2920}\label{sec:AR2920}

{\begin{figure}
\centering
\caption{{\sf SHARP} Parameters 2920}
\renewcommand{\tabcolsep}{0.01\textwidth}
\begin{tabular}{rr}
\includegraphics*[angle=90,height=0.15\textheight,width=0.48\textwidth]{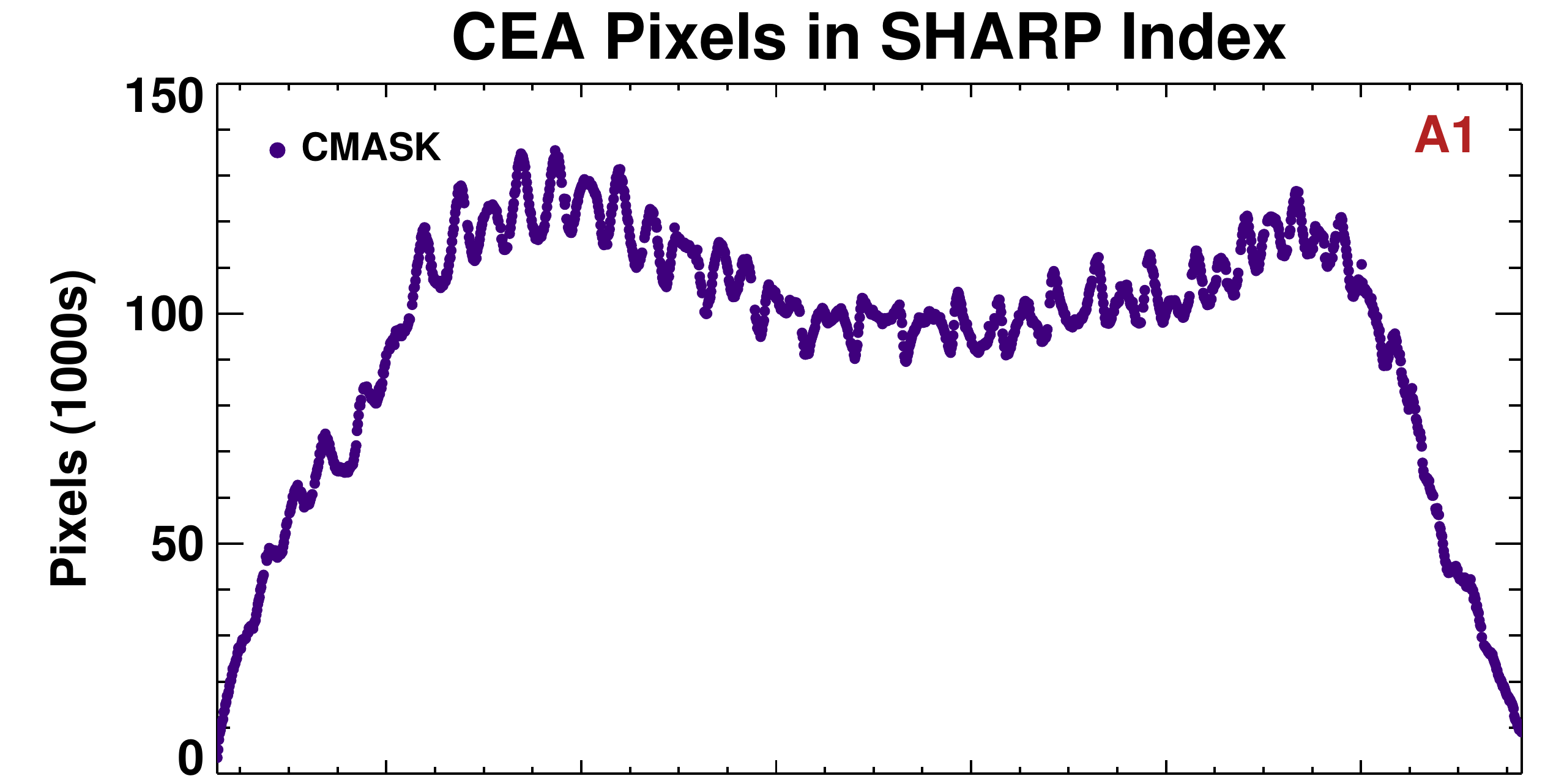}  &
\includegraphics*[angle=90,height=0.15\textheight,width=0.48\textwidth]{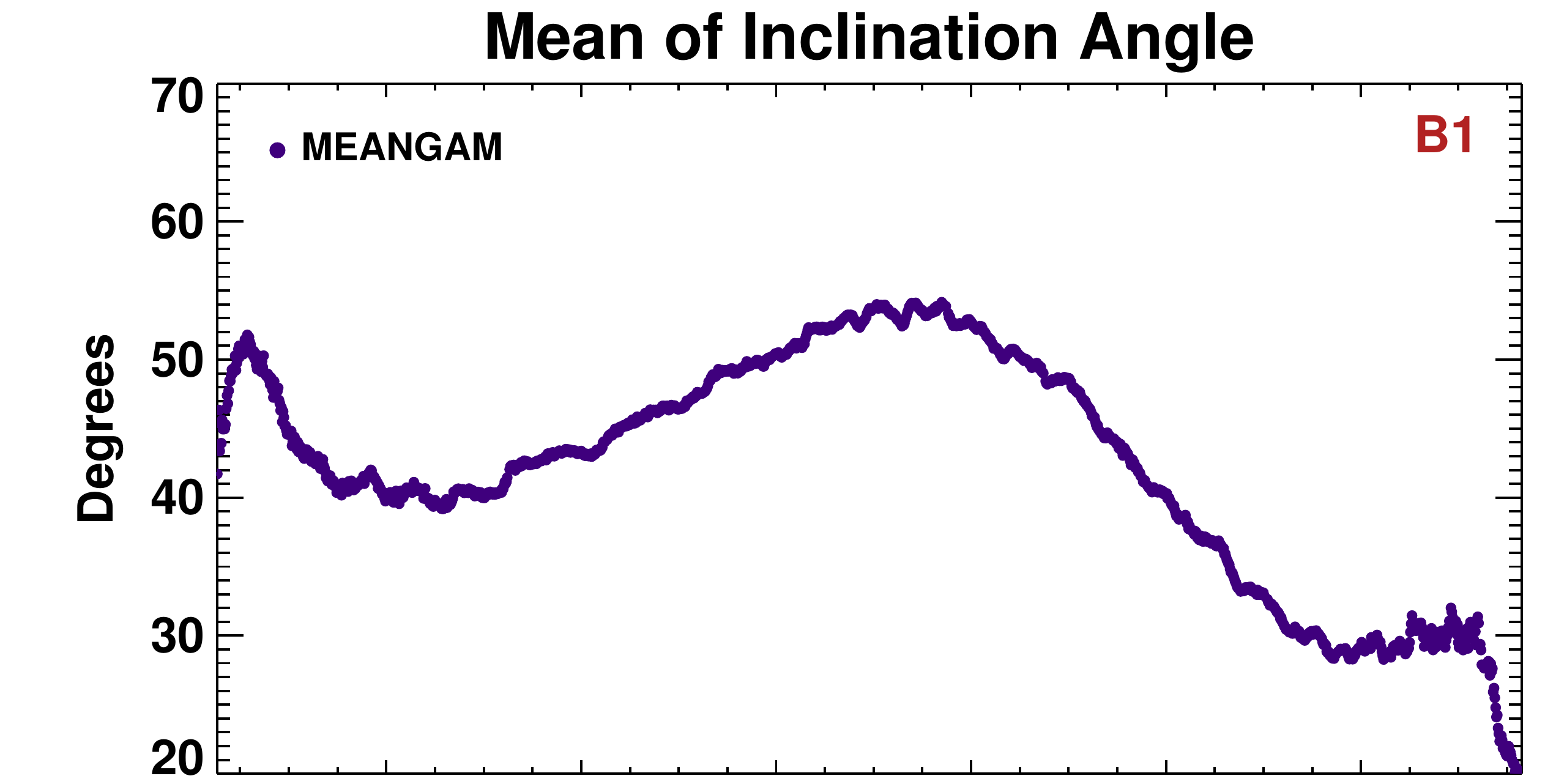} \\
\includegraphics*[angle=90,height=0.15\textheight,width=0.48\textwidth]{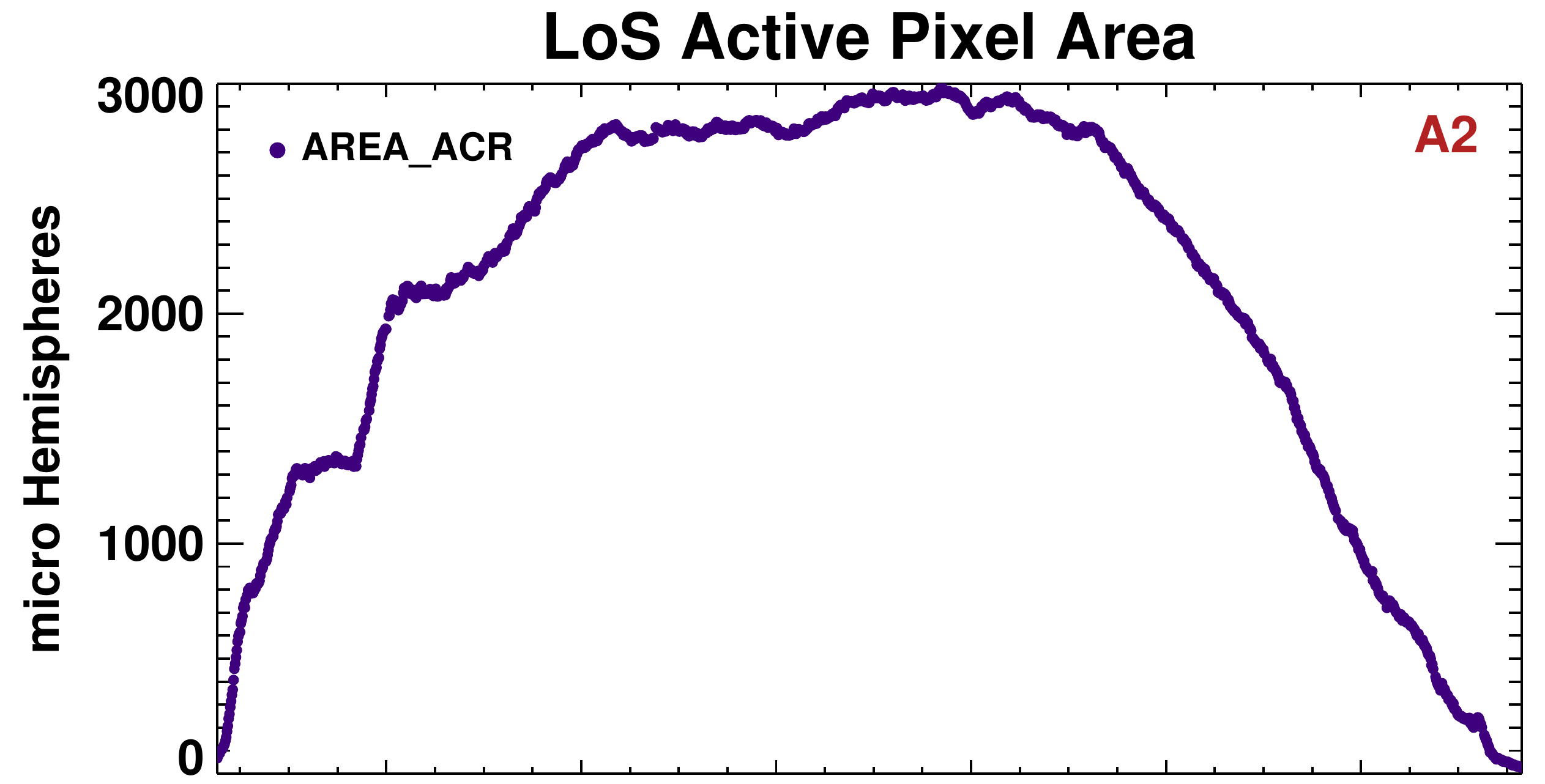} &
\includegraphics*[angle=90,height=0.15\textheight,width=0.48\textwidth]{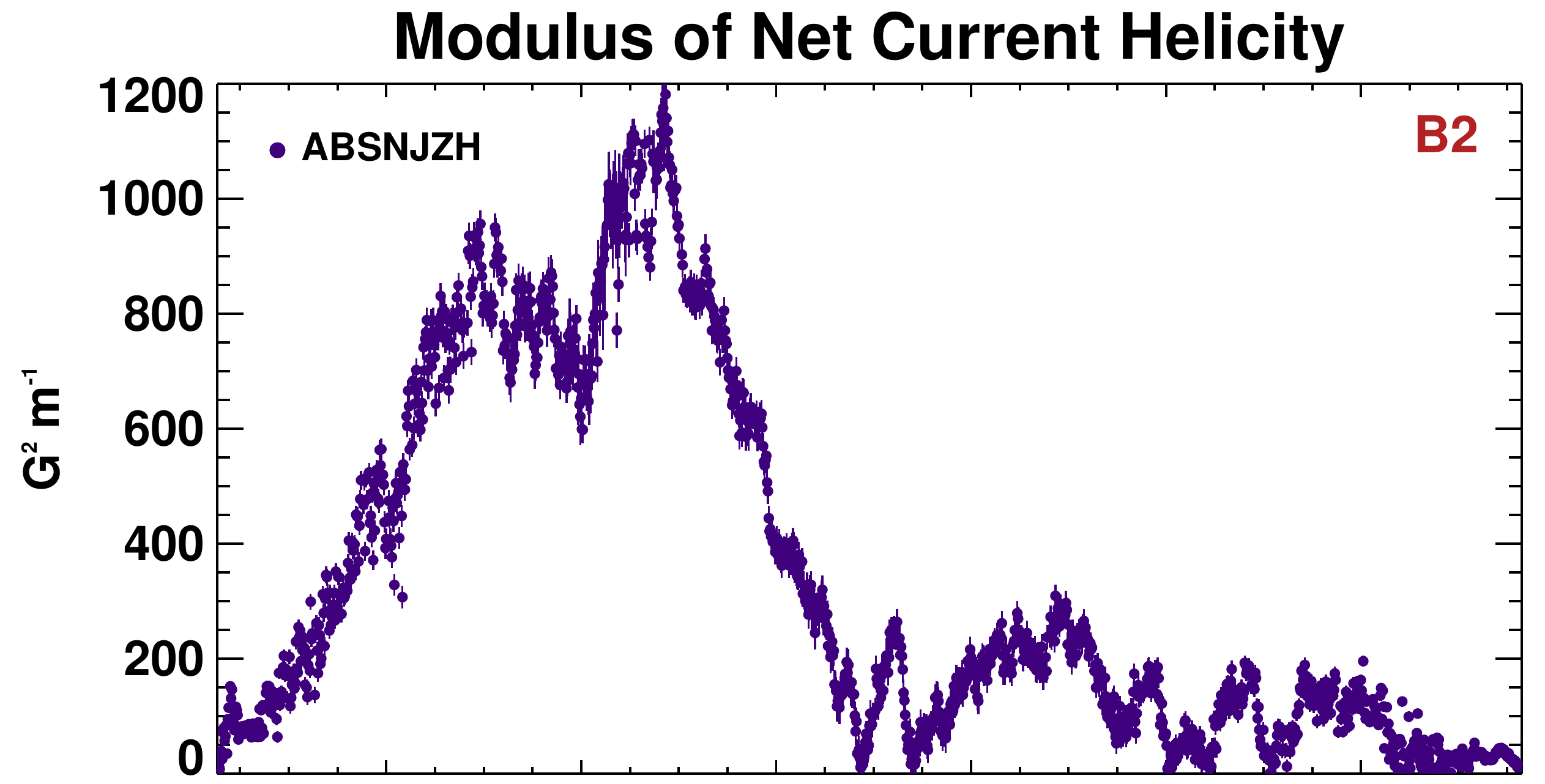} \\
\includegraphics*[angle=90,height=0.15\textheight,width=0.48\textwidth]{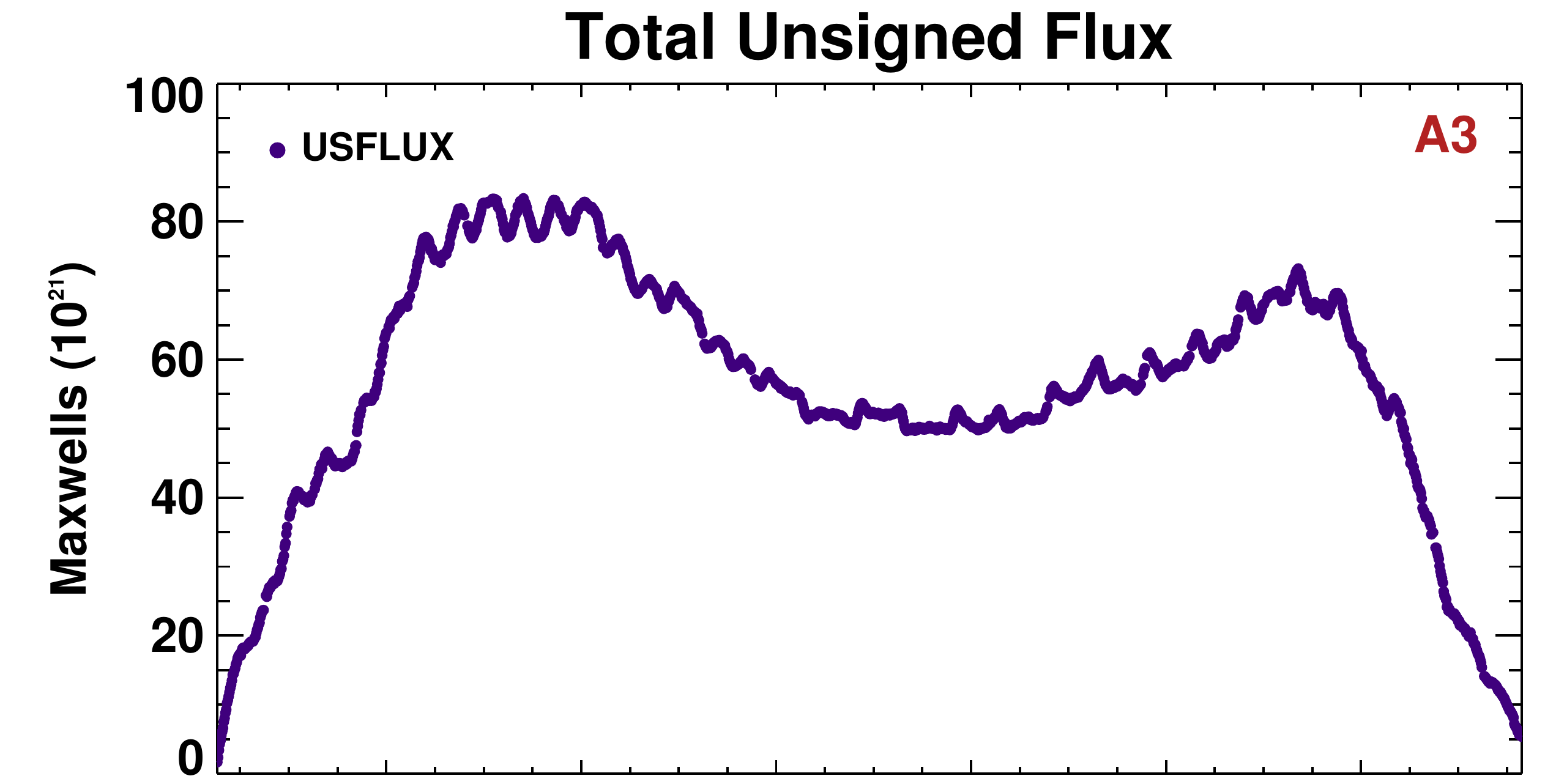} &
\includegraphics*[angle=90,height=0.15\textheight,width=0.48\textwidth]{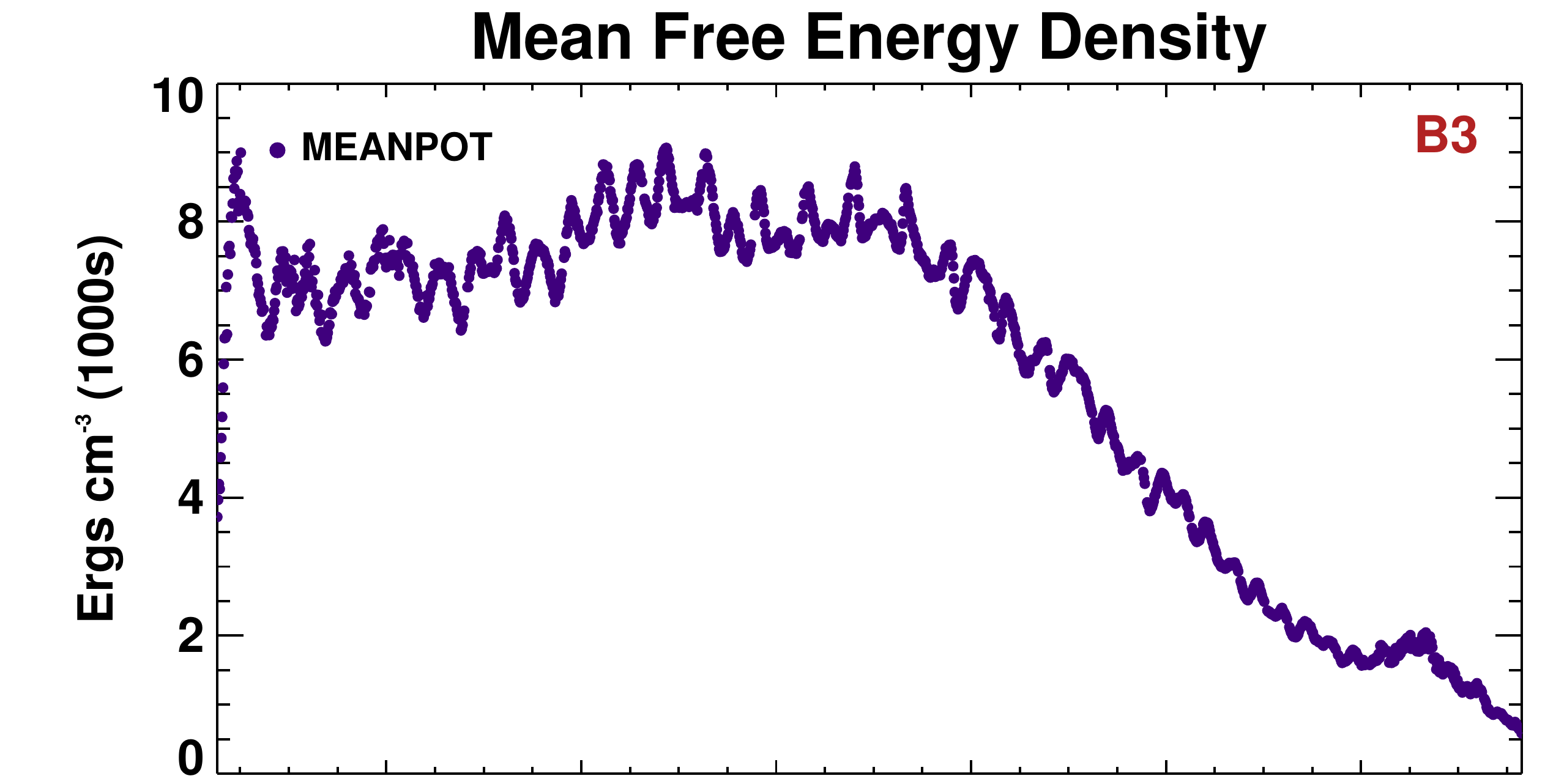} \\
\includegraphics[angle=90,height=0.15\textheight,width=0.48\textwidth]{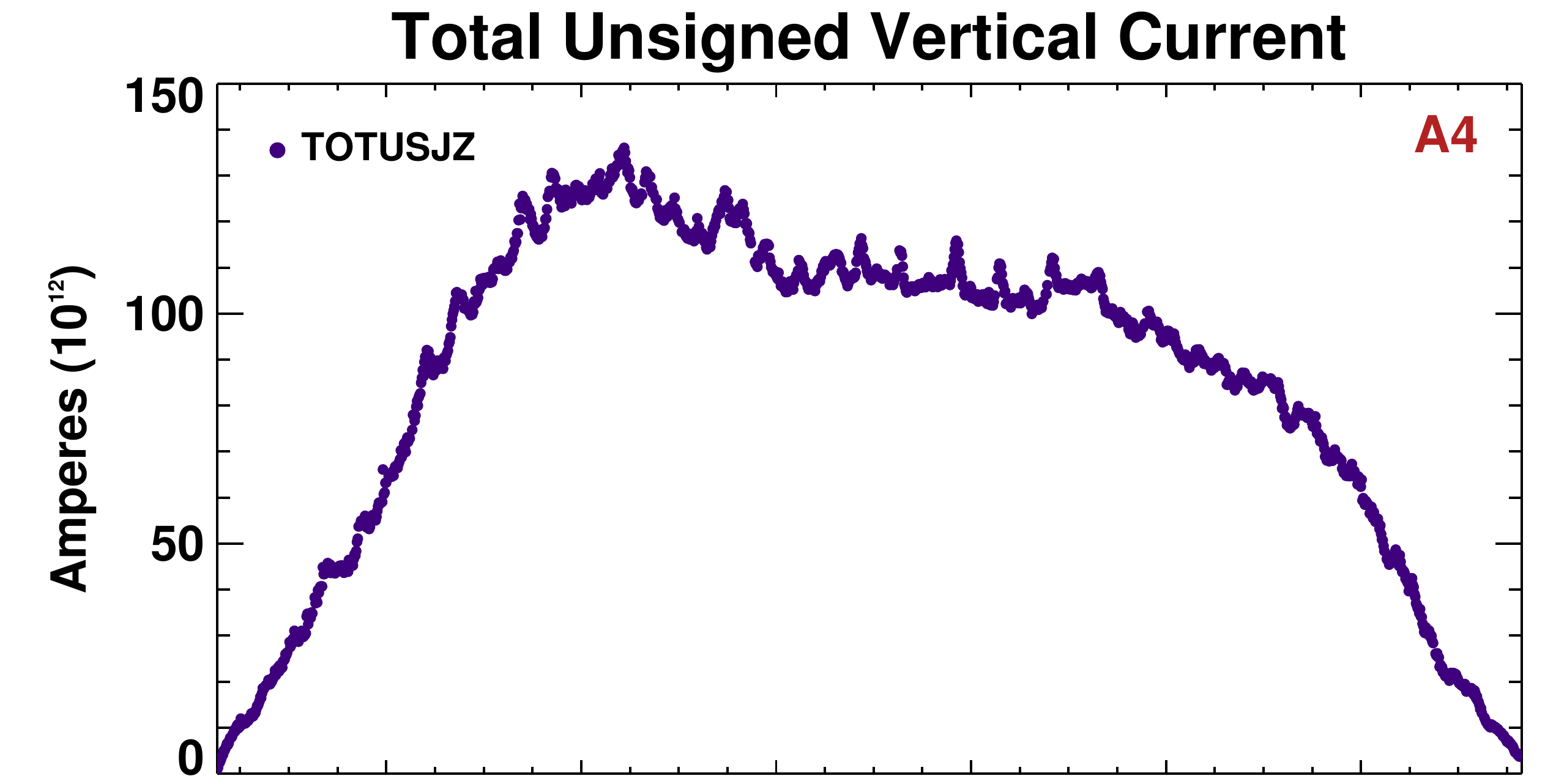} &
\includegraphics[angle=90,height=0.15\textheight,width=0.48\textwidth]{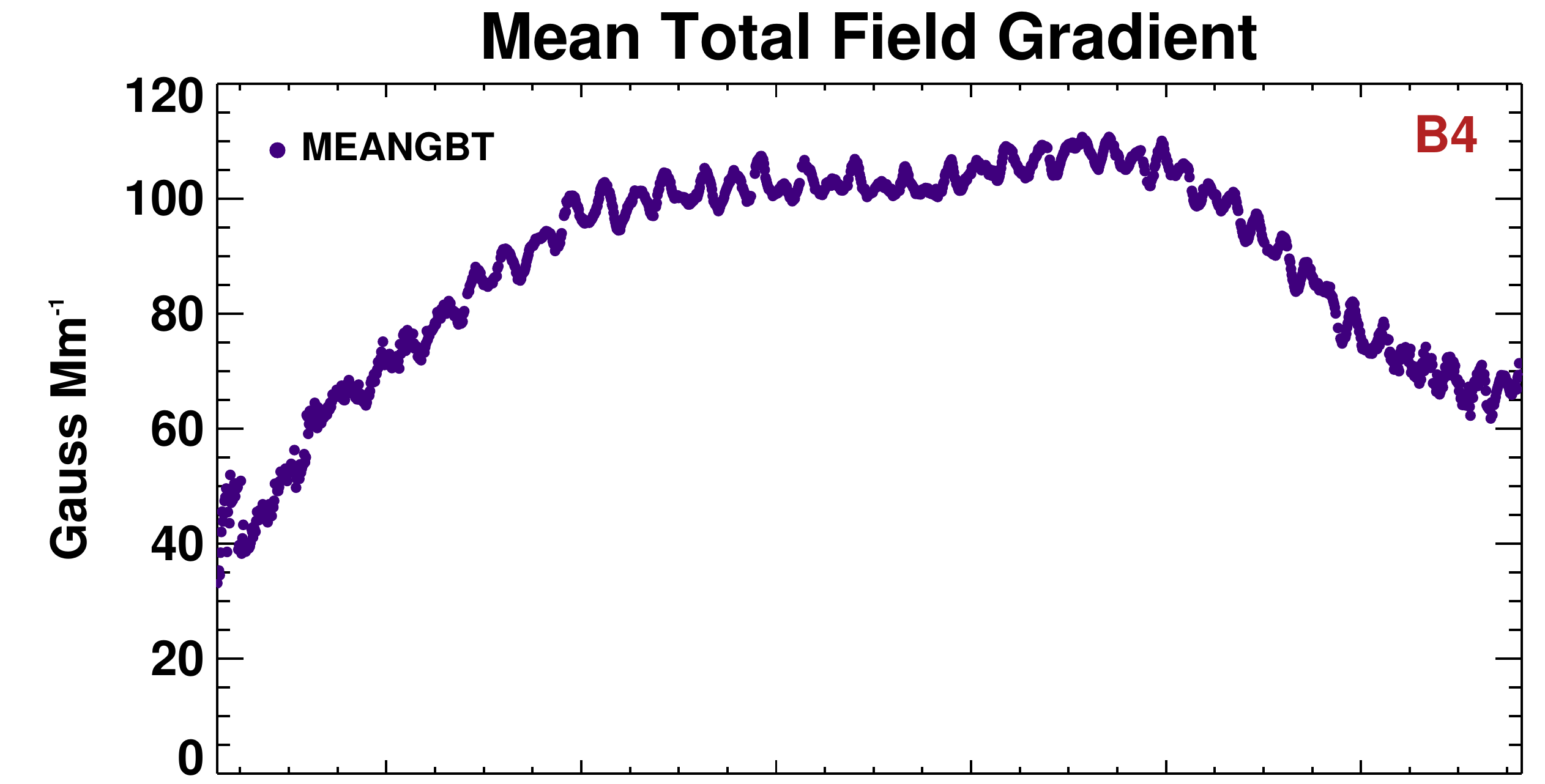} \\
\vspace{12pt}
\end{tabular}
\caption*{{\sf SHARP} Active Region Parameters for HARP 2920, 1\,--\,14 July, 2013.
Column A on the left shows four quantities: Panel A1, {\sc cmask}; A2, {\sc acr\_area};
A3, {\sc usflux}; and A4, {\sc totusjz}. 
Column B on the right shows four quantities: 
Panel B1, {\sc meangam}; B2, {\sc absnjzh}; B3, {\sc meanpot}; and B4, {\sc meangbt}.
}
\label{fig:params2920}
\end{figure}
}

Considering a single active-region complex does not provide sufficient context to understand how regions
differ from each other or how much of the variation in a quantity depends on disk position or other typical evolutionary
characteristics.
To illustrate the differences between regions, Figure \ref{fig:params2920}
shows selected {\sf SHARP} indices for HARP 2920 from the time that it first rotated onto
the disk on 1 July 2013 through its final disappearance on 14 July. 
HARP 401 was energetic and large, but had reasonably simple large-scale topology.
HARP 2920 was larger and more complex, ultimately including three NOAA regions: 
11785, 11787, and 11788. HARP\,2920 produced numerous C-class flares; the largest, class 
M\,1.5, occurred at 07:18 UT on 3 July while the region was still near the east limb.
Figure \ref{fig:params2920} Panel A1
({\sc cmask}, upper left) shows the number of high-confidence CEA pixels that contribute
to the indices. Panel A2 shows the area associated with strong pixels, {\sc area\_acr}.
The region grows as it rotates onto the disk and then on 3 and 4 July its
size nearly doubles from about 1400 microhemispheres on 2 July to 2100 on 3 July 
as a second activity complex (AR\,11787) rotates over the limb
and then to 2800 by the end of 4 July as new flux emerges. 
In the NRT HARP this appearance and nearby emergence results in the merger of two regions. 
The size of the region remains fairly stable as it continues to rotate across the disk. 
The active pixel area [{\sc area\_acr}] starts to diminish on 10 March, but the size of 
the high-confidence pixel area [{\sc cmask}] only begins to 
decrease rapidly starting on 12 July as the HARP rotates off the limb. 
Contrast this with the strong emergence of new flux within the existing flux 
system seen in HARP 401 on 8\,--\,9 July. 

The evolution of the total unsigned flux [{\sc usflux}] appears in Figure \ref{fig:params2920} Panel A3. 
The change in {\sc cmask} pixel number creates broad peaks near $60^\circ$ from central meridian 
on 4 July and 12 July in the {\sc usflux}.
The variations of {\sc cmask} and {\sc usflux} were also correlated for HARP 401, 
but the evolution across the disk was much different. 
The trend also seems to be reflected in an inverse fashion in the mean inclination angle [{\sc meangam}]
plotted in Figure \ref{fig:params2920} Panel B1 (top right). A similar inverted trend appears, with a broad peak near central
meridian on 8\,--\,9 July, in the measures of shear angle and the mean vertical current density
(not shown). The similarity of the {\sc meangam} profile for 401 and 2920 confirms that
significant effects due to the relative noise levels in Stokes $Q\,U\,V$ are important. 

Figure \ref{fig:params2920} Panel B2 shows the modulus of the net current
helicity [{\sc absnjzh}]. There is a strong rise on 2\,--\,4 July and again
on 5 July followed by a sharp decline on 6 and 7 July. The mean-current-helicity,
net-current-per-polarity, and mean-twist parameters (not shown) have a similar
profile. Contrast this with the weaker and relatively less volatile behavior
of HARP 401 (note the difference in plot scale) 
even though 401 was emerging much more new flux. 
The mean free energy density [{\sc meanpot}, Figure \ref{fig:params2920}, Panel B3] remains
fairly stable at 7000 ergs\,cm$^{-3}$ from the time the region appeared until
a steady decrease begins on 9 July. 
The mean free energy density of HARP 401 was significantly greater and 
increased by $\approx30$\,\% during its disk passage before beginning a similar decline.
The variations of the total unsigned vertical current [{\sc totusjz}, 
Figure \ref{fig:params2920} Panel A4) are representative of the total
unsigned current helicity and integrated free energy density proxy. Unlike
HARP 401, these quantities in HARP 2920 do not follow the evolution of the
unsigned flux or the area. There is an interesting small excursion in the 
vertical current on 6 July just after the helicity measures reach their peak
and begin their rapid decline. No similar relationship is seen in HARP 401. 

Finally, Figure \ref{fig:params2920} Panel B4 plots the mean of the horizontal 
gradient of the total field strength [{\sc meangbt}] which is indicative of the 
evolution of the mean gradients of the other field components.
The broad hump on the {\sc meangbt} curve that occurs on 9\,--\,10 July is not 
apparent in any of the indices unrelated to field strength gradients.
Otherwise the evolution is very smooth, much smoother than for HARP 401. 
All gradient indices exhibit a short-term (12-hour) variation that is related to the sensitivity of
the vector-field measurement to the orbital velocity of the spacecraft \cite{hoeksema2013}. 
The general profile of the mean gradient of the horizontal-field component (not shown)
for HARP 2920 has a broad peak near central meridian passage, as does the area of the strong-field elements. 
The mean gradients of the total and vertical field (not shown) follow more closely the flatter shape of the total area, with additional 
broad increases appearing near $60^\circ$ from central meridian 
associated with the increase in the number of weak and
intermediate strength pixels, though both start to fall off steadily on 10 July.

\section{Definitive and Near-Real-Time (NRT) SHARPs}\label{sec:NRT}

The definitive HARP processing module groups and tailors the identified regions
according to their complete life history. The definitive HARP geometry is
determined only after an active-region patch has crossed the face of the disk. At
each time step the rectangular bounding box of a definitive HARP on the CCD encloses the
fixed heliographic region that encompasses the greatest geometric extent attained
by the patch during its entire lifetime. The temporal life of a definitive HARP starts
when it rotates onto the visible disk or two days before an emerging magnetic feature
is first identified in the photosphere. The HARP expires two days after the
feature decays or when it rotates completely off the disk. The 
center of the HARP at central meridian passage is uniformly tracked at the
differential rotation rate appropriate for its latitude, given in keyword {\sc omega\_dt}. There
is necessarily a delay of about five weeks before definitive {\sf SHARP}s can be created.

Operational space-weather forecasting requires more timely data and would need to 
rely on the HMI NRT data stream. We outline below three primary differences between 
the NRT data and definitive {\sf SHARP} data. 
Note that the {\sc harpnum} for a particular
region will be different for the definitive and NRT {\sf SHARP} series.
The NRT {\sf SHARP}s are offered ``as is'', \textit{i.e.} there is no plan to necessarily 
correct the NRT data series when updates are made to the definitive {\sf SHARP}s. 
The NRT {\sf SHARP} archive begins 14 September 2012, but because of the inferior quality of 
the NRT data, we strongly recommend against
use of the NRT data except for forecasting and development 
of forecasting tools.

\begin{enumerate}

\item The NRT and definitive observables input data differ in completeness and calibration. 
Roughly 4\,\% of the data are delayed more than one hour; delays tend to be more clustered than random.
Calibrations and corrections to the NRT data rely on predicted conditions or on calibration information that may be increasingly
out of date as the day progresses. Effects of cosmic rays are not corrected.
The differences are generally minor or localized. For a detailed summary
of calibration procedures and the differences between the NRT and definitive
input data, see \inlinecite{hoeksema2013}.

\item NRT HARP geometry is determined as soon as possible, before the full
life cycle of the region is known. For that reason the photospheric area
enclosed by the box bounding the active region can grow (but will never
shrink) with time. In addition, the heliographic center of the NRT HARP
bounding box may shift in time as a region evolves. In general the size and
shape of the patch itself is the same in NRT and definitive HARPs. It is
important to note that NRT HARPs may merge, resulting in the termination of
one HARP and the continuation of another HARP, but augmented by the content
of the terminated HARP. This will typically cause a major discontinuity
in the NRT {\sf SHARP} indices at that time step. 
The {\sc h\_merge} keyword is set
when such a merge occurs, so that merging can be taken into account when the
discontinuities are observed. The {\sc h\_merge} keyword is also 
carried over into the
definitive HARPs, but in this case the region
configuration is consistent before and after the merge (the entire future
of all regions is available), so for definitive HARPs, the relic {\sc h\_merge}
keyword is not particularly significant. At least one merger occurred during
the lifetime of 494 of the first 3213 HARPs. Note again that the NRT and
definitive {\sc harpnum} will not be the same.

\item For NRT processing the annealing parameters for the disambiguation
code are adjusted to enable faster computation \cite{Barnes2013} and
a smaller buffer outside the HARP is used to compute the potential-field starting point.
The keyword {\sc ambnpad} gives the size of the buffer and is reduced to 50 currently 
for NRT {\sf SHARP}s from the 500 used for definitive processing. To
investigate how these input parameters affect the active-region indices,
we disambiguated a five-day cube of inverted data for HARP 401 using the two
different sets of disambiguation parameters. The resulting active-region
indices generally differ by less than a percent. For example, the typical difference
in the total field gradient was less than 0.05\,\% with a maximum difference of 0.3\,\%.
Starting on 15 January 2014, the definitive {\sf SHARP}s rely on full-disk 
rather than patch-wise disambiguation.

\end{enumerate}

\inlinecite{hoeksema2013} present for HARP 2920 a detailed comparison between
the definitive and quick-look total unsigned flux parameter and find that
the typical difference is about 1\,\% (see their Figure 5). The differences
have some systematic periodic components, likely attributable to differences
in calibration. The differences increase to a few percent when {\sf SHARP}s are 
near the limb. By far the largest difference ($\approx30\,\%$) is due to a merger.

\section{Sources of Uncertainty}\label{sec:Uncertainty}

The vector-magnetogram data used in this study have uncertainties and
limitations that are discussed at length by \inlinecite{hoeksema2013}. Many
of these issues are more significant in weak-field regions, which
do not contribute directly to the computation of active-region parameters, 
except that in intermediate field-strength regions near the noise threshold
the number of pixels can change appreciably. 
Systematic errors remain, the largest associated with the
daily variation of the radial velocity of the spacecraft inherent to the
geosynchronous orbit (\textit{e.g.} small periodic variations in 
Figures~\ref{fig:params401}\,--\,\ref{fig:arprofiles}). 
For each index we characterize the formal random
error in the computed active-region parameter. The inversion code provides estimates
of uncertainties at each pixel, including ${\chi}^{2}$, the computed standard deviations,
and certain correlation coefficients of the errors in the derived parameters. They
effectively provide a way to estimate a lower limit on the uncertainties. 
We use the uncertainty determined for each component of the vector magnetic field and 
formally propagate these error estimates per pixel per unit time per quantity for each 
{\sf SHARP} index. The uncertainty keyword is listed in the last column of 
Table \ref{tab:SpaceweatherFormulae}. 
To test the results, we verified our formal error propagation via a Monte Carlo analysis in
which we varied the input Stokes parameters according to the error estimates, 
a relatively early stage in the vector field pipeline. The variability found in the
final {\sf SHARP} indices is consistent with the formal error propagation results.

\section{Sample Data and Discussion}
\label{s:analysis}

{\begin{table}
\caption{Flare-Producing Active Regions}
\renewcommand{\arraystretch}{1.5}
\renewcommand{\tabcolsep}{0.2cm}
\begin{tabular}{ l r r c c }
Flare Peak [TAI] & Class & HARP & NOAA AR & \parbox{1.7cm}{(Lat., Lon.)\\in Degrees} \\
\hline
2011.02.15\_01:56:00&X2.2&377&11158& (-20.20, 12.77) \\
2011.03.09\_23:23:00&X1.5&401&11166& (8.86, 10.30) \\
2011.09.06\_22:20:00&X2.1&833&11283& (15.13, 14.19) \\
2012.03.07\_00:24:00&X5.4&1449&11429& (17.72, -25.90)\\
2012.11.21\_15:30:00&M3.5&2220&11618& (7.88, -5.19) \\
2012.11.27\_21:26:00&M1.0&2227&11620& (-13.40, 41.18) \\
2013.01.13\_00:50:00&M1.0&2362&11652& (19.49, 12.28) \\
2013.02.17\_15:50:00&M1.9&2491&11675& (12.43, -22.75) \\
2012.12.25\_06:43:00&C1.8&2314&11635& (11.07, 6.60) \\
2013.01.01\_09:06:00&C1.2&2337&11640& (27.21, -0.38) \\
2013.01.31\_04:34:00&C1.1&2420&11663& (-10.96, 9.63) \\
2013.02.03\_18:01:00&C1.5&2433&11665& (10.66, -2.94) \\
\end{tabular}
\caption*{The following active regions that produced X-, M-, and C-class
flares were used in our sample data. In the table, we list the time and
position of the active region during the GOES X-Ray flux peak; however,
we analyzed a five-day time series of data per active region. The latitude and
longitude are given in Stonyhurst coordinates and correspond to the latitude
and longitude of the flux-weighted center of active pixels at the time of the GOES X-Ray
flux peak. These correspond to keywords {\sc lat\_fwt} and {\sc lon\_fwt}.}
\label{tab:ar}
\end{table}} 

{\begin{figure}
\centering
\caption{Active Region Profiles}
\renewcommand{\tabcolsep}{0.0018\textwidth}
\begin{tabular}{cc}
 \includegraphics[angle=90,width=0.496\textwidth]{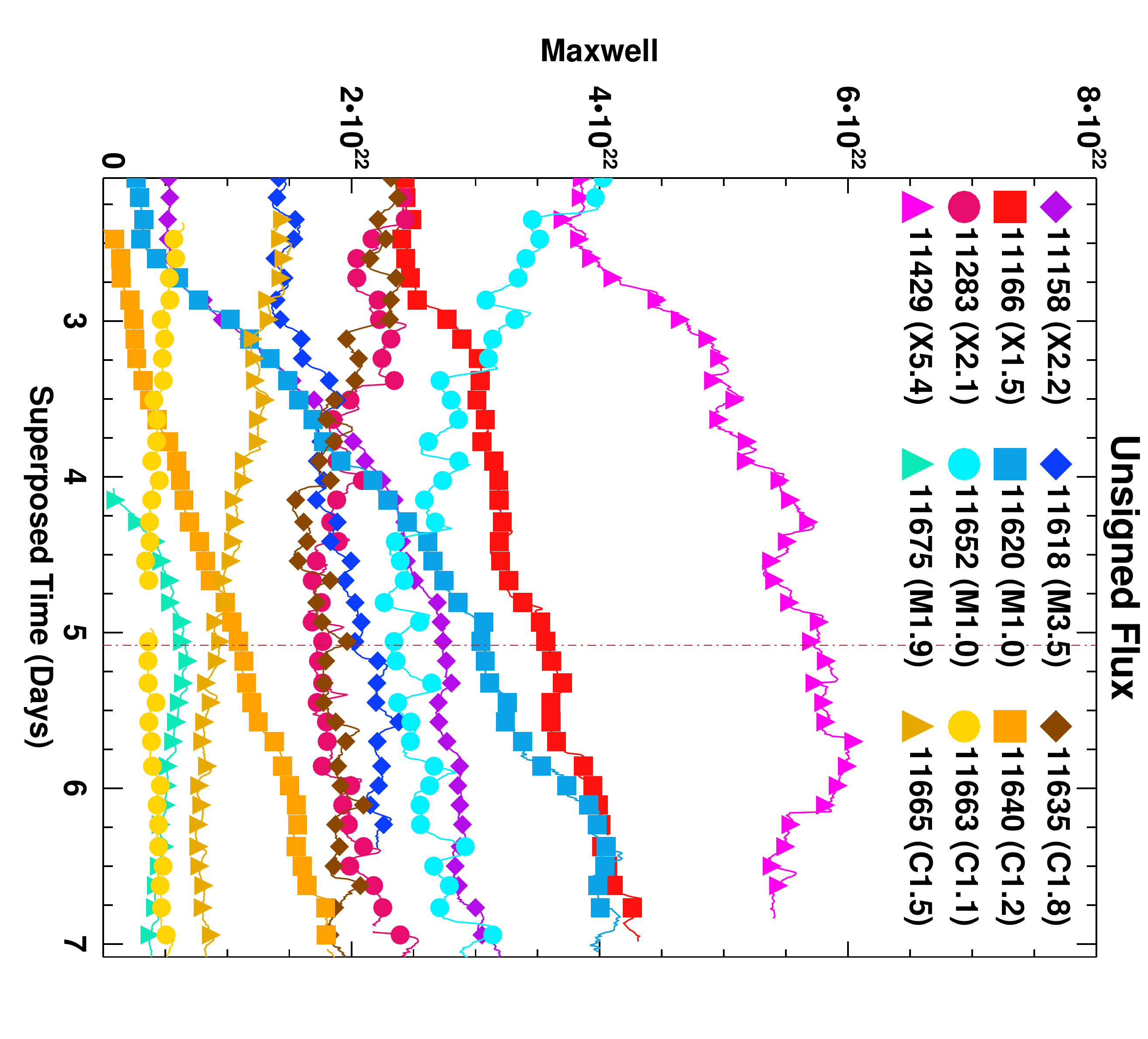} &
 \includegraphics[angle=90,width=0.496\textwidth]{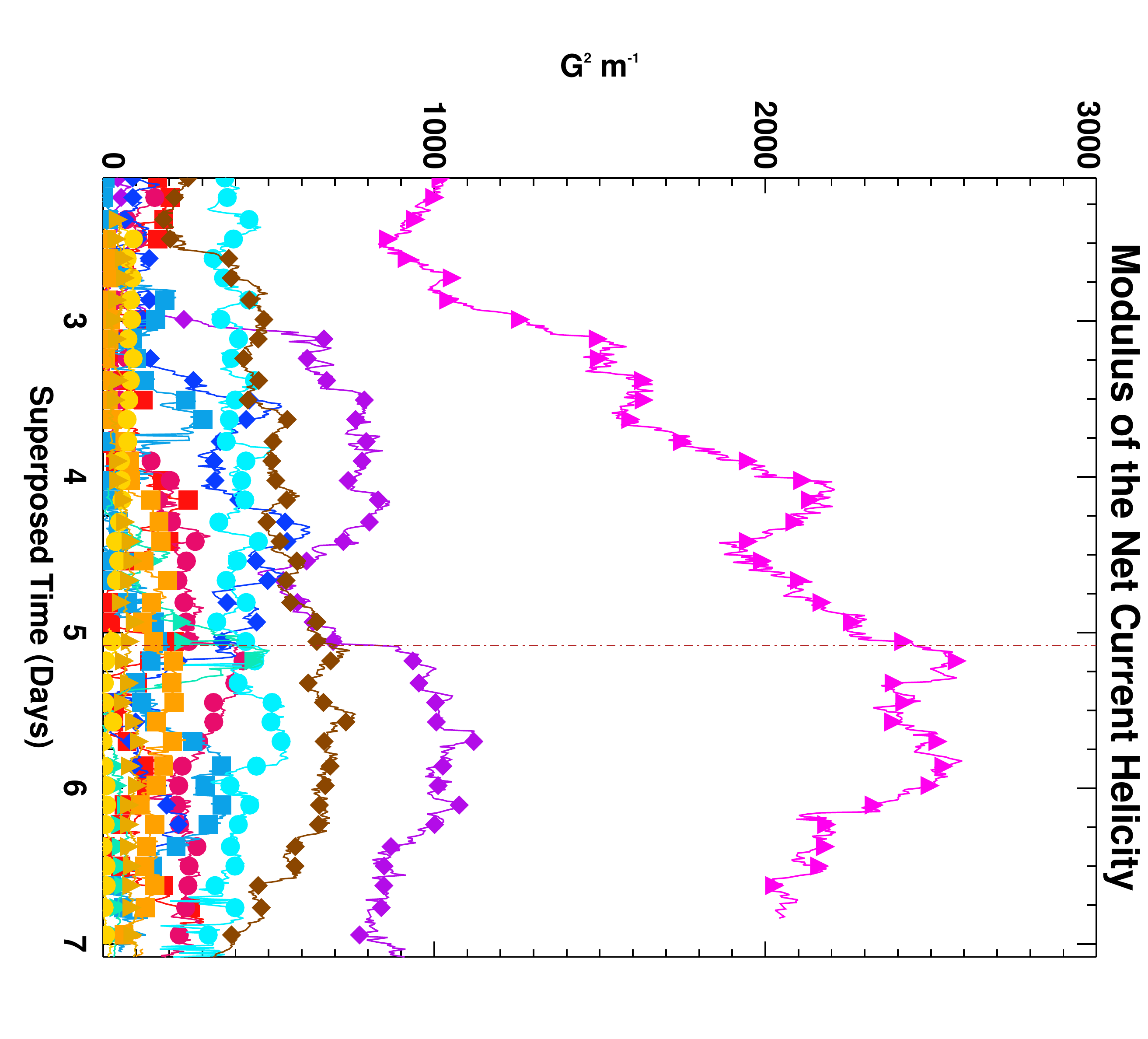} \\
 \includegraphics[angle=90,width=0.496\textwidth]{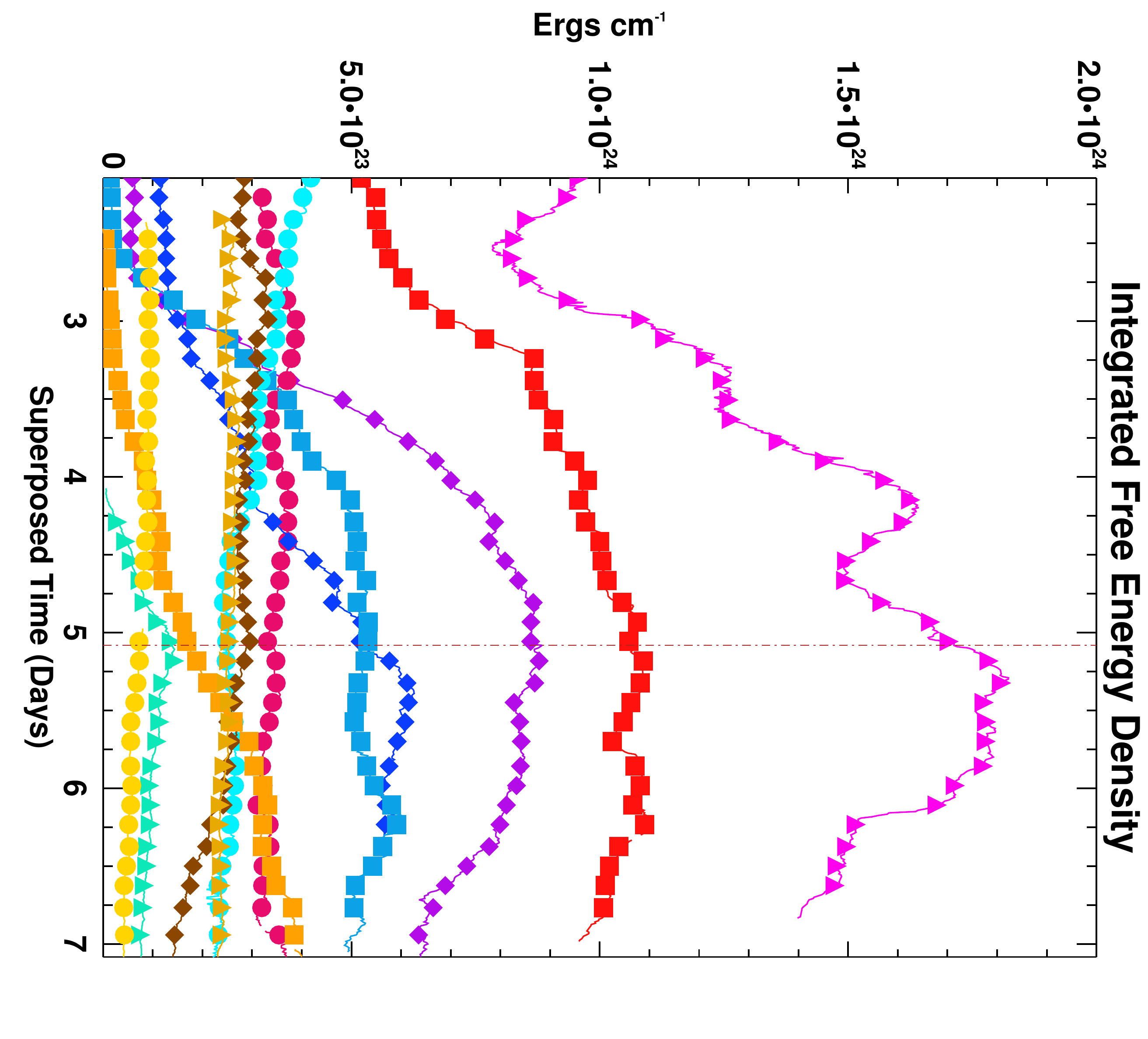} &
 \includegraphics[angle=90,width=0.496\textwidth]{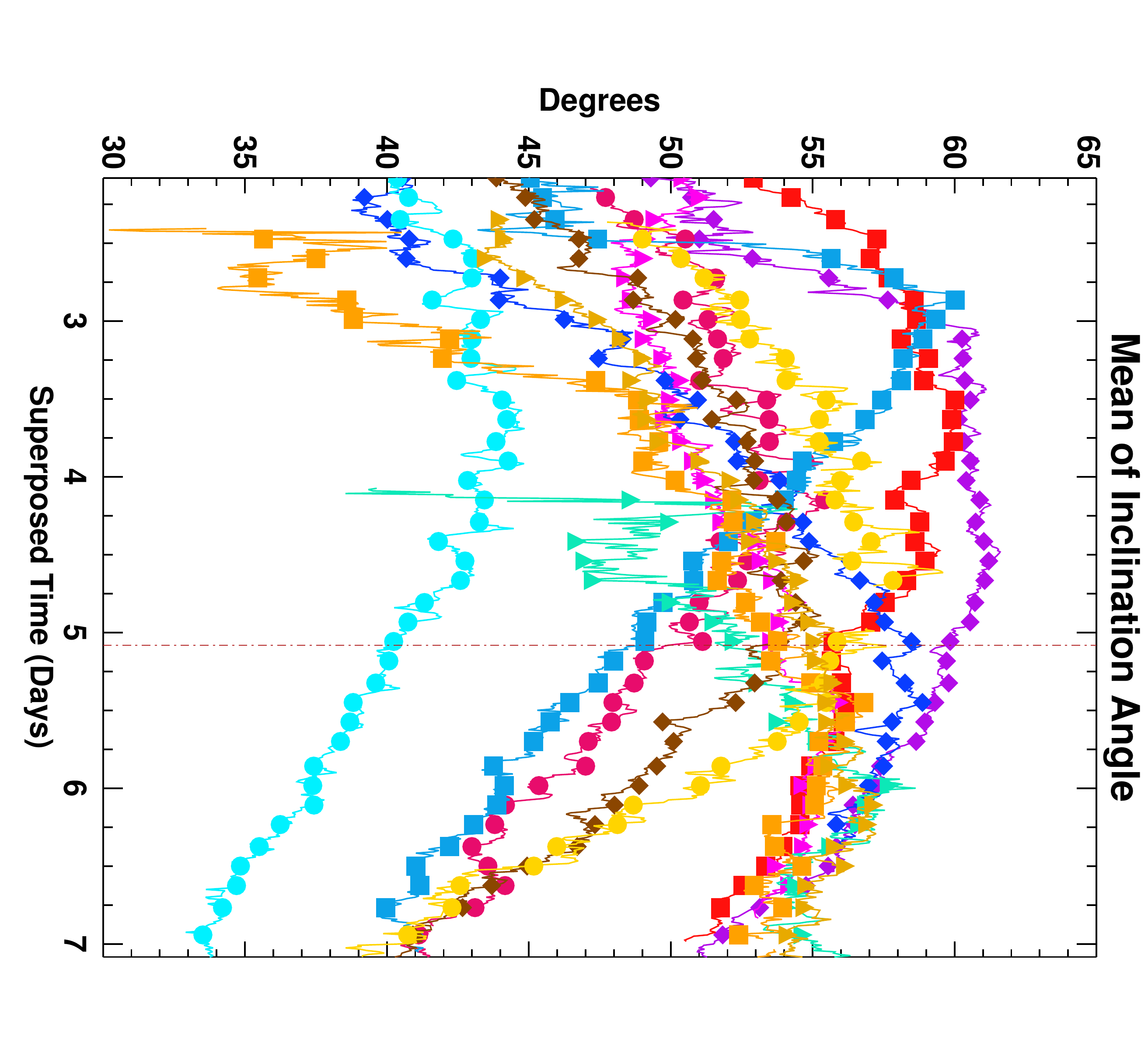} \\
\end{tabular}
\caption*{Clockwise from top left, temporal profiles of the total unsigned 
flux [{\sc usflux}], the modulus of net the current helicity [{\sc absnjzh}], the mean value 
of the inclination angle [{\sc meangam}], and the integrated total free-energy
density per active region [{\sc totpot}]. The entire sample is color coded: Active regions
associated with X-class flares are represented with red-purples, M-class by
blue-greens, and C-class by yellow-browns.
For clarity a larger symbol is plotted every three hours, \textit{i.e.} every 15th point.
The legend is in the top-left panel. The time profiles are adjusted to align the flare peaks a little after the start
of Day 5, as denoted by the red dotted--dashed line. Error bars are
plotted for all points; however, in most cases, they are smaller than the
point size. Scatter in the active-region parameters for NOAA AR\,11429 
for a few points following the flare peak is due to poor data quality
following an eclipse: thermal changes in the HMI front window affect
the focus. Periodicities in some of the parameters, most prominently in some
temporal profiles of unsigned flux, are systematic effects due to the daily variation of the
radial velocity of the spacecraft inherent to the geosynchronous orbit.}
\label{fig:arprofiles}
\end{figure}}

For illustrative purposes, Figure \ref{fig:arprofiles} shows the evolution
of a few {\sf SHARP} parameters for selected active regions 
associated with X-, M-, and C-class flares (Table \ref{tab:ar}).
A more complete analysis with comprehensive statistics is left for a future publication.
Region selection was based on the following criteria. i) To minimize the effects of the
increased noise in limb-ward data, we require that (a) the active region
must be within 45 degrees of central meridian during the GOES X-Ray flux
peak, and (b) for active regions that produced multiple flares, we chose
the flare that occurred while the region was closest to disk center. ii)
In some cases the identification and extraction algorithm \cite{Turmon2013}
identifies as one coherent magnetic structure -- \textit{i.e.}, one HARP -- a region
associated with multiple NOAA active regions. For simplicity such HARPs were
excluded from this sample. iii) We selected the largest flare class associated
with that active region (\textit{e.g.} a multi-flaring active region chosen for a
C-class flare would not be associated with an M- or X-class flare). From
that list we then arbitrarily selected four regions of each 
flare class to show as a demonstration of the presently available {\sf SHARP} parameters.

Figure \ref{fig:arprofiles} shows temporal profiles for each active region, 
color-coded by flare class, for the unsigned flux, the absolute value of the net 
current helicity, the mean of the absolute value of the inclination angle, 
and a proxy for the total free-energy density. 
These and other active region parameters appear
as keywords in the {\sf SHARP} data series 
and so can be displayed, retrieved, or used in a query with the JSOC 
data-handling tools without having to retrieve the image data.  A link to examples 
that can be used interactively with the JSOC {\sf lookdata} program can 
be found at the magnetic field portal (see Table \ref{tab:urls}).
The temporal profiles are adjusted to
align the flare occurrence time to a little after the start of Day 5, as indicated by
the red dotted-dashed line.  The {\sf SHARP} data can be used to create 
temporal profiles of the parameters for any active region since 1 May 2010.
Note that at the time of writing, the HMI analysis pipeline is running as fast as 
practical to close the remaining gap in {\sf SHARP} coverage by mid-2014.

We chose the four parameters in Figure \ref{fig:arprofiles} to suggest possible
uses of {\sf SHARP} indices for quickly and easily comparing regions of interest.
Magnetic flux has been well correlated with flaring activity
(\textit{e.g.} \opencite{barnesleka2008}; \opencite{komm2011}; \opencite{welsch}; 
and \opencite{g2012}), although the line-of-sight magnetic field data are
known to suffer from bias. Region 11429 was much greater in both total unsigned flux 
(upper left panel of Figure \ref{fig:arprofiles}) and in flare magnitude (Class X\,5.4).
Small flux regions showed little flare activity. It is easy to track the growth 
rate of total flux, \textit{e.g.} region 11620 grows rapidly during its disk transit.
Statistical studies of flare-related magnetic
field configurations, including the best determinations 
of the true total magnetic flux, have
been performed with vector magnetic data (\textit{e.g.} \opencite{lekabarnes2007},
\opencite{barnesleka2008}, \opencite{barnesleka2007}), albeit with the
recognized limitations of ground-based data sources, many of which are now
ameliorated with the SDO/HMI {\sf SHARP} series. Several studies use line-of-sight
magnetogram data to show that the photospheric magnetic field can store up
to 50\,\% of the total magnetic energy (\textit{e.g.} \opencite{ForbesPriest2002} and
references therein); however, this percentage may change when considering
the transverse component of the vector magnetic field. 
The integrated free-energy density {\sc totpot}, shown in the lower-left panel,
seems to increase significantly for most, but not all, of the large-flare regions; the exception was region 11283.
\inlinecite{fan} and \inlinecite{fang} suggest that some eruptive flares result in an imbalance
of magnetic torque at the photosphere; this may have implications for the
photospheric current helicity. 
Two of the largest regions, 11429 and 11158, had a large net current helicity and showed 
abrupt changes at the time of their X-class flares (upper right panel). C\,1.8-class region 11631 also had 
reasonably high net current helicity. A more comprehensive analysis is required to see if a significant 
relationship exists.
\inlinecite{hudson} noted that explosive
events should decrease coronal magnetic energy and thus lead the coronal
field to contract, increasing the inclination angle or the angle between
the vertical and horizontal photospheric field. Indeed, several studies
(\opencite{liudelta}; \opencite{Petrie2012}, \citeyear{Petrie2013}; \opencite{sun}; \opencite{wangpil})
show that the horizontal component of the magnetic field changes within select areas of an active 
region -- in particular, near the polarity inversion line.
However, the mean inclination angles shown in the lower-right panel give no indication of an obvious
systematic relationship to flare size or timing.
Such field changes may not be detectable in 
the large-scale {\sf SHARP} averages shown in Figure \ref{fig:arprofiles}.

We have implemented an interface to automatically submit {\sf SHARP} parameters, as
well as HARP geometry and location keywords, to the Heliophysics Events
Knowledgebase (HEK; \opencite{hurlburt2012}). The HEK is a web-based tool designed
to aid researchers in finding features and events of interest. Various
features extracted or extrapolated from HMI data, such as the location of
sunspots, polarity-inversion lines, and non-linear force-free numerical
models, are already available in the HEK (see Sections 13\,--\,15 of
\opencite{martens2012}).

The list of active-region parameters in the {\sf SHARP} data series is by no means exhaustive. We
plan to include additional parameters, including those that characterize
polarity-inversion lines and field morphologies of varying complexity.
Several studies show a relationship between
flaring activity and properties of the polarity-inversion line. For example,
\inlinecite{schrijverR} defined a parameter [$R$] that measures the flux contribution
surrounding polarity-inversion lines. After determining
$R$ for 289 active regions using line-of-sight magnetograms from the 
\textit{Solar and Heliospheric Observatory's Michelson Doppler Imager} 
(SOHO/MDI), he found that ``large flares, without exception, are associated with pronounced
high-gradient polarity-separation lines." \inlinecite{mason} developed
a similar parameter, called the Gradient-Weighted Inversion Line Length
(GWILL), applied it to 71\,000 MDI line-of-sight magnetograms of 1075 active
regions, and found that GWILL shows a 35\,\% increase during the 40 hours prior
to an X-class flare. \inlinecite{falconer56} devised a similar parameter
[WL$_\textit{sg}$] and computed it for 56 vector magnetic field measurements of
active regions. Using WL$_\textit{sg}$, they could predict CMEs with a 75\,\%
success rate.

Two additional approaches have been widely used to characterize active regions
in the context of energetic-event productivity. One is to model the coronal
magnetic field from the observed photospheric boundary and parametrize the results in
order to gauge the coronal magnetic field complexity and morphology. Examples
of relevant parameterizations include descriptions of the magnetic connectivity
({\it e.g.} $\phi_{ij}$ from \opencite{lekabarnes2006}, and $B_{\rm eff}$ from
\opencite{grust}), and topological descriptions (\opencite{lekabarnes2006};
\opencite{barnes2007}; \opencite{uu2007}; \opencite{cook2009}). The results 
are fairly convincing that parameters
based on models of the coronal magnetic field can add unique information
to what is otherwise available from characterizing the photosphere. 
Secondly, the fractal spectrum and related parameterizations
of the photospheric field provide additional measures of the magnetic
complexity, although the event-predictive capabilities of such measures require
additional research. While \inlinecite{mcateer2005} and \inlinecite{abra}
found a relation between fractal dimension and the range of multifractality
spectra and flare productivity, respectively, \inlinecite{g2012} found that
``both flaring and non-flaring active regions exhibit significant fractality,
multifractality, and non-Kolmogorov turbulence, but none of the three tested
parameters manages to distinguish active regions with major flares from the
flare-quiet ones." More study is required using these analysis approaches.
As the database of {\sf SHARP} active-region parameters grows, it will include
parameters derived from these and other relevant studies.


\section{Summary}
\label{s:conclusions}

The four {\sf SHARP} data series provide a systematic active-region database of patches of
photospheric vector magnetic field, Doppler velocity, continuum intensity,
and line-of-sight magnetic field extracted and tracked to mitigate
cumbersome handling of full-disk data.
At each 12-minute time step the {\sf SHARP} pipeline module automatically calculates sixteen indices 
that characterize active regions. The parameters have been chosen because they
are representative examples of the types of quantities linked to active-region flare 
productivity in the the literature.
These and other keywords can be used to identify and select regions of interest.
Definitive data are available a few weeks after regions complete their passage across the disk;
quick-look data for forecasting purposes are available within a few hours of being observed.
We compare temporal profiles of four {\sf SHARP} indices for 16 selected regions at the times 
of flares of various classes. We expect to add several more parameters to the database.
The {\sf SHARP} database can enable a more thorough investigation of these parameters as statistics accumulate.


\section*{Acknowledgments}
We thank the many team members who have contributed to the success of the SDO mission and
particularly to the HMI instrument. This work was supported by NASA Contract NAS5-02139 (HMI) to
Stanford University. Some of the research described here was carried out by staff of the Jet
Propulsion Laboratory, California Institute of Technology.
Efforts at NWRA were also supported through NASA Contracts NNH09CF22C and NNH12CG10C
and by NNG12PP28D/C\# GS-23F-0197P from NASA/Goddard Space Flight Center.
The authors thank Huned Botee for development of the {\sf SHARP} Data Viewer. 


\appendix

\section{SHARP Data Segment Descriptions}\label{s:Appendix}

The {\sf hmi.sharp\_720s} and {\sf hmi.sharp\_720s\_nrt} data series, 
which are in CCD coordinates, include 31 data arrays, or segments. 
Table \ref{tab:obsseg} describes the
segments associated with maps of the line-of-sight HMI observables. Table
\ref{tab:harpseg} describes the {\sc bitmap} segment associated with geometry of the
HARP data series. Table \ref{tab:invseg} describes the segments
associated with the vector-field inversion, including the vector
magnetic-field data as inclination, disambiguated azimuth, and field
strength. Segments that provide estimates of the uncertainties are listed
separately in Table \ref{tab:errors}.  Table \ref{tab:disseg} describes
two data segments associated specifically with the disambiguation module.
Table \ref{tab:infomap} describes three segments that
contains bits set by either the inversion or disambiguation module.

Table \ref{tab:ceaseg} describes the eleven map segments associated with the \textsf{hmi.sharp\_cea\_720s} and \linebreak \textsf{hmi.sharp\_cea\_720s\_nrt} data series.  The CEA data-series segments have been remapped to heliographic Cylindrical  Equal-Area coordinates centered on the patch. Several of the segments, such as plasma parameters from the inversion module, are not included in the CEA data series.

The keywords for the {\sf SHARP}-computed active region quantities and their associated 
uncertainties are described in Table \ref{tab:SpaceweatherFormulae}. 

Each of the {\sf SHARP} data series includes more than 300 keywords that provide information 
about HARP geometry, disk position, upstream processing, data statistics, \textit{etc.} 
A few are described in Table \ref{tab:otherkeywords}.
Additional documentation can be found on the JSOC wiki (see Table~\ref{tab:urls}).

\renewcommand*\thetable{\Alph{section}.\arabic{table}}
\setcounter{table}{0}

{\begin{table}
\caption{Line-of-Sight Observables}
\begin{flushleft}
\begin{tabular}{ l l p{10.2cm}}
Segment Name & Unit & Description \\\hline
{\sc magnetogram} & Mx\,cm$^{-2}$ & The {\sc magnetogram} segment contains HARP-sized line-of-sight magnetic field strength data from the series
{\sf hmi.M\_720s}.
 \\
{\sc dopplergram} & m\,s$^{-1}$ & The {\sc dopplergram} segment contains HARP-sized line-of-sight velocity data from the series {\sf hmi.V\_720s}.
 \\
{\sc continuum} & DN\,s$^{-1}$ & The {\sc continuum} segment contains HARP-sized computed continuum intensity data from the series {\sf hmi.Ic\_720s}. \\
\end{tabular}
\end{flushleft}
\label{tab:obsseg}
\end{table}}

{\begin{table}
\caption{HARP {\sc bitmap} Information}
\begin{flushleft}
\begin{tabular}{ l p{12.8cm}}
Name & Description \\\hline
{\sc bitmap} & The dimensionless {\sc bitmap} segment defines the bounding box and identifies which pixels 
are located within the HARP, 
and which are above the HARP noise threshold, by labeling each pixel with the following: 
\vspace{6pt}
\begin{description}
 \setlength{\itemsep}{0pt}
 \setlength{\parskip}{0pt}
\begin{itemize}
 \item[0] Off-disk.
 \item[1] Weak field, outside the HARP.
 \item[2] Strong field, outside the HARP.
 \item[33] Weak field, inside the HARP.
 \item[34] Strong field, inside the HARP.
\end{itemize}
\end{description}
\\
\end{tabular}
\end{flushleft}
\label{tab:harpseg}
\end{table}}

{\begin{table}
\caption{Inversion Module Outputs}
\begin{flushleft}
\begin{tabular}{ l l p{10.1cm}}
Segment Name & Unit & Description \\\hline
{\sc inclination} & Degree & The {\sc inclination} segment contains the magnetic field inclination with respect to the line-of-sight.
 \\
{\sc azimuth} & Degree & The {\sc azimuth} segment contains the magnetic field azimuth. Zero corresponds to the up direction of a column of pixels
on the HMI CCD; values increase counter-clockwise. The {\sc azimuth} in the {\sf SHARP} series has been disambiguated.
Keyword {\sc crota2} give the angle between up on the CCD and North on the Sun.  \\
{\sc field} & Mx\,cm$^{-2}$ & The {\sc field} segment contains the magnetic flux density. 
Currently, the filling factor is set equal to unity, so this quantity is also representative of the 
average magnetic field strength. 
The uncertainty (see {\sc field\_err} in Table \ref{tab:errors}) accounts for noise in both the line-of-sight and transverse field components.
Values of $\approx$220 Mx\,cm$^{-2}$ or less (2$\sigma$) are generally considered to be noise.
\\
{\sc vlos\_mag} & cm\,s$^{-1}$ & The {\sc vlos\_mag} segment contains the velocity of the 
plasma along the line-of-sight from the {\sf VFISV} inversion. Positive means redshift. 
[Note: These data are in cm\,s$^{-1}$, whereas the Dopplergram data are in m\,s$^{-1}$.]
\\
{\sc dop\_width} & m\AA\ & The {\sc dop\_width} segment contains the Doppler width of the spectral line, computed as if it were assumed to be a Gaussian.
\\
{\sc eta\_0} & & The {\sc eta\_0} segment contains the center-to-continuum absorption coefficient.
\\
{\sc damping} & m\AA\ & The {\sc damping} segment contains the electron dipole oscillation approximated as a simple harmonic oscillator. In the current version of
the {\sf VFISV} code, this parameter is constant and set to 0.5.
\\
{\sc src\_continuum} & DN\,s$^{-1}$ & The {\sc src\_continuum} segment contains the source function at the base of the photosphere. In the Milne--Eddington approximation, the source function varies linearly with optical depth.
\\
{\sc src\_grad} & DN\,s$^{-1}$ & The {\sc src\_grad} segment contains gradient of the source function with optical depth. By definition, {\sc src\_continuum} + {\sc src\_grad} = observed continuum intensity.
\\
{\sc alpha\_mag} & & The segment {\sc alpha\_mag} is defined as the portion of the resolution element that is filled with magnetized plasma. In the current version of
the {\sf VFISV} code, this parameter is constant and set to unity.
\\
\end{tabular}
\end{flushleft}
\caption*{In order to solve the inverse problem of inferring a vector
magnetic field from polarization profiles, the Very Fast Inversion of the
Stokes Vector ({\sf VFISV}) module solves a set of differential equations that
fit the parameters below. }
\label{tab:invseg}
\end{table}} 

{\begin{table}
\caption{Errors}
\begin{flushleft}
\begin{tabular}{ l p{10.6cm}}
Segment Name & Unit or Description \\\hline
{\sc inclination\_err} & Degree \\
{\sc azimuth\_err} & Degree \\
{\sc field\_err} & Mx\,cm$^{-2}$ \\
{\sc vlos\_err} & cm\,s$^{-1}$ \\
{\sc alpha\_err} & Filling factor error, currently set to unity by {\sf VFISV}. \\
{\sc field\_inclination\_err} & Cross correlation of errors in field strength and inclination.\\
{\sc field\_az\_err} & Cross correlation of errors in field strength and azimuth.\\
{\sc inclin\_azimuth\_err} & Cross correlation of errors in inclination and azimuth.\\
{\sc field\_alpha\_err} & Cross correlation of errors in field strength and filling factor (set to unity).\\
{\sc inclination\_alpha\_err} & Cross correlation of errors in inclination angle and filling factor (set to unity).\\
{\sc azimuth\_alpha\_err} & Cross correlation of errors in azimuth and filling factor (set to unity).\\
{\sc chisq} & A measure of how well the profiles are fit in the {\sf VFISV} least squares iteration. {\sc chisq} is not normalized. \\
\end{tabular}
\end{flushleft}
\caption*{The following segments contain formal computed standard deviations
and correlation coefficients of the uncertainties derived during the inversion that can be used to
determine the statistical errors of the vector magnetic field. The standard
deviations are the single-parameter quantities; the correlation coefficients are
the double-parameter entries. The calculated uncertainties and covariances
are only reliable if the {\sf VFISV} solution is close to an absolute minimum.
}
\label{tab:errors}
\end{table}} 
 
{\begin{table}
\caption{Disambiguation Module Segments}
\begin{flushleft}
\begin{tabular}{ l p{11.8cm}}
Segment Name & Description \\\hline
\sc{conf\_disambig} & The {\sc conf\_disambig} segment identifies the confidence assigned 
to the final disambiguation solution for each pixel. The confidence value nominally ranges
from 0\,--\,100 and depends on the field strength in the pixel compared to the estimated noise mask
or proximity to strong-field areas.
Currently only the values 90, 60, 50, and 0 are assigned.
For patch-wise disambiguated
{\sf SHARP}s only three values are assigned: 90, 60, or 0.
\vspace{4pt}
\begin{description}
 \setlength{\itemsep}{2pt}
 \setlength{\parskip}{2pt}
\item[90] Highest Confidence: Clusters of pixels with transverse field strength 
that exceeds the disambiguation noise threshold by {\sc doffset}.
\item[60] Intermediate Confidence: Pixels adjacent to strong-field regions. 
For patch-wise disambiguated {\sf SHARP}s (those based on {\sf hmi.Bharp\_720s} and all nrt HARPs), all 
pixels in the {\sf SHARP} that do not exceed the noise threshold are considered intermediate.  
For full-disk disambiguation ({\sf SHARP}s processed beginning 15 January 2014 that use {\sf hmi.B\_720s}), 
pixels within {\sc ambnpad}=5 of a strong-field pixel.
\item[50] Lower Confidence: In full-disk disambiguation only, the weak-field pixels
not within {\sc ambnpad}=5 pixels of a strong-field pixel in either $x$ or $y$.
\item[ 0] Not disambiguated, \textit{e.g.} off-disk pixels.
\end{description}
\\
\setlength{\parskip}{6pt}
\sc{disambig} & The {\sc disambig} segment encodes information about the
results of the disambiguation calculation in three bits. 
Each bit represents a different disambiguation solution in weak and some intermediate 
confidence pixels, as described below.
The three bits are identical for high-confidence pixels and for intermediate-confidence patch-wise 
disambiguations computed after August 2013.  A bit is set when $180^\circ$ needs to be
added to the {\sc azimuth} returned by the {\sf VFISV} {\tt fd10} inversion module. 
The {\sf SHARP} module has added $180^\circ$ to the reported azimuth value according to the
rules described below.

\vspace{6pt}
For all high-confidence pixels ({\sc conf\_disambig}$=90$)
the HMI pipeline determines the azimuth disambiguation using the minimum energy method and
records the result in Bit 0.

\vspace{6pt}
For the intermediate confidence pixels ({\sc conf\_disambig}$=60$), the minimum energy
disambiguation is determined and spatial smoothing is applied to the result and stored in Bit 0.

\vspace{6pt}
The {\sf SHARP}s use the results in Bit~0 to adjust the value in the {\sc azimuth} map segment for
high and intermediate confidence pixels.

\vspace{6pt}
For lower confidence pixels ({\sc conf\_disamb=50}) the results of three solutions are provided.
\begin{description}
 \setlength{\itemsep}{0pt}
  \setlength{\parskip}{0pt}
\item[Bit 0 (lowest bit)] gives the result of a potential field model solution.
\item[Bit 1 (middle bit)] assigns a random disambiguation for the pixel.
\item[Bit 2 (higher bit)] gives the radial-acute angle solution.
\end{description}

The results for the radial-acute angle solution (Bit 2) are used in the {\sf SHARP}s for lower confidence pixels.

\vspace{6pt}
In some cases for intermediate-confidence patch-wise {\sf SHARP}s disambiguated before August 2013 
Bits 1 and 2 of {\sc disambig} will include the results of the random or radial-acute angle solution.
Use of these bits for intermediate-condidence pixels is deprecated.
\\
\end{tabular}
\end{flushleft}
\label{tab:disseg}
\end{table}} 

{\begin{table}
\caption{Per pixel information about the status/quality of inversion or disambiguation processing}
\begin{flushleft}
\begin{tabular}{ l p{11.8cm}}
Segment Name & Description \\\hline
{\sc conv\_flag} &
These values are set by the {\sf VFISV} Code.
\vspace{4pt}
\begin{description}
 \setlength{\itemsep}{0pt}
  \setlength{\parskip}{0pt}
 \item[0] Reached convergence criteria (${\chi}^{2}_{old} - {\chi}^{2}_{new} < \epsilon$) 
 \item[1] Continuum intensity not above required threshold. Pixel not inverted.
 \item[2] Reached maximum number of iterations before convergence.
 \item[3] Reached maximum number of iterations and finished with too many consecutive non-improving
iterations (not used by the current {\tt fd10} code).
 \item[4] Not-a-Number in the computation of ${\chi}^{2}$. 
 \item[5] Not-a-Number in Singular Value Decomposition of Hessian matrix.
\end{description}
\\
{\sc confid\_map} &
The {\sc confid\_map} segment identifies the confidence index of the inversion output. 
The index value at each pixel will take the integer value from 0 (best) to 6 (worst), 
defined as the highest item number satisfying the following conditions: 
\begin{description}
 \setlength{\itemsep}{0pt}
  \setlength{\parskip}{0pt}
\item[0] No issue found in the input Stokes. 
\item[1] Signals for the transverse field component in the input Stokes parameters [$Q$ and $U$] were weak. 
\item[2] Signal for the line-of-sight field component in the input Stokes parameters [$V$] was weak. 
\item[3] Magnetic field signals of both LoS and transverse component were weak. 
\item[4] The ME-{\sf VFISV} inversion did not converge within the iteration maximum of 200. 
\item[5] If the difference between the absolute value of the line-of-sight 
field strength derived from magnetogram algorithm and the absolute value 
of the LoS component from the {\sf VFISV} inversion $|B$ cos(inclination)$|$ is greater than 500 Gauss, 
we expect the inversion did not solve the problem correctly. 
\item[6] One (or more) of the 24 input Stokes arrays had NaN value.
\end{description}
\\
{\sc info\_map} & The dimensionless {\sc info\_map} segment identifies the quality index of the inversion output at each pixel. 
The 16 bits in the top 4 hex digits are set by the inversion module, 
while the 16 bottom bits are updated during the disambiguation step. 
The meaning of the bits is defined as follows (a star indicates an arbitrary number):
\begin{description}
 \setlength{\itemsep}{0pt}
 \setlength{\parskip}{0pt}
\item Set by Disambiguation \vskip6pt
 \item[0x****0000] Not disambiguated.
 \item[0x****0001] Weak field, not annealed (only for full disk, filled with potential field, radial acute, or random solution).
 \item[0x****0003] Weak field, annealed.
 \item[0x****0007] Strong field, annealed.

\item \vskip6pt Set by {\sf VFISV} Inversion \vskip6pt
 \item[0x0000**** ] Pixel with no recorded inversion issue.
 \item[0x0\{0-5\}00****] Same as the three bits of the convergence index in {\sc conv\_flag}.
 \item[0x0800****] Bad pixel, defined using the same criteria as 5 of {\sc confid\_map}.
\item \vskip6pt The following bits do not necessarily indicate errors: \vskip6pt
 \item[0x1000****] Low Q or U signal: $\sqrt{(Q_0 + \ldots + Q_5)^2 + (U_0 + \ldots + U_5)^2}$ was smaller than $0.206 \sqrt{I_0
+ \ldots + I_5 }$ (the nominal photon noise level).
 \item[0x2000****] Low V signal: $|V_0| + |V_1| + \ldots + |V_5|$ was smaller than $0.206 \sqrt{I_0 + \ldots + I_5 }$.
 \item[0x4000****] Low $B_{\rm LoS}$ value: $|B_{\rm LoS}|$ from magnetogram algorithm was smaller than 6.2 Gauss (the nominal noise level).

 \item[0x8000****] Missing data. 
\end{description}
\\
\end{tabular}
\end{flushleft}
\label{tab:infomap}
\end{table}} 

{\begin{table}
\caption{Segments Unique to the CEA Series}
\begin{flushleft}
\begin{tabular}{ l l p{10.1cm}}
Segment Name & Unit & Description \\\hline
\sc{bp} & Mx\,cm$^{-2}$ &
$\phi$ (westward) component of the CEA vector magnetic field in the direction of solar rotation.
\\
\sc{bt} & Mx\,cm$^{-2}$ &
$\theta$ (southward) component of the CEA vector magnetic field.
 \\
\sc{br} & Mx\,cm$^{-2}$ &
Radial (out of photosphere) component of the CEA vector magnetic field.
 \\
\sc{bp\_err} & Mx\,cm$^{-2}$ &
Computed uncertainty (standard deviation) of the $\phi$ component of the CEA vector magnetic field. CEA uncertainties are determined at the nearest CCD pixel in the original computation.
 \\
\sc{bt\_err} & Mx\,cm$^{-2}$ &
Computed uncertainty (standard deviation) of the $\theta$ component of the CEA vector magnetic field.
 \\
\sc{br\_err} & Mx\,cm$^{-2}$ &
Computed uncertainty (standard deviation) of the radial component of the CEA vector magnetic field.
 \\
\sc{magnetogram} & Mx\,cm$^{-2}$ &
The {\sc magnetogram} segment contains HARP-sized line-of-sight magnetogram data from the series {\sf hmi.M\_720s}. The field is
remapped, but not transformed, \textit{i.e.} it is still the line-of-sight component relative to HMI.
 \\
\sc{dopplergram} & m\,s$^{-1}$ &
The {\sc dopplergram} segment contains HARP-sized Dopplergram data from the series {\sf hmi.V\_720s}. The Doppler velocity is
remapped, but not transformed, \textit{i.e.} it is still the line-of-sight component relative to HMI.
 \\
\sc{continuum} & DN\,s$^{-1}$ &
The {\sc continuum} segment contains HARP-sized computed continuum intensity from the series {\sf hmi.Ic\_720s}.
 \\
\sc{conf\_disambig} & &
The {\sc conf\_disambig} segment identifies the final disambiguation solution for each pixel with a value which maps to a
confidence level in the result (roughly a probability). 
The CEA value is the same as the value of the nearest un-remapped CCD pixel.
 \\
{\sc bitmap} & &
The {\sc bitmap} segment identifies the pixels located within the HARP.
The CEA value is the same as the value of the nearest un-remapped CCD pixel.
 \\
\end{tabular}
\end{flushleft}
\caption*{Map segments in the CEA {\sf SHARP}s.
The CEA magnetic-field values are represented differently, as
spherical vector field components $B_r$, $B_\theta$, and $B_\phi$ at each remapped
grid point. Statistical uncertainties are given for each field component,
but no cross-correlations are provided. The errors in $B_r$, $B_\theta$,
and $B_\phi$ at each remapped pixel are calculated from the variances of the
inverted magnetic field -- $B_{\rm Total}$, inclination, and azimuth --
and the covariances between them. The nearest-neighbor method is used to
get the values of the variances and covariances at the original CCD pixel
nearest the final remapped pixel. These values are then propagated to derive
the errors for $B_r$, $B_\theta$, and $B_\phi$. If nothing is specified in
the unit column, the quantity is dimensionless.}
\label{tab:ceaseg}
\end{table}} 

{\begin{table}
\caption{{\sf SHARP} Module Keywords}
\begin{flushleft}
\begin{tabular}{ l p{12.4cm}}
Keyword & Description \\\hline
{\sc harpnum} & The identifying number of the {\sf SHARP}, one of two prime keywords. \\
{\sc t\_rec} & The center time of the observation, the other prime keyword. \\
{\sc quality} & A specific bit in {\sc quality} is set when a specific problem exists for this observation. 
\textit{E.g.} bit 0x0100 is set during an eclipse. 
See the jsoc wiki entry \href{http://jsoc.stanford.edu/jsocwiki/Lev1qualBits}{{\sf Lev1qualBits}} referenced in Table \ref{tab:urls} for details. \\
{\sc date} & The time at which the {\sf SHARP} module was run. \\
{\sc codever7} & The software version number of the {\sf SHARP} code.
Code version numbers are given for several modules in other keywords. \\
{\sc wcsname} & World Coordinate System (WCS) coordinate system name. 
A number of keywords not listed in this table provide information about the coordinates \cite{thompson}. \\
{\sc dsun\_obs} & The distance from HMI to the Sun center in meters. \\
{\sc obs\_vr} & The radial velocity of HMI away from the Sun in m\,s$^{-1}$. \\
{\sc h\_merge} & Indicates if two NRT HARPs were merged at this time step. \\
{\sc omega\_dt} & Rotation rate of the region in degrees per day. \\
{\sc npix} & The number of CCD pixels in the patch. \\
{\sc cmask} & The number of pixels that contribute to the calculation of the {\sf SHARP} indices. \\
{\sc area} & The de-projected area of the patch in micro-hemispheres. \\
{\sc nacr} & The number of strong LoS magnetic field pixels in the patch. \\
{\sc mtot} & The sum of the absolute values of the LoS magnetic field in the patch. \\
{\sc mnet} & The sum of the LoS magnetic field in the patch. \\
{\sc t\_first} & The first {\sc t\_rec} of this {\sc harpnum} \\
{\sc t\_last} & The final {\sc t\_rec} of this {\sc harpnum} \\
{\sc lon\_fwt} & The Stonyhurst longitude of the LoS flux-weighted center of the patch. \\
{\sc lat\_fwt} & The Stonyhurst latitude of the LoS flux-weighted center of the patch. \\
{\sc noaa\_ar} & The NOAA Active Region first associated with the patch, if any. \\
{\sc noaa\_num} & The number of NOAA Active Regions associated with the patch. \\
{\sc noaa\_ars} & List of the NOAA Active Regions associated with the patch. \\
\end{tabular}
\end{flushleft}
\caption*{Definitions of selected {\sf SHARP} keywords. 
See references in Table \ref{tab:urls} for links to more information.}
\label{tab:otherkeywords}
\end{table}
}
\clearpage

\newcommand{\adv}{    {\it Adv. Space Res.}} 
\newcommand{\annG}{   {\it Ann. Geophys.}} 
\newcommand{\aap}{    {\it Astron. Astrophys.}}
\newcommand{\aaps}{   {\it Astron. Astrophys. Suppl.}}
\newcommand{\aapr}{   {\it Astron. Astrophys. Rev.}}
\newcommand{\ag}{     {\it Ann. Geophys.}}
\newcommand{\aj}{     {\it Astron. J.}} 
\newcommand{\apj}{    {\it Astrophys. J.}}
\newcommand{\apjl}{   {\it Astrophys. J. Lett.}}
\newcommand{\apss}{   {\it Astrophys. Space Sci.}} 
\newcommand{\cjaa}{   {\it Chin. J. Astron. Astrophys.}} 
\newcommand{\gafd}{   {\it Geophys. Astrophys. Fluid Dyn.}}
\newcommand{\grl}{    {\it Geophys. Res. Lett.}}
\newcommand{\ijga}{   {\it Int. J. Geomagn. Aeron.}}
\newcommand{\jastp}{  {\it J. Atmos. Solar-Terr. Phys.}} 
\newcommand{\jgr}{    {\it J. Geophys. Res.}}
\newcommand{\mnras}{  {\it Mon. Not. Roy. Astron. Soc.}}
\newcommand{\nat}{    {\it Nature}}
\newcommand{\pasp}{   {\it Pub. Astron. Soc. Pac.}}
\newcommand{\pasj}{   {\it Pub. Astron. Soc. Japan}}
\newcommand{\pre}{    {\it Phys. Rev. E}}
\newcommand{\solphys}{{\it Solar Phys.}}
\newcommand{\sovast}{ {\it Soviet  Astron.}} 
\newcommand{\ssr}{    {\it Space Sci. Rev.}} 

\bibliographystyle{spr-mp-sola}
\bibliography{sw}  

\end{article} 
\end{document}